\documentclass[useAMS,usenatbib]{mn2e}

\pdfoutput=1

\usepackage{ifpdf}
\usepackage{amssymb,amsmath}
\usepackage{graphics}
\usepackage{subfigure}
\usepackage[dvips]{graphicx}
\usepackage{epstopdf}
\usepackage{textcomp}
\usepackage{hyperref}
\usepackage[english]{babel}
\usepackage{color}
\usepackage{verbatim}
\usepackage[normalem]{ulem}

\title{Star formation and molecular hydrogen in dwarf galaxies: a non-equilibrium view}

\author[Hu et al.]
{Chia-Yu Hu$^{1}$, Thorsten Naab$^{1}$, Stefanie Walch$^{2}$, Simon C. O. Glover$^{3}$, Paul C. Clark$^{4}$ \\
$^{1}$Max-Planck-Institut f\"ur Astrophysik,
Karl-Schwarzschild Strasse 1, D-85740 Garching, Germany\\
$^{2}$Physikalisches Institut der Universit\"at zu K\"oln, 
Z\"ulpicher Strasse 77, D-50937 K\"oln, Germany\\
$^{3}$Zentrum f\"ur Astronomie der Universit\"at Heidelberg, 
Institut f\"ur Theoretische Astrophysik, Albert-Ueberle-Str. 2, 69120 Heidelberg, Germany\\
$^{4}$School of Physics and Astronomy, Cardiff University, 5 The Parade, 
Cardiff CF24 3AA, Wales, UK\\
}

\begin{document}
\maketitle

\begin{abstract}
	
We study the connection of star formation to atomic (HI) and molecular  
hydrogen (H$_2$) in isolated, low metallicity dwarf galaxies with  
high-resolution ($m_{\rm gas}$ = 4 M$_\odot$, $N_{\rm ngb}$ = 100) SPH simulations. The  
model includes self-gravity, non-equilibrium cooling, shielding from  
a uniform and constant
interstellar radiation field, the chemistry of H$_2$ formation,  
H$_2$-independent star formation, supernova feedback and metal  
enrichment. We find that the H$_2$ mass fraction is sensitive to the  
adopted dust-to-gas ratio and the strength of the interstellar  
radiation field, while the star formation rate is not. Star formation  
is regulated by stellar feedback, keeping the gas out of thermal  
equilibrium for densities $n <$ 1 cm$^{-3}$.
Because of the long chemical timescales, the H$_2$ mass remains out of  
chemical equilibrium throughout the simulation. Star formation is  
well-correlated with cold ( T $\leqslant$ 100 K ) gas, but this dense and cold  
gas - the reservoir for star formation - is dominated by HI, not H$_2$.  
In addition, a significant fraction of H$_2$ resides in a diffuse, warm  
phase, which is not star-forming. The ISM is dominated by warm gas  
(100 K $<$ T $\leqslant 3\times 10^4$ K) both in mass and in volume. The scale height of  
the gaseous disc increases with radius while the cold gas is always  
confined to a thin layer in the mid-plane. The cold gas fraction is  
regulated by feedback at small radii and by the assumed radiation  
field at large radii. The decreasing cold gas fractions result in a  
rapid increase in depletion time (up to 100 Gyrs) for total gas  
surface densities $\Sigma_{\rm HI+H_2} \lesssim$ 10 M$_\odot$pc$^{-2}$, in  
agreement with observations of dwarf galaxies in the Kennicutt-Schmidt  
plane.

\end{abstract}

\begin{keywords}
methods: numerical,  galaxies: ISM, galaxies: evolution
\vspace{-1.0cm}
\end{keywords}

\section{Introduction}

Dwarf galaxies are thought to be the building blocks of larger galaxies in the hierarchical picture of galaxy formation.
While they contribute little to the total mass budget of the galaxy population,
they are the most numerous type of galaxy in the local Universe.
Dwarf galaxies tend to contain fewer heavy elements compared to more massive ones 
\citep{2004ApJ...613..898T,2012AJ....144..134H}.
As such, they are ideal laboratories for studying physical processes of the interstellar medium (ISM) under chemically simple conditions,
which can be quite different from those found in normal spiral galaxies such as the Milky Way.

On kpc-scales, spatially resolved observations of nearby star-forming spiral galaxies have demonstrated that the surface density of the star formation rate has a clear correlation with the molecular hydrogen (H$_2$) surface density and little correlation with the atomic hydrogen (HI) surface density
\citep{2008AJ....136.2846B,2011ApJ...730L..13B}.
In light of the observational evidence,
H$_2$-dependent sub-resolution recipes for star formation have been widely adopted in hydrodynamical simulations as well as semi-analytic models,
where the H$_2$ abundances are calculated either by directly incorporating non-equilibrium chemistry models (e.g. \citealp{2009ApJ...697...55G, 2012MNRAS.425.3058C}) or by analytical ansatz assuming chemical equilibrium (e.g. \citealp{2010MNRAS.409..515F,2012ApJ...749...36K, 2014ApJ...780..145T, 2014MNRAS.445..581H,2015arXiv150300755S}).
The implicit assumption is that star formation only takes place in H$_2$-dominated clouds,
and the HI-to-H$_2$ transition has been thought to be responsible for the so-called star formation threshold, the surface density of gas below which star formation becomes extremely inefficient.

However,
recent theoretical studies have cast doubt on the causal connection between H$_2$ and star formation
\citep{2011ApJ...731...25K, 2012MNRAS.421....9G}, based on the insensitivity of the gas thermal properties to the H$_2$ abundances.
Radiative cooling from H$_2$ is negligible when the gas temperature drops below a few hundred Kelvin.
The formation of carbon monoxide (CO), which is indeed an efficient coolant at very low temperatures, does require the presence of H$_2$.
Nevertheless,
singly ionized carbon (C$^{+}$) provides cooling that is almost 
equally efficient as CO 
at all but the lowest temperatures through fine structure line emission.
It is therefore possible for gas devoid of both H$_2$ and CO to cool down to low enough temperatures and form stars by gravitational collapse.
In this picture,
the correlation between H$_2$ and star formation originates from the fact that both H$_2$ formation and star formation take place in regions well-shielded from the interstellar radiation field (ISRF) instead of the former being a necessary condition for the latter.
Such a correlation is expected to break down in low-metallicity environments where 
the H$_2$ formation timescales are much longer.
Indeed,
\citet{2012ApJ...759....9K} predicts that star formation would proceed in HI-dominated clouds if the gas metallicity is below a few percent of solar metallicity (Z$_{\odot}$) based on timescale arguments.
\citet{2012MNRAS.426..377G} also examine this issue, using simulations of isolated clouds, 
and show that the star formation rate of the clouds is insensitive to their molecular
content and that for metallicities ${\rm Z} \sim 0.1 \: {\rm Z_{\odot}}$ and
below, the star-forming clouds are dominated by atomic gas.

Observationally,
the detection of H$_2$ in star-forming dwarf galaxies has proven to be very challenging.
CO emission, as the standard approach to derive the H$_2$ abundance, tends to be very faint in these galaxies.
\citet{2012AJ....143..138S} observed 16 nearby star-forming dwarf galaxies and
only detected CO successfully in five galaxies with oxygen abundance 12 + log$_{10}$ (O/H) $\gtrsim$ 8.0 with very low CO luminosities per star formation rate compared to those found in massive spiral galaxies.
CO was not detected in the other 11 galaxies with 12 + log$_{10}$ (O/H) $\lesssim$ 8.0
even if stacking techniques were used.
The interpretation was that the CO-to-H$_2$ conversion factor is significantly higher in low metallicity environment, 
under the assumption that there should be much more H$_2$ given their star formation rate (i.e., assuming constant H$_2$ depletion time $\approx$ 2 Gyr).
The dependence of the CO-to-H$_2$ conversion factor on metallicity is already observed in Local Group galaxies  using dust modeling to estimate the H$_2$ mass \citep{2011ApJ...737...12L} 
and has considerable theoretical support \citep{2010ApJ...716.1191W, 2011MNRAS.412..337G, 2011MNRAS.412.1686S, 2011MNRAS.415.3253S, 2013ARA&A..51..207B}.
However, an alternative interpretation is that H$_{2}$ is also rare and star formation in these
systems is taking place in regions dominated by atomic hydrogen
\textbf{
(see \citealp{2015A&A...582A..78M} who adopted such an interpretation for their observed galaxies).
}

In this paper we conduct numerical simulations to study the ISM properties in an isolated star-forming dwarf galaxy.
Our model includes gravity, hydrodynamics, non-equilibrium cooling and chemistry, shielding from the ISRF, an H$_2$-independent star formation recipe, stellar feedback and metal enrichment in a self-consistent manner.
We investigate the relationship between H$_2$ and star formation and explore the effects of varying the strength of ISRF and the dust-to-gas mass ratio (DGR).
In Section \ref{sec:method} we describe the details of our numerical method.
In Section \ref{sec:ISM_in_eq} we show the ISM properties when it is in thermal and chemical equilibrium.
In Section \ref{sec:sim} we present the results of our numerical simulations.
In Section \ref{sec:discussion} we discuss the implications of our results and the potential caveats.
In Section \ref{sec:summary} we summarize our work.

\section{Numerical Method}\label{sec:method}
\subsection{Gravity and Hydrodynamics}
We use the {\sc gadget-3} code \citep{2005MNRAS.364.1105S} where the collisionless dynamics of gravity is solved by a tree-based method \citep{1986Natur.324..446B}, while the hydrodynamics is solved by the smoothed particle hydrodynamics (SPH) method \citep{1977AJ.....82.1013L, 1977MNRAS.181..375G}.
We have implemented a modified version of SPH, called SPHGal, in {\sc gadget-3} which shows a significantly improved numerical accuracy in several aspects \citep{2014MNRAS.443.1173H}.
More specifically,
we adopt the pressure-energy formulation of SPH\footnote{
\textbf{
We use the pressure-energy SPH instead of the pressure-entropy SPH as used in \citep{2014MNRAS.443.1173H} for reasons described in Appendix \ref{app:pesph}.}
}
\citep{2010MNRAS.405.1513R,2013ApJ...768...44S} which is able to properly follow fluid instabilities and mixing without developing severe numerical artifacts commonly found in traditional SPH (e.g. \citealp{2007MNRAS.380..963A}).
The so-called `grad-h' correction term is included following \citet{2013MNRAS.428.2840H} to ensure the conservation properties when the smoothing length varies significantly (e.g. at strong shocks).
The smoothing length is set such that there are $N_{\rm ngb}$ = 100 particles within a smoothing kernel. 
We use a Wendland $C^4$ kernel function that has been shown to be stable against pairing instability for large $N_{\rm ngb}$ \citep{2012MNRAS.425.1068D},
which is necessary for reducing the `E$_0$ error' \citep{2010MNRAS.405.1513R} and therefore improving the numerical convergence rate \citep{2012MNRAS.422.3037R,2012MNRAS.425.1068D}.
We adopt artificial viscosity to properly model shocks with an efficient switch that only operates at converging flows, similar to the prescriptions presented in \citet{1997JCoPh.136...41M} and \citet{2010MNRAS.408..669C}.
We also include artificial thermal conduction \citep{2008JCoPh.22710040P} but only in converging flows to smooth the thermal energy discontinuities, 
which can lead to severe noise at strong shocks when the pressure-energy formulation is used.
The viscosity and conduction coefficients are varied in the range of [0.1,1] and [0,1] respectively.
Our SPH scheme shows satisfactory behaviors and accuracies in various idealized numerical tests presented in \citet{2014MNRAS.443.1173H}.

\subsection{Chemistry Model}\label{sec:chem}
Our model of the chemistry and cooling follows closely the implementation in the SILCC project \citep{2015MNRAS.454..238W,2015arXiv150806646G},
based on earlier work by \citet{1997ApJ...482..796N}, \citet{2007ApJS..169..239G} and \citet{2012MNRAS.421..116G}.
We track six chemical species: H$_2$, H$^+$, CO, H, C$^+$, O and free electrons.
Only the first three species are followed explicitly, 
i.e., their abundances are directly integrated based on the rate equations in our chemistry network.
The fractional abundance of neutral hydrogen is given by 
\begin{equation}
	x_{\rm H^0} = 1 - 2 x_{\rm H_2} - x_{\rm H^+},
\end{equation}
where $x_i$ denotes the fractional abundance of species $i$; note that all fractional abundances quoted here are relative to the number of H nuclei.
Silicon is assumed to be present in singly ionized form (i.e.\ as Si$^{+}$) throughout the simulation, while carbon and oxygen may either be present as C$^{+}$ and O, or in the form
of CO,
which leads to
\begin{eqnarray}
\begin{aligned}
	x_{\rm C^+} = x_{\rm C, tot} - x_{\rm CO}, \\
	x_{\rm O} = x_{\rm O, tot} - x_{\rm CO},
\end{aligned}
\end{eqnarray}
where $x_{\rm C, tot}$ and $x_{\rm O, tot}$ are the abundances of the total carbon and oxygen respectively.
Finally, the abundance of free electron is given by 
\begin{equation}
	x_{\rm e^{-}} = x_{\rm H^+} + x_{\rm C^+} + x_{\rm Si^+}.
\end{equation}

A list of our chemical reactions for H$_2$ and H$^+$ is summarized in Table 1 of \citet{2012MNRAS.421.2531M}.
The H$^+$ is formed via collisional ionization of hydrogen with free electrons and cosmic rays, and is 
depleted through electron recombination in both the gas phase and on the surfaces of dust grains.
The H$_2$ is formed on the surfaces of dust grains and is destroyed via ISRF photo-dissociation, cosmic ray ionization, and collisional dissociation (with H$_2$, H and free electrons).

Our chemical model also includes a treatment of carbon
chemistry, following the scheme introduced in \citet{1997ApJ...482..796N}.
It assumes that the rate-limiting step of CO formation is the process ${\rm C^+} + {\rm H_2} \rightarrow {\rm CH_2^+}$.
The CH$_2^+$ may either react with atomic oxygen and form CO or be destroyed via ISRF photo-dissociation.
However, we will not go into detailed investigation about CO in this work,
as a proper modeling for CO formation is beyond our resolution limit. 
This is especially so in low metallicity environments where CO only resides 
in regions of very high density \citep{2012MNRAS.426..377G}.

\subsection{Cooling/Heating Processes}\label{sec:cool}
We include a set of important non-equilibrium cooling and heating processes.
The term 'non-equilibrium' refers to the fact that the processes depend not only on the local density and temperature of the gas but also on its chemical abundances of species, which may not be in chemical equilibrium.
Cooling processes include fine-structure lines of C$^+$, O and Si$^+$, the rotational and vibrational lines of H$_2$ and CO, the hydrogen Lyman-alpha line, the collisional dissociation of H$_2$, the collisional ionization of H, and the recombination of H$^+$ in both the gas phase and on the surfaces of dust grains. Heating processes include photo-electric effects from dust grains and polycyclic aromatic hydrocarbons (PAHs), ionization by cosmic rays, the photo-dissociation of H$_2$, the UV pumping of H$_2$ and the formation of H$_2$.

We do not follow the non-equilibrium cooling and heating processes in high temperature regimes. 
For T $ > 3 \times 10^4$ K we adopt a cooling function presented in \citet{2009MNRAS.393...99W}, which assumes that the ISM is optically thin and is in ionization equilibrium with a cosmic UV background from \citet{2001cghr.confE..64H}.
The total cooling rate depends on the temperature and density of the gas as well as the abundance of heavy elements.
We trace eleven individual elements (H, He, C, N, O, Ne, Mg, Si, S, Ca and Fe) for both gas and star particles based on the implementation in \citet{2013MNRAS.434.3142A}.

\subsection{Shielding of the Interstellar Radiation Field}\label{sec:shielding}
Shielding from the ISRF affects both the chemistry and the cooling/heating processes.
For the hydrogen chemistry,
the H$_2$ ISRF photo-dissociation rate, $R_{\rm pd, H_2}$, is attenuated by both the dust shielding and the H$_2$ self-shielding:
\begin{equation}\label{eq:Ratepd}
	R_{\rm pd, H_2} = f_{\rm sh} R_{\rm pd, H_2, thin} ,
\end{equation}
where $R_{\rm pd, H_2, thin} = 3.3\times 10^{-11} G_0$ s$^{-1}$ is the unattenuated photo-dissociation rate,
$G_0$ is the strength of the ISRF relative to the Solar neighborhood value estimated by \citet{1968BAN....19..421H}.
The total attenuation factor is
$f_{\rm sh} = f_{\rm dust, H_2} f_{\rm self, H_2}$ where
$f_{\rm dust, H_2}$ and $f_{\rm self, H_2}$ are the attenuation factors
by dust extinction and by H$_2$ self-shielding, respectively.
We adopt 
\begin{equation}
	f_{\rm dust, H_2} = {\rm exp}(- D \sigma_{\rm dust} N_{\rm H,tot}),
\end{equation}
where $D$ is the DGR relative to the Milky Way value ($\sim$1 \%) and $\sigma_{\rm dust} = 2\times 10^{-21}$ cm$^2$ is the averaged cross section of dust extinction.
The H$_2$ self-shielding is related to the H$_2$ column density $N_{\rm H_2}$ using the relation given by \citet{1996ApJ...468..269D}.
A similar treatment is used for the carbon chemistry. 
The CO photo-dissociation rate is attenuated by the dust extinction, H$_2$ shielding and CO self-shielding.
A more detailed description can be found in \citet{2015MNRAS.454..238W}.

Dust extinction also reduces the photo-electric heating rate by blocking the radiation in the energy range between 6 and 13.6 eV.
We adopt the heating rate given by \citet{1994ApJ...427..822B} and \citet{2004ApJ...612..921B} of:
\begin{equation}\label{eq:photoheat}
	\Gamma_{\rm pe} = 1.3\times 10^{-24} \epsilon D G_{\rm eff}  n  ~{\rm erg~ s^{-1} cm^{-3}},
\end{equation}
where $G_{\rm eff} = G_0 {\rm exp}(-1.33\times 10^{-21} D N_{\rm H,tot})$ is the attenuated radiation strength, $n$ is the number density of the gas and $\epsilon$ is the heating efficiency defined as
\begin{equation}
	\epsilon = \frac{0.049}{1 + (0.004\psi^{0.73})} 
	+ \frac{0.037 (T/10000)^{0.7}}{1 + 2\times 10^{-4} \psi}
\end{equation}
where $\psi = G_{\rm eff} T^{0.5}/n_{\rm e^-}$ and $n_{\rm e^-}$ is the electron number density.

To calculate the column densities relevant for shielding we have incorporated the {\sc TreeCol} algorithm \citep{2012MNRAS.420..745C} into our version of {\sc Gadget-3}.
\textbf{
Unlike the extragalactic UV background that is external to the simulated galaxy,
the sources of the ISRF are the young stars embedded in the ISM.
This means that the column densities should not be integrated over the entire disc,
but have to be truncated at certain length scales.
Ideally, one should integrate the column densities up to individual stars that contribute to the 
local radiation field separately, but this would entail performing a computationally expensive 
radiative transfer calculation on every timestep, which is impractical. Instead, we make the 
simplifying assumption that the material beyond a predefined shielding
length $L_{\rm sh}$ from a local gas particle is not relevant for the shielding (see e.g. \citealp{2008MNRAS.389.1097D} or \citealp{2014MNRAS.441.1628S}, who make similar approximations in their galactic simulations). 
We justify this assumption by noting that for gas particles in dense clouds -- the only ones which are significantly shielded -- the dominant contribution to the shielding generally comes
from gas and dust in the cloud itself or in its immediate vicinity, rather than from the
diffuse gas between the clouds. Therefore, provided we adopt a value for $L_{\rm sh}$ that
is larger than the typical size of the dense clouds, our results should be insensitive
to the precise value chosen for $L_{\rm sh}$
\footnote{
\textbf{
Along a few lines of sight which intercept other dense clouds, taking a small $L_{\rm sh}$ causes us to potentially underestimate the shielding.
However, the local radiation field will always be dominated by other lines of sight with low column density.}
}.
}

For each gas particle, {\sc TreeCol} defines $N_{\rm pix}$ equal-area pixels using the
{\sc healpix} algorithm (Gorski \& Hivon 2011) and computes the column density
within each pixel out to $L_{\rm sh}$. We set $N_{\rm pix} = 12$  and $L_{\rm sh} = 50$~pc
throughout this work. 
\textbf{
However, we explore in Appendix \ref{app:Lsh} the effect of varying
$L_{\rm sh}$ within the range 20--200~pc and show that as expected our results are 
insensitive to the precise value of $L_{\rm sh}$ within this range.
}

\subsection{Star Formation Model}\label{sec:sfrmodel}
Unlike cloud-scale simulations, our mass resolution is not sufficient to follow the gravitational collapse of the gas to densities where it will inevitably end up in stars.
Instead, we define an instantaneous star formation rate for each gas particle to estimate how much gas is expected to be converted into stars:
${\rm SFR_{gas}} = \epsilon_{\rm sf}  m_{\rm{gas}} / t_{\rm ff}$,
where $t_{\rm ff} = (4\pi G \rho_{\rm gas})^{-0.5}$ is the local free-fall time, $\epsilon_{\rm sf}$ is the star formation efficiency, and $\rho_{\rm gas}$ is the volumetric density of gas.  
We set $\epsilon_{\rm sf}$ = 0.02 to account for the inefficient star formation which might originate from the sub-resolution turbulent motions \citep{2005ApJ...630..250K}.
We assume the gas is `star-forming' (${\rm SFR_{gas}} > 0$) only if the gas has $n_{\rm H} \geq n_{\rm H,th}$, $T \leq T_{\rm th}$ and negative velocity divergence.
We choose $n_{\rm H,th}$ = 100 cm$^{-3}$ 
\textbf{
as this is the typical densities of the giant molecular clouds in our Galaxy, the reservoir gas for star formation for which $\epsilon_{\rm sf}$ = 0.02 is defined.
We also set $T_{\rm th}$ = 100 K to ensure that we do not attempt to form stars in hot dense gas which has a high Jeans mass. In practice, most gas with $n_{\rm H} > n_{\rm H,th}$ would actually be colder than 100 K (cf. Fig. \ref{fig:PD_plus_Coolrates}).
}
Our definition of `star-forming gas' is very simplistic.
In reality, stars may still form out of a gas cloud with $n_{\rm H} < n_{\rm H,th}$ if the sub-resolution density structure is very clumpy.
\textbf{
Note that the appropriate choice of $n_{\rm H,th}$ and $\epsilon$ also depends on resolution.
With higher resolutions,
one should be able to follow the gravitational collapse to smaller scales and denser environments (e.g. individual molecular cores),
and therefore both $n_{\rm H,th}$ and $\epsilon$ should be set higher.
}

We adopt the commonly used stochastic approach of star formation:
in each timestep of size $\Delta$t,
a star-forming gas particle has a probability of $\epsilon_{\rm sf} \Delta t / t_{\rm ff}$ to be converted into a star particle of the same mass.
Since $\Delta t \ll t_{\rm ff} / \epsilon_{\rm sf}$ almost always holds,
the conversion test in each timestep is a rare event Bernoulli trial and the number of 'success' events during a time period of $t_{\rm ff} / \epsilon_{\rm sf}$ for a given ${\rm SFR_{gas}}$ follows a Poisson distribution with the parameter $\lambda$ = 1 (which is the expectation value),
i.e., one gas particle would be converted into a star particle in a time period of $t_{\rm ff} / \epsilon_{\rm sf}$ on average.
Note that the ratio of the standard deviation to the expectation value for a Poisson distribution is $\lambda^{-0.5}$,
so the actual star formation timescale can deviate from the input by $\approx$ 100\% if we look at a single particle.
Only when we look at the averaged star formation rate of a group of particles would the random fluctuation be reduced to a satisfying level.

The instantaneous star formation rate of gas particles, however, is not an observable.
In fact, it is merely an estimate for the star formation that will happen in the next timesteps.
Therefore, we measure the star formation rate in a given region by the total mass of newly formed star particles with age less than $t_{\rm SF}$ in that region divided by $t_{\rm SF}$, which is set to be 5 Myr in this work.
Such a definition is more compatible with what is measured in observations than the instantaneous star formation rate assigned to the gas particles based on the adopted star formation model.
These two definitions of star formation rate give almost identical results if we sum over a large enough region (e.g. the total star formation rate of the galaxy),
although locally (both in space and in time) they can be quite different (cf. Section \ref{sec:KS}).
Throughout this work we will adopt this definition when we present the star formation rate (except for Fig. \ref{fig:KS_vary_pix_g1d01} where both definitions are shown).

\subsection{Stellar Feedback and Metal Enrichment}

\subsubsection{Supernova type II (SNII)}\label{sec:feedback}
We assume each star particle represents a stellar population of mass $m_{\rm star}$ and calculate the corresponding mass $\delta m_{\rm SNII}$ that will end up in SNII. 
For our adopted Kroupa initial mass function \citep{2001MNRAS.322..231K}, 
we have $\delta m_{\rm SNII} \simeq 0.12 m_{\rm star}$.
When the age of a star particle reaches 3 Myr, 
we return $\delta m_{\rm SNII}$ of mass to the ISM with enriched metal abundances according to the metallicity dependent yields given by \citet{1995ApJS..101..181W}.
The returned mass is added to the nearest 100 neighboring gas particles weighted by the smoothing kernel.
The energy budget for a $m_{\rm star}$ stellar population is $N_{\rm SNII} E_{51}$ where $E_{51} = 10^{51}$ erg is the typical energy for a single SN event and 
$N_{\rm SNII} = m_{\rm star} / 100 {\rm M}_\odot$
is the number of SN events.

Physically, a supernova remnant (SNR) should first go through a non-radiative phase where momentum is generated (the Sedov phase) until radiative cooling kicks in and the total momentum gradually saturates 
(e.g. \citealp{1988RvMP...60....1O, 1998ApJ...500..342B}).
However, numerically, the resolution requirement for modeling the correct evolution of a SNR is very demanding (see e.g. \citealp{2015MNRAS.451.2757W}).
It has long been recognized that insufficient resolution leads to numerical over-cooling: 
most of the injected energy is radiated away before it can significantly affect the ISM.
As shown in \citet{2012MNRAS.426..140D}, this occurs when the cooling time is much shorter than the response time of the gas for a given resolution.
For usual implementations in SPH simulations where $m_{\rm star} = m_{\rm gas}$,
this means that the returned mass $\delta m_{\rm SNII}$ is always much smaller than $m_{\rm gas}$ and the SNR would be poorly resolved.
The situation would not be alleviated by reducing $m_{\rm gas}$ as long as $m_{\rm star} = m_{\rm gas}$ is assumed.
\citet{2012MNRAS.426..140D} circumvent this issue by injecting the energy to fewer neighboring particles and, if necessary, stochastic injection which groups several SN events into a single energetic one.
By doing so they guarantee that the gas would always be heated to the local minimum of the cooling function and has more time to develop the blast wave.

Although grouping several SN events into one energetic explosion enhances the dynamical impact of SN feedback on the ISM,
it also coarsens the granularity (both spatial and temporary) of the SN events.
As shown in \citet{2015ApJ...802...99K},
a single energetic explosion over-produces both the momentum and energy compared to a series of spatially coherent explosions with the same amount of total energy.
The difference would probably be even more severe if the explosions occur at different locations.
Therefore, for a given energy budget, coarser SN sampling can over-estimate the impact of SN feedback.
With the SN sampling as a free parameter,
the effect of feedback becomes tunable or even arbitrary.
Indeed, in large-scale cosmological simulations with necessarily compromised resolutions,
the feedback has to be calibrated by fitting to observations \citep{2015MNRAS.446..521S} and thus can only be regarded as a phenomenological model.
However,
if one's goal is to directly resolve individual blast waves without tunable parameters,
as we try to do in this work,
then the SN sampling should be set to the physical one by making sure that each SN event has the canonical energy of $10^{51}$ erg.

As the star particle mass reaches $m_{\rm star} < 100$ M$_\odot$, the energy budget for a star particle becomes smaller than $E_{51}$ (i.e. $N_{\rm SNII} < 1$), which is also unphysical.
In this work we adopt an ansatz similar to the stochastic injection in {\citet{2012MNRAS.426..140D}.
We let star particles with the appropriate age explode as SNII with a probability of $m_{\rm gas}$/100 M$_{\odot}$ and with an energy of $E_{51}$.
In this work, $m_{\rm gas}$ = 4 M$_{\odot}$ and therefore each star particle
has a 4\% chance of producing a type II SN.
Our resolution is close to the $m_{\rm gas}$ = 1 M$_{\odot}$ requirement for a reasonably converged energy and momentum evolution of an SNR as shown in the resolution study in Appendix \ref{app:fbtest}.

Note that despite the fact that our particle mass (4 M$_\odot$) is
comparable to the mass of a single star,
the star particles should still be considered as stellar populations instead of individual stars.
A star particle, once formed, is only a representative point of the collisionless distribution function that describes the stellar system of the galaxy.
With this interpretation,
there is no conceptual issue even when the particle mass reaches sub-solar scales,
though the system might be over-resolved.
The collisional dynamics of star clusters is suppressed by our gravitational softening (2 pc), and thus could only be included in a sub-resolution model,
which is not considered in this work.

\subsubsection{Supernova type Ia (SNIa) and asymptotic-giant-branch (AGB) stars}
We include feedback by SNIa and AGB stars based on the implementation presented in \citet{2013MNRAS.434.3142A}.
For SNIa,
we adopt a delay time distribution (DTD),
the SN rate as a function of time for a given stellar population formed in a single burst.
The DTD has a power-law shape $\sim t^{-1}$ where $t$ is the stellar age,
with the normalization of 2 SNIa events per 1000 M$_\odot$ \citep{2012PASA...29..447M}.
The amount of mass returned to the ISM is calculated by sampling the DTD with a 50 Myr time bin,
and the metal yields based on \citet{1999ApJS..125..439I}.
Similarly,
the mass returned by the AGB stars is calculated from the metal yields presented in \citet{2010MNRAS.403.1413K} with the same time bin as SNIa.
Assuming an outflow velocity $v$ = 3000 km/s and 10 km/s for SNIa and AGB stars, respectively,
we return energy of 0.5 $\delta m v^2$ into the ISM where $\delta m$ is the returned mass in a given time bin,
though in our simulations their effect is sub-dominant compared to the SNII feedback.

\subsection{Timestep Limiter}
We include a timestep limiter similar to \citet{2009ApJ...697L..99S} and \citet{2012MNRAS.419..465D} to correctly model the strong shocks.
For each active particle $i$,
we identify any neighboring particles within its smoothing kernel whose timesteps are four times longer than the timestep of particle $i$ and force them to be active at the next synchronization point.
In addition,
we re-calculate the hydrodynamical properties for the particles which we inject the feedback energy into and make them active such that their timesteps will be assigned correctly at the next synchronization point.
Note that by modifying a particle's timestep before it completes its full kick-drift-kick procedure we necessarily violate energy conservation,
but to a much lesser extent than a numerical scheme without the timestep limiter.

\subsection{Numerical Resolution}\label{sec:res}
The mass resolution of an SPH simulation depends not only on the particle mass ($m_{\rm gas}$) but also the number of particles within a smoothing kernel ($N_{\rm ngb}$).
For a given kernel function and $m_{\rm gas}$,
using more particles in a kernel means smoothing over more mass and hence worse resolution.
However,
because of the low-order nature of SPH,
a relatively large $N_{\rm ngb}$ is required to reduce the so-called 'E$_0$-error', a zeroth order numerical noise induced by particle disorder (see e.g., \citealp{2010MNRAS.405.1513R, 2012MNRAS.425.1068D}).
We adopt $N_{\rm ngb}$ = 100 as a compromise between suffering too much from the E$_0$-error ($N_{\rm ngb}$ too small) and over-smoothing ($N_{\rm ngb}$ too large).
It seems reasonable to regard the kernel mass $M_{\rm ker} \equiv N_{\rm ngb} m_{\rm gas}$ as the resolution of SPH simulations.
However,
different kernel functions entail different extent of smoothing and different scales of compact support ($H$).
A Gaussian kernel is an extreme example with infinite $N_{\rm ngb}$ and $H$ while its smoothing scale is obviously finite.
The same $N_{\rm ngb}$ (and hence $H$) can therefore mean very different resolutions depending on the adopted kernel.
A more physical meaningful way is required to define a length scale which reflects the true extent of smoothing.
\citet{2012MNRAS.425.1068D} proposed to define the smoothing scale as $h = 2 \sigma$, where $\sigma$ is the second moment of the kernel function (or 'standard deviation').
Following such a definition,
$h \approx 0.55 H$ for the commonly used cubic spline kernel and $h \approx 0.45 H$ for our adopted Wendland $C^4$ kernel.
Therefore,
the mass resolution would be $N_{\rm ngb} m_{\rm gas} (h/H)^3 \approx 0.1 M_{\rm ker}$.
For $m_{\rm gas} = 4 M_\odot$ in our simulations (see Section \ref{sec:IC}),
this means that $40 M_\odot$ is the mass scale for which we define the local density of a particle.
Everything below $40 M_\odot$ is blurred by smoothing.

\subsubsection{Jeans mass criterion}

In hydrodynamical simulations that include self-gravity (as we do in this work),
one important scale to resolve is the Jeans mass such that 
\begin{equation}\label{eq:MJ}
	M_{\rm J} \geq N_{\rm J} M_{\rm ker},
\end{equation}
where $M_{\rm J}$ is the Jeans mass and $N_{\rm J}$ is a prescribed number to ensure $M_{\rm J}$ is well-resolved \citep{1997MNRAS.288.1060B,2008ApJ...680.1083R}.
\textbf{
When the Jeans mass is not resolved ($M_{\rm J} < N_{\rm J} M_{\rm ker}$),
perturbations on all scales down to the resolution limit ($\sim M_{\rm ker}$) should collapse physically.
However, perturbations can be created}
by numerical noise, triggering gravitational collapse, 
which tends to occur when the perturbations are only marginally resolved.
Eq. \ref{eq:MJ} makes sure that perturbations near the resolution limit are physically stable and all perturbations that collapse are well-resolved and are of physical origin rather than numerical noise.
The choice of $N_{\rm J}$ is somewhat arbitrary, as it is difficult to judge whether a collapse is physical or numerical.
Commonly suggested values in the literature are 
in the range of $N_{\rm J}$ = 4 $-$ 15 (e.g. \citealp{2008ApJ...680.1083R,
2011MNRAS.417..950H,2015arXiv150608829R}),
which seems to be quite stringent considering the resolution is about 0.1 $M_{\rm ker}$.
This may be related to the smaller $N_{\rm ngb}$ commonly used which is more prone to noise.
As such,
we take $N_{\rm J}$ = 1, which means that we require one kernel mass to resolve the Jeans mass.
According to the gas distribution in the phase diagram (see Fig. \ref{fig:PD_plus_Coolrates}),
our maximum number density satisfying $M_{\rm J} \geq M_{\rm ker}$ is about 200 cm$^{-3}$,
which corresponds to the smoothing length $H \approx$ 2 pc.
This motivates us to choose our gravitational softening length to be also 2 pc.

A commonly adopted approach to ensure that the Jeans mass is formally resolved throughout the simulations is to introduce a 'pressure floor' which artificially stabilizes the gas (by adding pressure by hand) when it becomes Jeans-unresolved (violating Eq. \ref{eq:MJ}) (e.g. \citealp{2008ApJ...680.1083R,2011MNRAS.417..950H,2013MNRAS.436.1836R,2014MNRAS.444.1518V,2015MNRAS.446..521S}).
However,
physically, 
a Jeans-unresolved perturbation should collapse.
In fact,
the only way to properly follow the evolution of the Jeans-unresolved
gas is to increase the resolution so that it becomes Jeans-resolved.
When a gas cloud becomes Jeans-unresolved,
the credibility of gravitational effect is lost irrespective of whether a pressure floor (or any other similar approach that makes the equation of state artificially stiffer) is used or not.
Besides,
the pressure floor makes the gas artificially stiff and thus also sabotages the accuracy of hydrodynamical effects.
We therefore refrain from using a pressure floor and simply try to keep the majority of gas Jeans-resolved by using high enough resolution.

\section{The ISM in Equilibrium}\label{sec:ISM_in_eq}
Before delving into the non-equilibrium evolution of the ISM and the complicated interplay of all the physical processes, 
in this section we investigate the ISM properties in both thermal and chemical equilibrium
under typical conditions in dwarf galaxies.
To see how a system evolves towards chemical and thermal equilibrium in a uniform and static medium,
we run the code with only the cooling and chemistry modules while turning the gravity and SPH solvers off. 
The density and column density are directly specified as input parameters for each particle rather than calculated with {\sc SPHGal} or {\sc TreeCol}.
The initial temperature is set to be $10^4$ K. 
We set the metallicity Z = 0.1 Z$_\odot$ and the cosmic ray ionization rate $\zeta = 3\times 10^{-18}$ s$^{-1}$.

The parameters that directly affect H$_2$ formation and destruction are $G_0$ and $D$ (Eq. \ref{eq:timeH2form} and \ref{eq:t_pd}). 
The strength of the ISRF in the diffuse ISM is expected to correlate with the star formation rate density.
As will be shown in Section \ref{sec:sim}, the total star formation rate of our simulated galaxy is $\approx 10^{-3}$ M$_{\odot}$ yr$^{-1}$ and the radius of the star-forming region is $\approx$ 2 kpc.
Thus, the star formation rate surface density is $\approx 8 \times 10^{-5}$ M$_{\odot}$ yr$^{-1}$ kpc$^{-2}$,
which is about ten times smaller than the value in the solar neighborhood.
Assuming the correlation between star formation rate surface density and $G_0$ is linear (e.g. \citealp{2010ApJ...721..975O}),
$G_0$ = 0.17 would be a plausible choice.
On the other hand,
the emergent radiation from the less dusty star-forming regions in low-metallicity environments seems to be stronger than in metal-rich ones due to the higher escape fraction of UV photons (see e.g. \citealp{2011ApJ...741...12B,2015A&A...578A..53C}). 
Therefore, the resulting $G_0$ in dwarf galaxies might be as strong as that in the solar neighborhood.
The observed DGR of a galaxy has been shown to scale linearly with its metallicity (though with significant scatter).
This implies $D$ = 0.1 for our adopted metallicity Z = 0.1 Z$_\odot$.
However,
below Z $\approx 0.2$ Z$_\odot$, the DGR starts to fall below the linear DGR-Z relation \citep{2014A&A...563A..31R}.

In this work we explore three different combinations of $G_0$ and $D$. We will use the following naming convention:
\begin{itemize}
	\item \textit{G1D01}:  $G_0$ = 1.7, $D$ = 0.1
	\item \textit{G1D001}: $G_0$ = 1.7, $D$ = 0.01
	\item \textit{G01D01}: $G_0$ = 0.17, $D$ = 0.1
\end{itemize}

\subsection{Chemical Equilibrium}\label{sec:chem_eq}
\subsubsection{The H$_2$ formation/destruction timescale}\label{sec:timescale}
We adopt the H$_2$ formation rate:
\begin{equation}\label{eq:form}
	\dot{n}_{\rm H_2} = R_{\rm form} D n_{\rm H} n_{\rm HI}
\end{equation}
where $R_{\rm form}$ is the rate coefficient \citep{1979ApJS...41..555H}, 
and $n_{\rm H_2}$, $n_{\rm HI}$ and $n_{\rm H}$ 
are the number densities of molecular hydrogen, atomic hydrogen, and
hydrogen nuclei, respectively.
$D$ is the DGR normalized to the Milky Way value such that $D$ = 1 means the DGR = 1\%.
At low temperatures ($T \lesssim$ 100 K), the rate coefficient $R_{\rm form} \approx 3\times 10^{-17}$ cm$^3$ s$^{-1}$ is relatively insensitive to temperature variations,
and the solution to Eq. \ref{eq:form} is $n_{\rm H_2} / n_{\rm H} = 1 - \exp(-t/t_{F})$ with the formation timescale:
\begin{equation}\label{eq:timeH2form}
	t_F = ( R_{\rm form} D n_{\rm H} )^{-1} \approx \frac{1}{n_1 D} ~{\rm Gyr}, 
\end{equation}
where $n_1 \equiv n_{\rm H} / 1 {\rm cm^{-3}}$.

Meanwhile,
the dominant process of H$_2$ destruction is photo-dissociation by the ISRF in the Lyman and Werner bands (11.2 $-$ 13.6 eV).
From Eq. \ref{eq:Ratepd},
the destruction timescale is
\begin{equation}\label{eq:t_pd}
	t_D = (R_{\rm pd, H_2})^{-1} \approx \frac{1}{f_{\rm sh} G_0} ~{\rm kyr}.
\end{equation}
Therefore, H$_2$ clouds exposed to an unattenuated radiation field will be quickly destroyed, independent of the gas density.
However,
the photo-dissociation itself is also an absorption process,
which attenuates the UV radiation efficiently and thus lengthens $t_D$ as the H$_2$ column density accumulates (H$_2$ self-shielding).

To illustrate the formation and destruction of H$_2$ for different ISM parameters,
we set up hydrogen column densities from $N_{\rm H,min} = 10^{16}$ cm$^{-2}$ to $N_{\rm H,max} = 10^{24}$ cm$^{-2}$ sampled by 32768 points equally spaced in $\log_{10} N_{\rm H}$ and evolve the chemistry network including cooling and heating.
This corresponds to a one-dimensional absorbing slab in a homogeneous medium.
The H$_2$ column density $N_{\rm H_2}$ is obtained by direct integration:
\begin{equation}
	N_{\rm H_2}(N_{\rm H}) = \int_{N_{\rm H,min}}^{N_{\rm H}} x_{\rm H_2}(N_{\rm H}') d N_{\rm H}'.
\end{equation}
We define the H$_2$ mass fraction $f_{\rm H_2} \equiv 2 x_{\rm H_2} = 2 n_{\rm H_2} / n_{\rm H}$ and therefore $f_{\rm H_2}$ = 1 means that the hydrogen is fully molecular.

In the upper three rows in Fig. \ref{fig:chem_eq} we show the time evolution of $f_{\rm H_2}$ in a uniform medium of $n_{\rm H}$ = 100 cm$^{-3}$ in \textit{G1D01}, \textit{G1D001} and \textit{G01D01} from top to bottom.
Initially, the medium is totally atomic ($f_{\rm H_2}$ = 0) in the left panels and totally molecular ($f_{\rm H_2}$ = 1) in the right panels.
The formation times are $t_F$ = 100 Myr, 1000 Myr and 100 Myr for \textit{G1D01}, \textit{G1D001} and \textit{G01D01} respectively.
If the medium is initially atomic,
in a few $t_F$,
gas at sufficiently high $N_{\rm H}$ becomes totally molecular and the system reaches chemical equilibrium,
which is consistent with Eq. \ref{eq:timeH2form}.
If the initial medium is molecular,
the gas would be photo-dissociated starting from the surface (low $N_{\rm H}$).
In the unattenuated region,
Eq. \ref{eq:t_pd} suggests rapid destruction on time scales of kyr, much shorter than $t_F$.
However,
only the thin surface directly exposed to the UV radiation would be destroyed very promptly.
Due to H$_2$ self-shielding,
the dissociating front propagates slowly into high $N_{\rm H}$ and the system reaches chemical equilibrium on time scales much longer than 1 kyr.
On the other hand,
the free-fall time at $n_{\rm H}$ = 100 cm$^{-3}$ is $t_{\rm ff} \approx$ 5 Myr.
Therefore,
in the ISM of dwarf galaxies,
the timescale to reach equilibrium, either staring from an atomic or molecular medium, is quite long compared to the local free-fall time.


\subsubsection{The H$_2$ fraction in chemical equilibrium}
An unperturbed cloud will eventually reach a steady state where the H$_2$ formation rate equals its destruction rate:
\begin{equation}\label{eq:chemEq}
	\dot{n}_{\rm H_2} = R_{\rm form} D n_{\rm H} n_{\rm HI} - R_{\rm pd, H_2} n_{\rm H_2} = 0.
\end{equation}
Together with $f_{\rm H_2} = 2 n_{\rm H_2} / n_{\rm H}$ and $n_{\rm H} = n_{\rm HI} + 2 n_{\rm H_2}$ (assuming the ionization fraction of hydrogen is zero),
the ratio of H$_2$ to HI at chemical equilibrium can be written as
\begin{equation}\label{eq:xeq}
	r_{\rm H_2/HI} \equiv \frac{f_{\rm H_2}}{1 - f_{\rm H_2}} = 1.8\times 10^{-6} \frac{n_1 D}{f_{\rm sh} G_0}.
\end{equation}
In unshielded regions, $f_{\rm H_2}\approx r_{\rm H_2/HI}$ tends to be very low for typical conditions.
Only in well-shielded environments where very few H$_2$-dissociating photons are present can $f_{\rm H_2}$ grow to a significant value.
The H$_2$ profile  is dictated by the value of $n_1 D / f_{\rm sh}G_0$,
which can be viewed as a dimensionless parameter representing the capability of H$_2$ formation relative to its destruction.
In the bottom row of Fig. \ref{fig:chem_eq} we show the equilibrium H$_2$ profile ($f_{\rm H_2}$ vs. $N_{\rm H}$) with different $G_0$ and $n_1$.
For a given value of $D$,
increasing (decreasing) $G_0$ by a factor of 10 has the same effect as decreasing (increasing) $n_1$ by the same factor.
The equilibrium H$_2$ profile is thus a self-similar solution depending on the value of $n_1 / G_0$ \citep{1988ApJ...332..400S, 2010ApJ...709..308M, 2014ApJ...790...10S}.
Note that although the equilibrium profile is self-similar,
the time required for reaching the equilibrium ($t_F$ or $t_D$) is not.


\begin{figure}
	\centering
	\includegraphics[trim = 20mm 0mm 0mm 0mm, clip, width=3.6in]{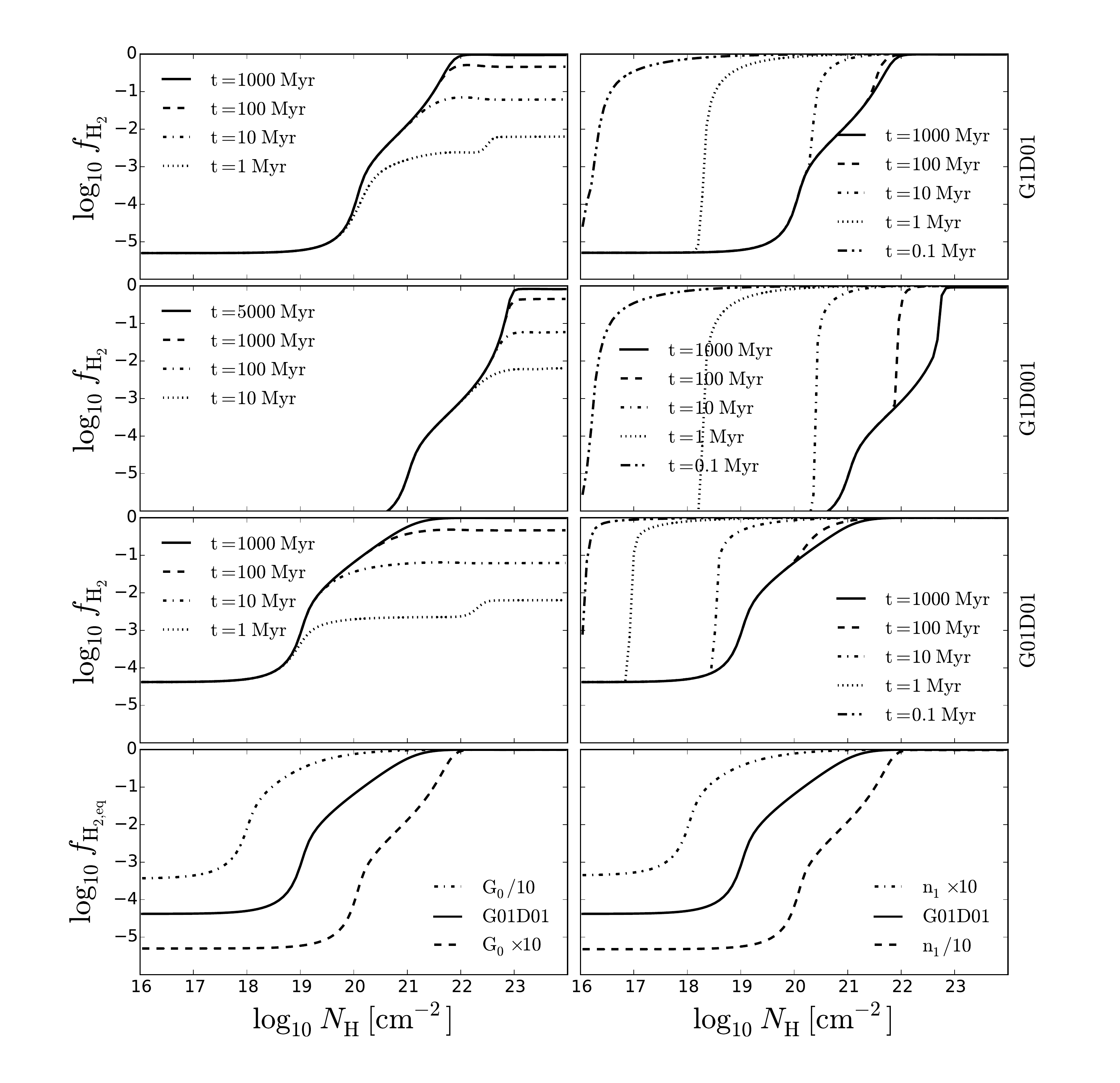}
	\caption{
		\textit{Upper three rows}: time evolution of $f_{\rm H_2}$ in a uniform medium of $n_{\rm H}$ = 100 cm$^{-3}$ in \textit{G1D01} (1st row), \textit{G1D001} (2nd row) and \textit{G01D01} (3rd row).
		The medium is initially totally atomic ($f_{\rm H_2}$ = 0) in the left panels and totally molecular ($f_{\rm H_2}$ = 1) in the right panels.
		\textit{Bottom row}: the equilibrium H$_2$ profile ($f_{\rm H_2}$ vs. $N_{\rm H}$) with different $G_0$ (left panel) and $n_1$ (right panel).
		For a given $D$, the equilibrium H$_2$ profile is self-similar depending on the dimensionless parameter $n_1 / G_0$.
		}
	\label{fig:chem_eq}
\end{figure}


\subsection{Thermal Equilibrium}\label{sec:thermal_eq}
We set up the hydrogen number density $n_{\rm H}$ from $10^{-4}$ to $10^{4}\ {\rm cm}^{-3}$ sampled by 32768 points equally spaced in $\log_{10} n_{\rm H}$ and let the system cool from $10^4$ K until it reaches thermal equilibrium.
Here we assume the column density to be $N_{\rm H}$ = $n_{\rm H} H$ where $H$ is the SPH smoothing length.
The H$_2$ column density is obtained by the approximation $N_{\rm H_2}$ = $x_{\rm H_2} N_{\rm H}$.
This is the minimum column density that would be assigned to a gas particle for the given density in our simulations if there were no other contributing material integrated by {\sc TreeCol}.
The smoothing length is calculated assuming the same mass resolution ($m_{\rm gas}$ = 4 M$_\odot$,$N_{\rm ngb}$ = 100) as in our simulations in Section \ref{sec:sim}.

In the upper three panels in Fig. \ref{fig:PD_thermal_eq} we show the individual cooling and heating processes in thermal equilibrium in \textit{G1D01}, \textit{G1D001} and \textit{G01D01}.
The 4th panel shows the equilibrium temperature vs. density, and the 5th panel shows the equilibrium pressure vs. density.
For almost the entire range of density,
the photo-electric effect is the main heating mechanism.
In \textit{G1D001}, the photo-electric effect becomes less efficient (due to the low DGR), and thus cosmic ray ionization dominates at $n_{\rm H} < 10^{-1}$ cm$^{-3}$.
The dust-gas collision rate rises with $n_{\rm H}$ but only becomes an important cooling mechanism at the highest densities which rarely occur in our simulations.
The most dominant coolant in the range of $n_{\rm H} = 10^{-1}$ - $10^3$ cm$^{-3}$ is the C$^+$ fine structure line emission.
Below $10^{-1}$ cm$^{-3}$, as the temperature increases, the OI fine structure emission and the hydrogen Lyman-alpha emission start to dominate over C$^+$.
Above $10^3$ cm$^{-3}$ the gas becomes optically thick to the UV radiation and so the photo-electric heating rate decreases. 
Meanwhile, the C$^+$ cooling is taken over by the CO cooling as most carbon is in the form of CO in this regime.
The H$_2$ cooling is unimportant in all cases 
as it is typically only abundant at densities where the temperature is too low to excite H$_2$.
In \textit{G1D001} and \textit{G01D01},
the photo-electric heating is less efficient than in \textit{G1D01} due to lower DGR and weaker ISRF respectively (cf. Eq. \ref{eq:photoheat}).
Therefore, the equilibrium temperatures are the highest in \textit{G1D01} (bottom panel of Fig. \ref{fig:PD_thermal_eq}).
Note that in low metallicity and low density regimes,
the time required to reach thermal equilibrium can become quite long \citep{2012ApJ...759....9K,2014MNRAS.437....9G}.
Therefore, the gas is expected to be out of thermal equilibrium at low densities (cf. Fig. \ref{fig:PD_plus_Coolrates}).

The maximum density with an equilibrium temperature about $10^4$ K is $n_{\rm cool}$ = 0.25, 0.03 and 0.1 cm$^{-3}$ in \textit{G1D01}, \textit{G1D001} and \textit{G01D01}, respectively.
As will be shown in Fig. \ref{fig:profile_SFRsc_3by1},
$n_{\rm cool}$ determines the maximum radius for star formation as it marks the onset of thermal-gravitational instability \citep{2004ApJ...609..667S}.

\begin{figure} 
	\centering
	\includegraphics[trim = 0mm 0mm 0mm 0mm, clip, width=90mm]{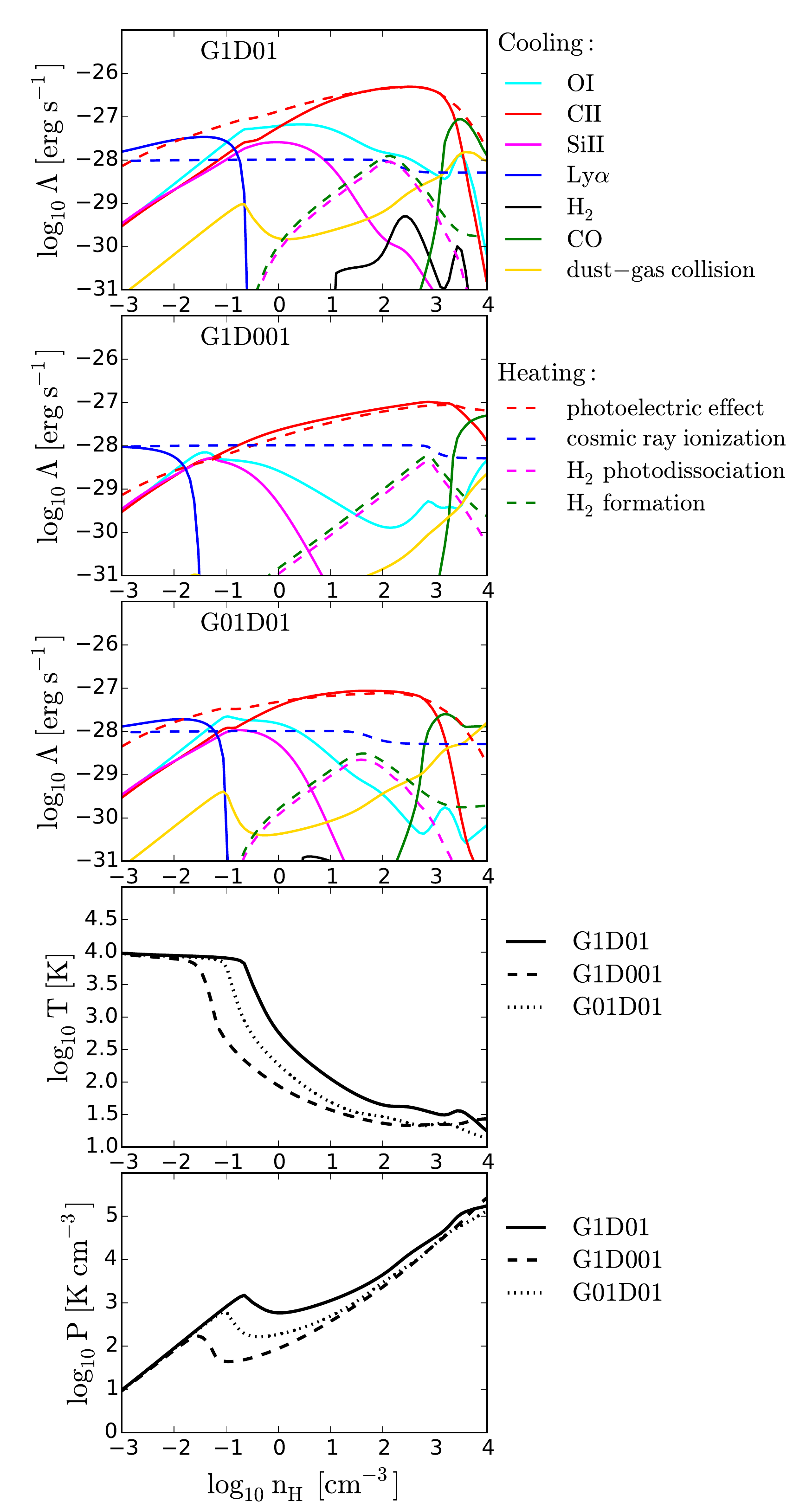}
	\caption{
		The cooling and heating rates for gas in thermal equilibrium in \textit{G1D01} (1st panel), \textit{G1D001} (2nd panel) and \textit{G01D01} (3rd panel).
		The 4th panel shows the equilibrium temperature vs. density, and the 5th panel shows the equilibrium pressure vs. density.
		The maximum density with an equilibrium temperature about $10^4$ K is $n_{\rm cool}$ = 0.25, 0.03 and 0.1 cm$^{-3}$ in \textit{G1D01}, \textit{G1D001} and \textit{G01D01}, respectively.
		}
	\label{fig:PD_thermal_eq} 
\end{figure}

\section{Simulations}\label{sec:sim}

\subsection{Initial Conditions}\label{sec:IC}
We set up the initial conditions using the method developed in \citet{2005MNRAS.364.1105S}.
The dark matter halo has a virial radius $R_{\rm vir}$ = 44 kpc and a virial mass $M_{\rm vir}$ = $2\times 10^{10} {\rm M}_\odot$, and follows a Hernquist profile with an NFW-equivalent \citep{1997ApJ...490..493N} concentration parameter $c$ = 10. 
The spin parameter for the dark matter halo is $\lambda$ = 0.03.
A disc comprised of both gas and stellar components is embedded in the dark matter halo.
The total mass of the disc is $6\times 10^7 {\rm M}_\odot$ with a gas fraction of 66 \%.
The small baryonic mass fraction (0.3 \%) is motivated by the results of abundance matching \citep{2010ApJ...710..903M,2013MNRAS.428.3121M}.
The disc follows an exponential profile with scale-length of 0.73 kpc.
The scale-length is determined by assuming the disc is rotationally supported and the total angular momentum of the disc is 0.3 \% of that of the dark matter halo,
which leads to a one-to-one relation between $\lambda$ and the scale-length.
Since dwarf galaxies are usually expected to have a relatively thick disc (e.g. \citealp{2015arXiv150304370E}),
we set the scale-height of the disc to be 0.35 kpc.
The initial metallicity (gas and stars) is 0.1 Z$_\odot$ uniformly throughout the disc 
with the relative abundances of the various metals the same as in solar metallicity gas.
The cosmic ray ionization rate is $3\times 10^{-18}$ s$^{-1}$.
The particle mass is $m_{\rm dm} = 10^4\ {\rm M}_\odot$ for the dark matter and $m_{\rm gas} = m_{\rm disc} = 4\ {\rm M}_\odot$ for the baryons (both stars and gas). 
The gravitational softening length is 62 pc for dark matter and 2 pc for baryons.
As discussed in Section \ref{sec:ISM_in_eq}, we explore three different combinations of $G_0$ and $D$ as shown in Table \ref{table:parameters}.
Run \textit{G1D01\_noFB} has the same $G_0$ and $D$ as \textit{G1D01} while the stellar feedback is switched off.

We use cylindrical coordinates $R$ and $z$ to describe the simulations,
where $R$ is the galactocentric radius and $z$ is the rotation axis of the disc.
The origin is chosen at the center of mass of the stellar disc,
so that $z$ = 0 is the mid-plane of the disc.
We also use the spherical coordinate $r$ to indicate the distance from the origin to a certain radius.

\begin{table}
	\caption{Simulation runs and the corresponding setup}  
	\label{table:parameters}
	\begin{tabular}{| l | c | c | c | c |}    
		\hline\hline
		Name                   &  $G_0$     & $D$       & feedback\\
		\hline
		\textit{G1D01}                  &   1.7     &  0.1      & yes  \\
		\textit{G1D001}                 &   1.7     &  0.01     & yes  \\
		\textit{G01D01}                 &   0.17    &  0.1      & yes  \\
		\textit{G1D01\_noFB}            &   1.7     &  0.1      & no   \\
		\hline\hline
	\end{tabular}
\end{table}

\subsection{Morphology}

In Fig. \ref{fig:column_4by4_1} we show, from left to right, the column density maps of HI (1st column), H$_2$ (2nd column) and H$^+$ (3rd column) and temperature maps (slices, 4th column) at $t$ = 500 Myr, where $t$ is the simulation time.
The upper two rows are the face-on and edge-on views for run \textit{G1D01}, while the lower two rows are for run \textit{G1D001}.
Fig. \ref{fig:column_4by4_2} shows the same maps as Fig. \ref{fig:column_4by4_1} but for run \textit{G01D01} (upper two rows) and run \textit{G1D01\_noFB} (lower two rows).
The temperature maps are slices across the origin so that the face-on slices are at the mid-plane $z = 0$.
We use different ranges of the color bars for different chemical species for display purposes.

The ISM is strongly dominated by HI in all cases.
It has clumpy density structures with many SN-driven bubbles. 
Most H$_2$ resides in a thin layer in the mid-plane with little H$_2$ at large $|z|$.
The H$_2$ approximately traces dense gas.
The H$^+$ traces the shells of the SN-driven bubbles and is not confined to the mid-plane.
The warm (green) gas occupies most of the volume in almost all cases.
The hot (red) gas also fills up a non-negligible amount of volume (mainly in the SN bubbles)
while the cold (blue) gas occupies very little volume.

Comparing the runs with feedback,
H$_2$ is most abundant in \textit{G01D01} while almost non-existent in \textit{G1D001}.
This is expected since the strong ISRF and low DGR in \textit{G1D001} is a hostile environment for H$_2$ formation.
If we turn off feedback,
H$_2$ becomes more abundant than in \textit{G01D01} (cf. Fig. \ref{fig:newIC_sfr_sc_h2_time}, 
although it occupies little volume because the gas collapses into massive clumps.

From the face-on images, the region where the density structures can be found is the largest in \textit{G1D001} because the radius $R$ beyond which gas can not cool and collapse is the largest.
This will be shown more quantitatively in Fig. \ref{fig:profile_SFRsc_3by1}.

\begin{figure*}   
	\begin{minipage}[b]{1.0\linewidth}
		\begin{center}
			\includegraphics[trim = 15mm 0mm 0mm 0mm,clip,width=19.cm]{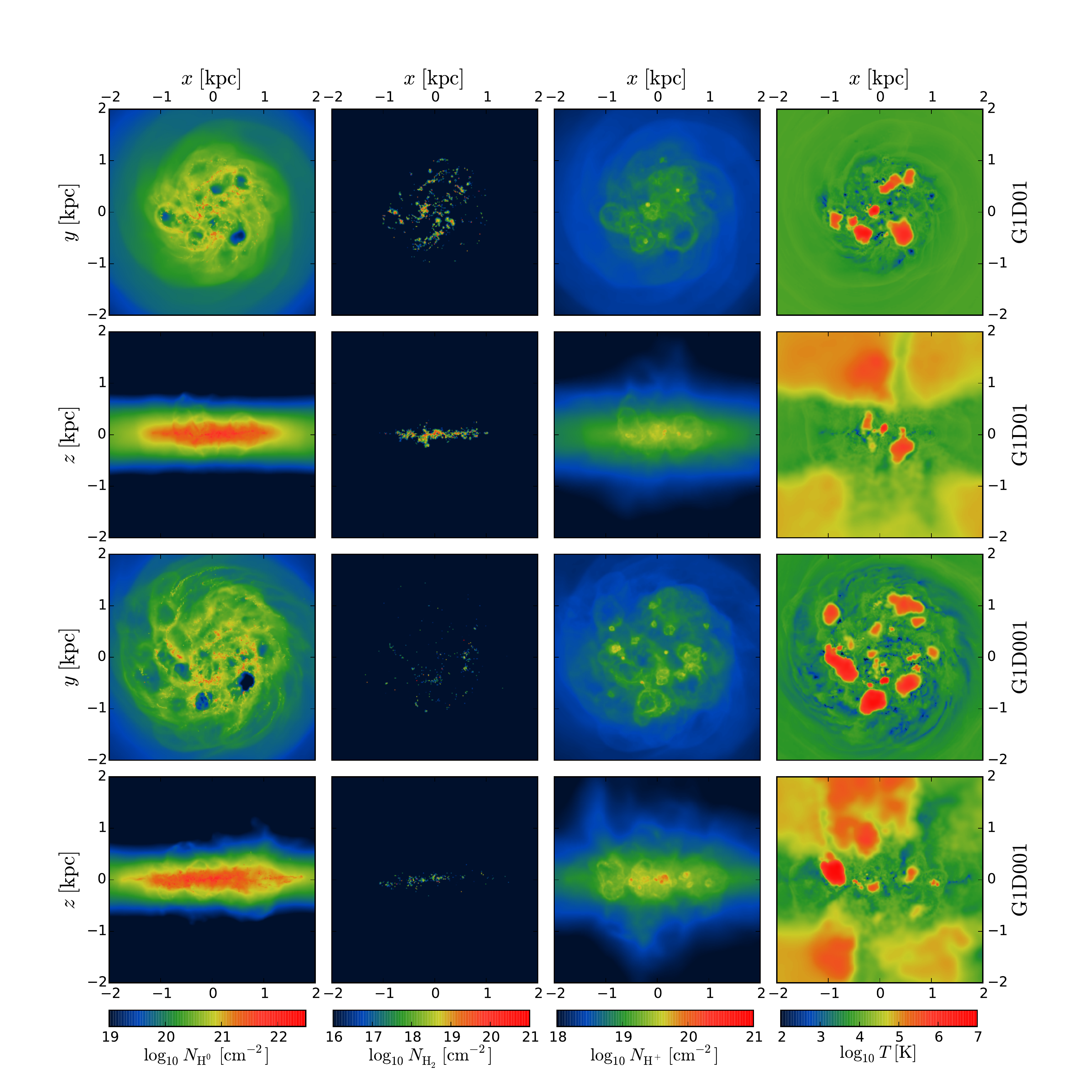}
		\end{center}
		\caption{
			Images of \textit{G1D01} and \textit{G1D001} at $t$ = 500 Myr.
			From left to right: the column density maps of HI (1st column), H$_2$ (2nd column) and H$^+$ (3rd column) and temperature maps (slices) (4th column).
			The top two rows are the face-on and edge-on views for run \textit{G1D01} while the bottom two rows are for run \textit{G1D001}.
			Note that we use different ranges of the color bars for different chemical species for display purpose, as the system is strongly dominated by HI.	
		}
		\label{fig:column_4by4_1}
	\end{minipage}
\end{figure*}

\begin{figure*}   
	\begin{minipage}[b]{1.0\linewidth}
		\begin{center}
			\includegraphics[trim = 15mm 0mm 0mm 0mm,clip,width=19.cm]{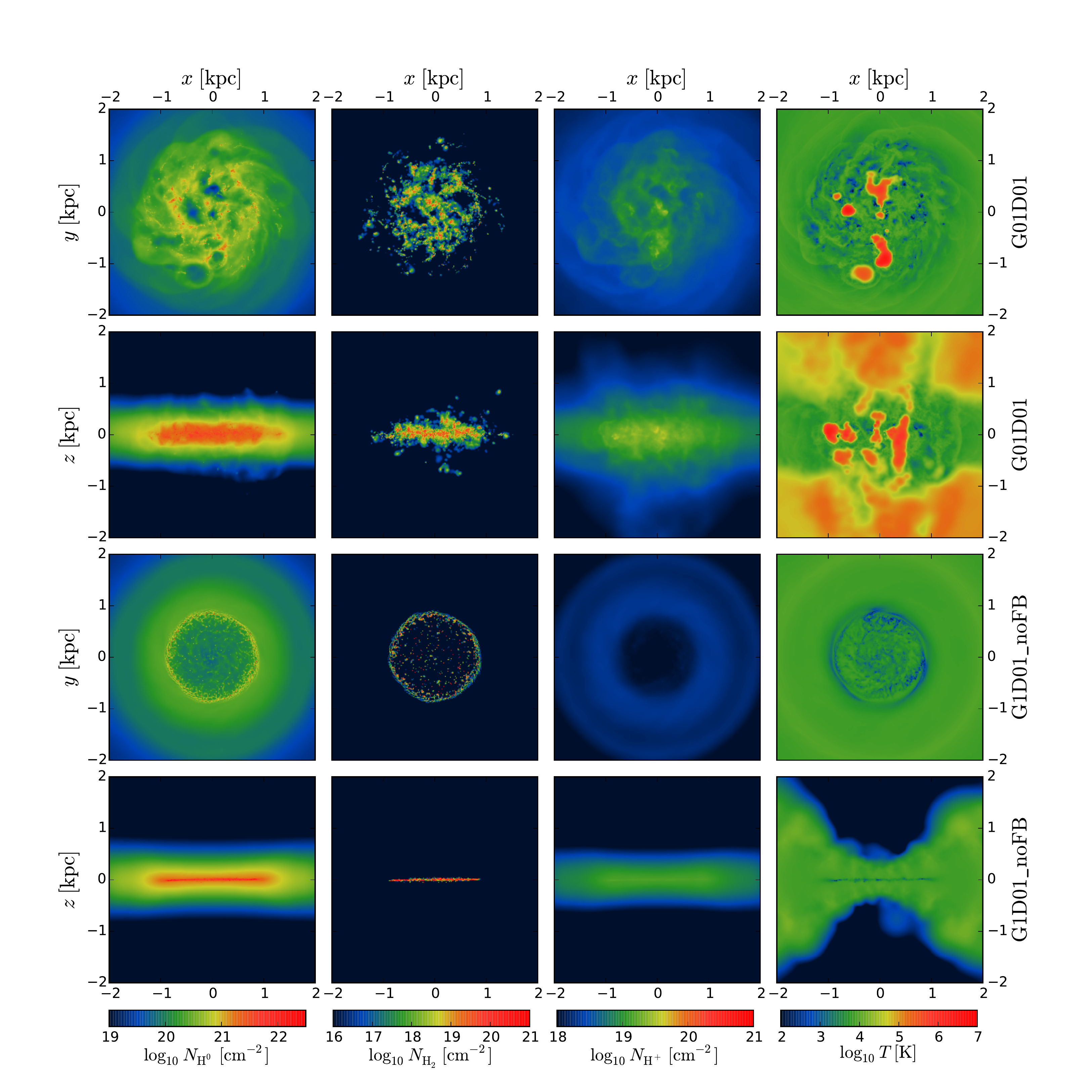}
		\end{center}
		\caption{
			Same as Fig. \ref{fig:column_4by4_1} but for run \textit{G01D01} (upper two rows) and run \textit{G1D01\_noFB} (lower two rows).
		}
		\label{fig:column_4by4_2}
	\end{minipage}
\end{figure*}

In Fig. \ref{fig:zoom_scale_2kpc200pc40pc_lines_01} we show the face-on column density for the \textit{G1D001} run at $t$ = 500 Myr with zoom-in's at different scales.
The central large panel shows the disc within R $<$ 2 kpc and the other four panels are zoom-in views of the central one.
The top left panel shows a filamentary structure about 300 pc long.
The bottom left panel is an example of a SN bubble with a size of about 200 pc.
The top right panel shows a region with plenty of dense clouds and the bottom right panel is a further zoom-in from the top right.
Note that the spatial resolution we have in dense gas is about 2 pc and therefore the clumps in the bottom right panel are expected to be still well-resolved.
The ISM is highly inhomogeneous with a complex density structure.

\begin{figure*}   
	\begin{minipage}[b]{1.0\linewidth}
		\begin{center}
			\includegraphics[trim = 40mm 0mm 0mm 0mm,clip,width=20cm]{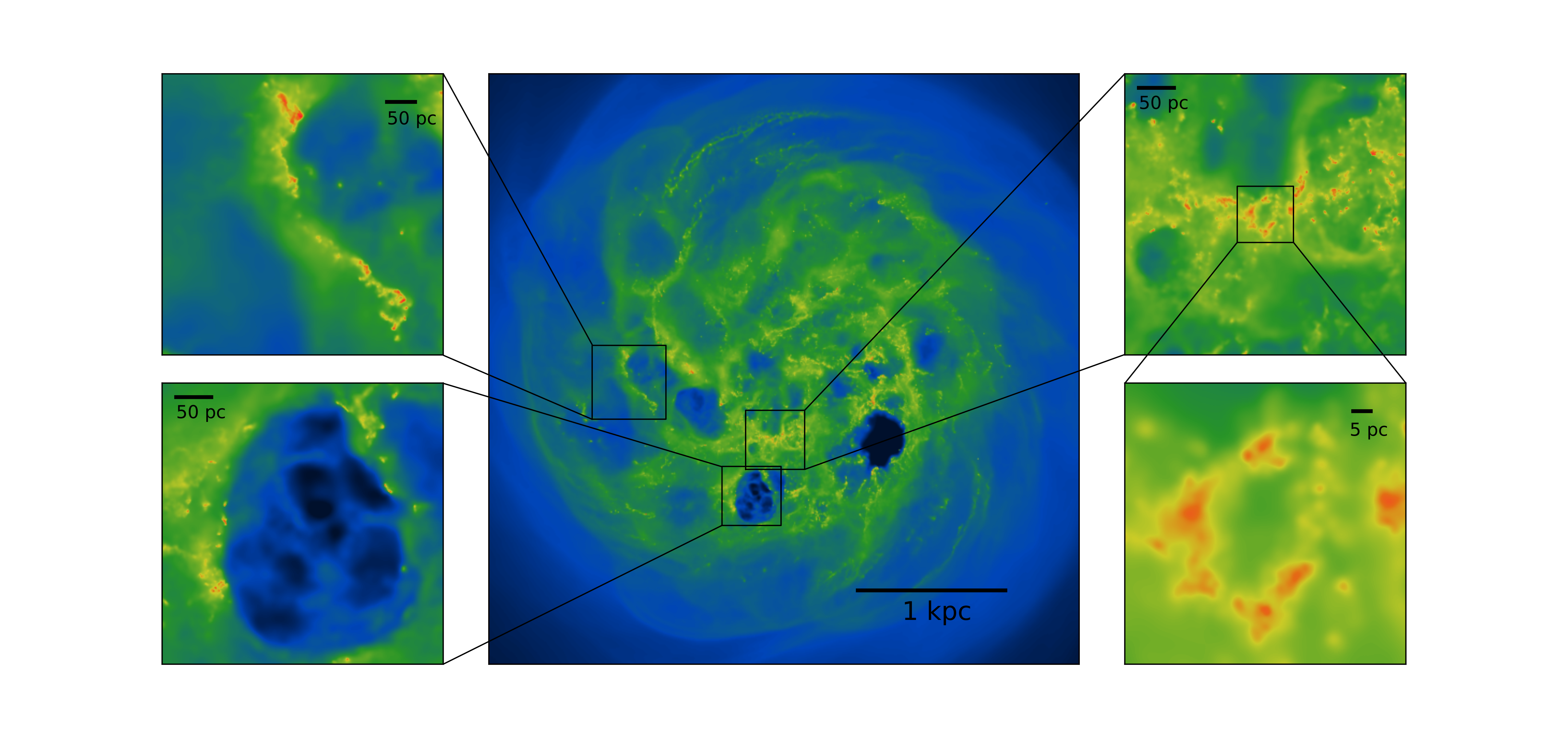}
		\end{center}
		\caption{
			The face-on column density maps of HI for the \textit{G1D001} run at $t$ = 500 Myr at different scales. 
			The color scale is the same as in Fig. \ref{fig:column_4by4_1} and \ref{fig:column_4by4_2}.
			\textit{Central panel}: the entire star-forming region (box size = 4 kpc).
			\textit{Top left}: a filamentary structure that is about 300 pc long.
			\textit{Bottom left}: a 200-pc scale bubble driven by SN feedback.
			\textit{Top right}: a group of dense clouds.
			\textit{Bottom right}: further zoom-in of the top right.
			The effective spatial resolution is about 2 pc (see Section \ref{sec:res}).
			The ISM is highly inhomogeneous with complex density structures.		
		}
		\label{fig:zoom_scale_2kpc200pc40pc_lines_01}
	\end{minipage}
\end{figure*}

\subsection{Time Evolution of Global Properties}\label{sec:global}

\subsubsection{Star formation rate and $H_2$ mass fraction}
In the upper panel of Fig. \ref{fig:newIC_sfr_sc_h2_time} we show the time evolution of the total star formation rate of the galaxy (SFR) for \textit{D1G01}, \textit{D1G001}, \textit{D01G01} and \textit{D1G01\_noFB}.
In the first 50 Myr the gas collapses onto a thin disc as it cools down and starts forming stars.
This phase of global collapse is quickly terminated by SN feedback and the system gradually settles to a steady state, with SFR $\approx 10^{-3}$ M$_\odot$ yr$^{-1}$.
The gas depletion time of the galaxy $t_{\rm dep} \equiv M_{\rm gas}$/SFR $\approx$ 40 Gyr is so long that the gas reservoir is able to sustain the SFR throughout the simulation time.
This is true even if galactic outflows are considered, as we will show in Section \ref{sec:outflows}.
The evolution of the SFRs is very similar in all runs with feedback, 
suggesting that the thermal properties of gas are not very sensitive to the strength of the ISRF or the DGR.
The only run that shows significant difference is the no-feedback one \textit{G1D1\_noFB}.
Here the SFR first gets as high as 0.07 M$_\odot$ yr$^{-1}$ at $t$ = 200 Myr, more than an order of magnitude higher than all the other runs,
and quickly declines afterwards.
This is due to a relatively short gas depletion time $t_{\rm dep}$ $\approx$ 500 Myr.
The gas reservoir is rapidly consumed by star formation.

In the lower panel of Fig. \ref{fig:newIC_sfr_sc_h2_time} we show the H$_2$ mass fraction of the galaxy $F_{\rm H_2}$ (solid lines),
defined as the total H$_2$ mass divided by the total gas mass in the ISM ($F_{\rm H_2} = \Sigma_i (f^{i}_{{\rm H_2}} m^{i}_{\rm gas}) / \Sigma_i m^{i}_{\rm gas}$, where $m^{i}_{\rm gas}$ and $f^{i}_{{\rm H_2}}$ are the mass and the H$_2$ mass fraction of individual particle $i$).
The ISM is defined as gas within the region $R <$ 2 kpc (which is roughly the edge of star formation activities, see Fig. \ref{fig:profile_SFRsc_3by1}) and $|z| <$ 1 kpc (to exclude the halo region).
The dashed lines show the corresponding (chemical-)equilibrium H$_2$ mass fraction $F_{\rm H_{2,eq}} = \Sigma_i (f^{i}_{{\rm H_2,eq}} m^{i}_{\rm gas}) / \Sigma_i m^{i}_{\rm gas}$,
where $f^{i}_{{\rm H_2,eq}}$ is calculated by Eq. \ref{eq:chemEq} (i.e. assuming all particles are in chemical equilibrium).
Note that the self-shielding factor is still obtained from the non-equilibrium $f_{\rm H_2}$ from the simulations.
Unlike the SFR, $F_{\rm H_2}$ differs significantly between different runs.
In \textit{G1D01} $F_{\rm H_2}$ is about 0.05\%.
Lowering $D$ by a factor of ten (\textit{G1D001}) decreases $F_{\rm H_2}$ by more than an order of magnitude, as the H$_2$ formation rate is directly proportional to $D$.
With ten times smaller $G_0$ (\textit{G01D01}) the difference is slightly smaller: $F_{\rm H_2}$ increases by about a factor of three.
The no-feedback run (\textit{G1D1\_noFB}) shows the highest $F_{\rm H_2}$ (up to 4\%) as the global gravitational collapse forms massive dense clumps which provides an ideal environment for forming H$_2$.
After $t$ = 200 Myr, $F_{\rm H_2}$ declines because most of the massive clumps have turned into stars.

Comparing $F_{\rm H_2}$ and $F_{\rm H_{2,eq}}$ in each run indicates that the equilibrium prediction systematically overestimates the H$_2$ mass fraction.
This difference is expected due to the slow H$_2$ formation rate:
$t_F$ = $(n_1 D)^{-1}$ Gyr.
In the ISM constantly stirred by the SN-driven turbulence,
it is unlikely for a gas cloud to stay unperturbed for a few $t_F$ and form H$_2$.
On the other hand,
the destruction rate can also be slow in well-shielded regions,
as shown in Section \ref{sec:timescale},
which also leads to some over-abundant H$_2$ gas that should have been destroyed if it were in chemical equilibrium.
The over-abundant and under-abundant H$_2$ gas compensate each other and therefore the global H$_2$ fraction $F_{\rm H_2}$ is only slightly lower than $F_{\rm H_{2,eq}}$,
although locally they are far out of chemical equilibrium (as we will show in Section \ref{sec:h2_noneq}).

In the no-feedback run (\textit{G1D01\_noFB}) $F_{\rm H_2}$ and $F_{\rm H_{2,eq}}$ agree pretty well.
This is partly due to the relatively 'quiescent' ISM (due to the lack of SN feedback) allowing for H$_2$ formation in a less disturbed environment. 
More importantly, the high-density clumps result in much shorter $t_F$, driving the system towards chemical equilibrium much faster.


\begin{figure}
	\centering
	\includegraphics[trim = 0mm 0mm 0mm 0mm, clip, width=3.2in]{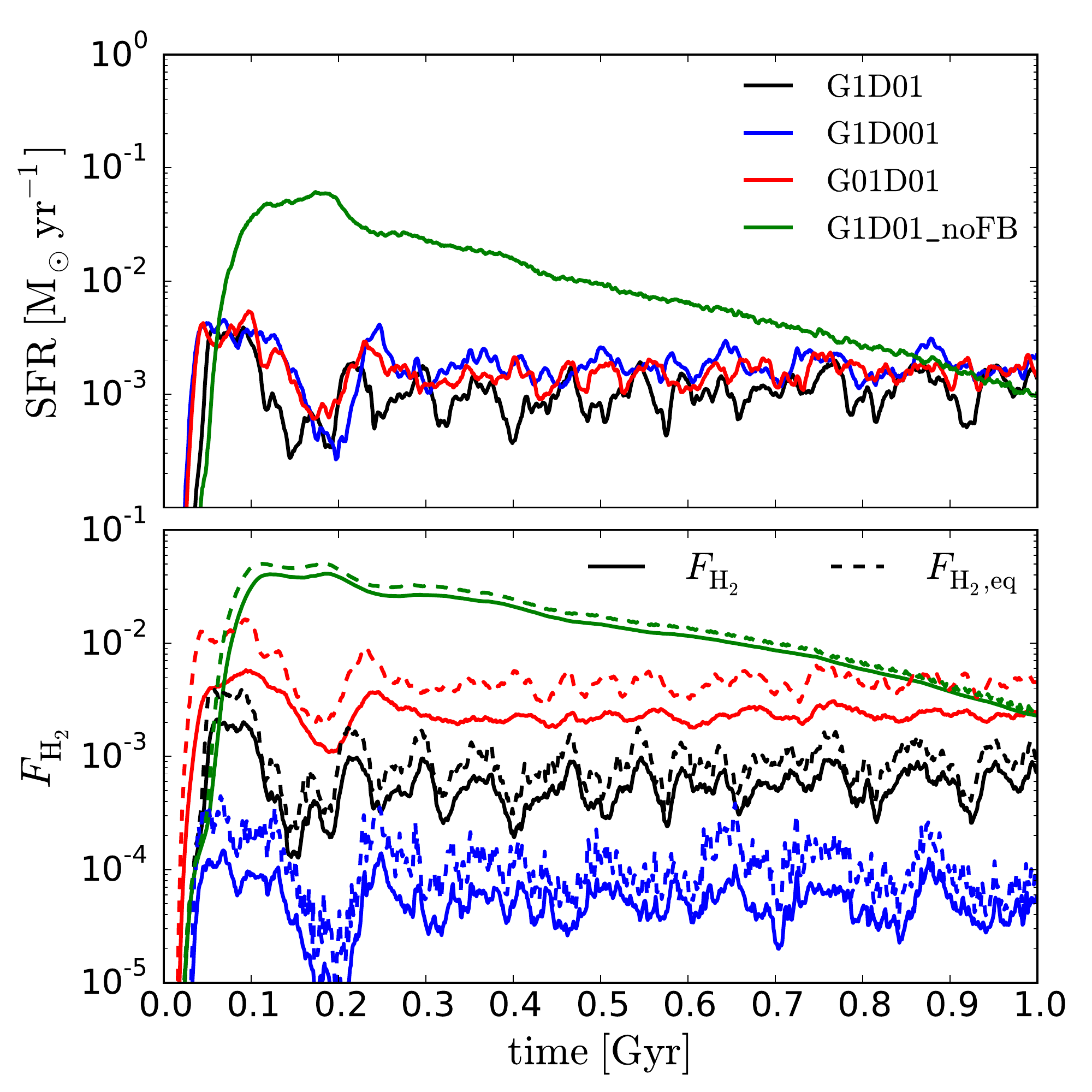}
	\caption{ Time evolution of global quantities of the galaxy.
		\textit{Upper panel}: total star formation rate (SFR). 		
		The SFR is relatively insensitive to the choice of $G_0$ and $D$ but differs strongly if stellar feedback is switched off. 
		\textit{Lower panel}: H$_2$ mass fraction in the ISM. 
		The ISM is defined as gas within the region $R <$ 2 kpc and $|z| <$ 1 kpc.
		Solid lines ($F_{\rm H_2}$) are from the simulations while dashed lines ($F_{\rm H_2,eq}$) are calculated assuming chemical equilibrium.
		The H$_2$ mass fraction is very sensitive to variations of $G_0$ and $D$.
		The equilibrium prediction $F_{\rm H_2, eq}$ systematically over-estimates the H$_2$ mass fraction.}
	\label{fig:newIC_sfr_sc_h2_time} 
\end{figure}

\subsubsection{Mass and volume fraction of different phases}\label{sec:MVfrac}

In Fig. \ref{fig:VMfrac_time} we show the time evolution of the mass fraction ($f_{\rm M}$, solid lines) and volume fraction ($f_{\rm V}$, dashed lines) of the ISM ($R <$ 2 kpc and $|z| <$ 1 kpc) in hot ($T > 3\times 10^4$ K), warm (100 K $< T \leq 3\times 10^4$ K) and cold ($T \leq 100$ K) phases.
The volume fraction is obtained by integrating the volume (estimated by the smoothing kernel size, $H^3$)-weighted temperature histogram for different phases.
The warm gas dominates the ISM both in mass ($f_{\rm M, warm}\approx 1$) and also in volume ($f_{\rm V, warm}\approx 0.9$).
As visualized in Fig. \ref{fig:column_4by4_1} and \ref{fig:column_4by4_2},
the hot phase occupies a non-negligible volume fraction ($f_{\rm V, hot}\approx 0.1$),
though it contributes only $\approx 10^{-3}$ in mass.
On the other hand, the volume occupied by the cold gas is negligibly small,
though its mass fraction can be as high as $f_{\rm M, cold}\approx 0.1$. 
The cold-gas fraction is a bit higher in \textit{G1D001} and \textit{G01D01} due to less efficient photo-electric heating in the low-DGR and weak-ISRF conditions (Eq. \ref{eq:photoheat}).
In principle, a lower DGR also results in less dust-shielding, which would potentially result in less cold gas.
However, dust-shielding only operates at 
$N_{\rm H,tot} > 1/ (D \sigma_{\rm dust})$,
which is rare in our simulations except for the densest gas.
There is also a slight increase in the cold-gas fraction over time as a result of metal enrichment,
which will be shown in Fig. \ref{fig:oifr}.
In the no-feedback run \textit{G1D01\_noFB},
there is almost no hot gas as the hot gas is mostly produced by SN blastwaves,
and the cold-gas fraction decreases over time due to the high SFR depleting the reservoir.

\begin{figure*}   
	\begin{minipage}[b]{1.\linewidth}
		\begin{center}
			\includegraphics[trim = 30mm 0mm 0mm 0mm, clip, width=180mm]{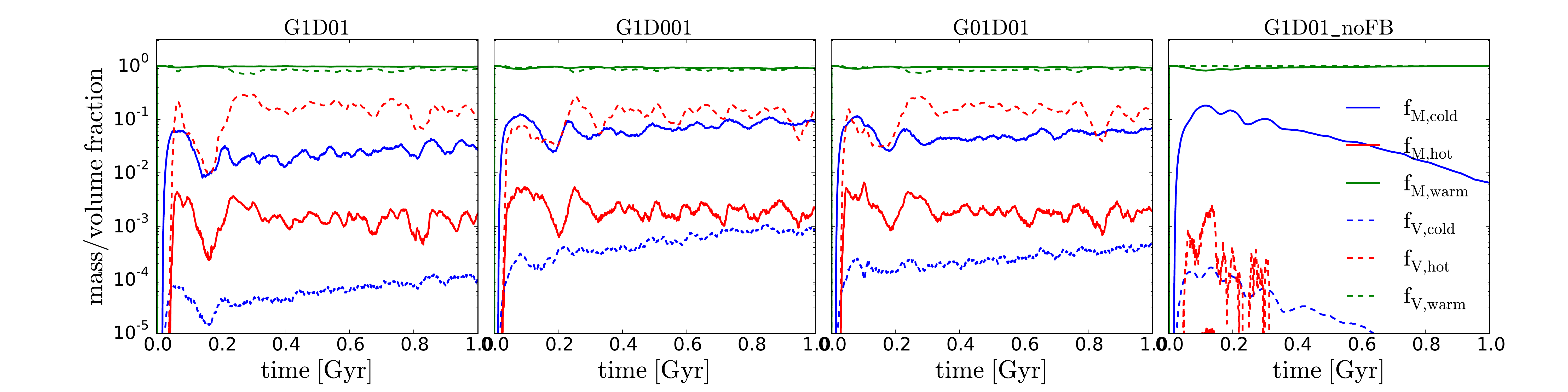} 
		\end{center}
		\caption{
			Time evolution of the mass fraction ($f_M$, solid lines) and volume fraction ($f_V$, dashed lines) of the ISM ($R <$ 2 kpc and $|z| <$ 1 kpc) in hot ($T > 3\times 10^4$ K), warm (100 K $< T \leq 3\times 10^4$ K) and cold ($T \leq 100$ K) phases. 
			The warm gas dominates the ISM both in mass and in volume. In all runs except for \textit{G1D01\_noFB}, the hot gas occupies a non-negligible volume ($f_{\rm V, hot}\approx 0.1$) but contains little mass, while the cold gas shows the opposite behavior and slightly increases over time due to metal enrichment. In \textit{G1D01\_noFB}, the hot gas is absent and the cold gas decreases over time due to the rapid depletion of gas by star formation.
		} \label{fig:VMfrac_time}
	\end{minipage}
\end{figure*}

\subsubsection{Feedback-driven galactic outflows}\label{sec:outflows}
Supernova explosions inject energy into the ISM and may push the gas out of the disc and drive galactic outflows.
We define the outflow rate as the mass flux integrated over a chosen surface area $S$:
\begin{equation}
	\dot{M}_{\rm out} \equiv \int_{S}{\rho \vec{v}\cdot\hat{n}} da  
\end{equation}
where $\rho$ is the gas density, $\vec{v}$ is the gas velocity and $\hat{n}$ is the normal direction of the surface (in the obvious sense of outwards from the disc).
In SPH simulations,
$\dot{M}_{\rm out}$ can be estimated as $\dot{M}_{\rm out} = \sum_{i} m^i_{\rm gas} \vec{v^i}\cdot\hat{n} / \Delta x $ where $m^i_{\rm gas}$ and $\vec{v^i}$ are the mass and velocity of particle $i$, and $\Delta x$ is a predefined thickness of the measuring surface. 
The summation is over particles within the shell of the surface with $\vec{v^i}\cdot\hat{n} >$ 0.
The choice of $\Delta x$ is a compromise between being too noisy ($\Delta x$ too small) and over-smoothing ($\Delta x$ too large).
The inflow rate $\dot{M}_{\rm in}$ is defined in the same way but with the direction of $\hat{n}$ reversed.

As a proxy for how much gas is leaving or entering the disc,
we measure the outflow rate $\dot{M}_{\rm out,2kpc}$ and inflow rate $\dot{M}_{\rm in,2kpc}$ at the planes $z = \pm 2$ kpc, truncated at $R < 6$ kpc. 
The choice of $z = \pm 2$ kpc is to avoid counting the thick disc material as outflows/inflows,
especially at larger radii where the gas disc flares,
(though the distinction between disc material and outflows/inflows is somewhat arbitrary).
The thickness of the planes $\Delta x$ is set to 0.2 kpc but the results are not sensitive to the chosen value.
To assess how much gas is escaping the dark matter halo,
we also measure the outflow/inflow rate at the virial radius $r$ = 44 kpc (denoted as $\dot{M}_{\rm out/in,vir}$).
Here the thickness of the sphere $\Delta x$ is set to 2 kpc due to the much lower density around the virial radius.
The mass loading factor $\eta$ is defined as $\dot{M}_{\rm out} / {\rm SFR}$.
There should be a time difference between when the star formation occurs and when the outflowing material causally linked to that star formation event reaches the measuring surface.
Our definition of the mass loading factor does not account for this effect and thus may seem inappropriate.
This is especially true for $\eta_{\rm vir}$ as it takes about 200 Myr for the gas to reach $r_{\rm vir}$ assuming the outflowing velocity of 200 km/s.
However,
as the SFR does not change dramatically throughout the simulations,
$\eta$ can still be a useful proxy for how efficient feedback is at driving outflows.

In Fig. \ref{fig:oifr} we show the time evolution of $\dot{M}_{\rm in,2kpc}$ in panel (a), $\dot{M}_{\rm out,2kpc}$ and $\dot{M}_{\rm out,vir}$ in panel (b), and $\eta_{\rm 2kpc}$ and $\eta_{\rm vir}$ in panel (c), respectively.
First we focus on the $z = \pm 2$ kpc planes.
Comparing with Fig. \ref{fig:newIC_sfr_sc_h2_time},
$\dot{M}_{\rm out,2kpc}$ rises sharply right after the first star formation occurs.
Since there is no gas above the disc in the initial conditions,
the initial outflows gush out freely into vacuum and thus the mass loading is very high, ($\eta_{\rm 2kpc} >$ 10).
This initial phase of violent outflow lasts about 100 Myr.
A significant fraction of gas falls back onto the disc due to the gravitational pull, which gives rise to the inflows starting from $t \approx$ 150 Myr.
From $t \gtrsim 200$ Myr,
the system reaches a quasi-steady state with a small but nonzero net outflow $\dot{M}_{\rm out,2kpc} - \dot{M}_{\rm in,2kpc}$.
The large fluctuations of $\dot{M}_{\rm out,2kpc}$ indicate that the outflows are intermittent on timescales of about 50 Myr.
On the timescale of 1 Gyr, however, $\dot{M}_{\rm out,2kpc}$ is rather constant and is slightly higher than the star formation rate ($\eta_{\rm out,2kpc} \gtrsim 1$).

Now we turn to the spherical surface at $r_{\rm vir}$ = 44 kpc.
Due to the time delay for gas to travel from the disc to $r_{\rm vir}$, 
the first nonzero $\dot{M}_{\rm out,vir}$ appears at $t \approx$ 200 Myr.
The peak value of $\dot{M}_{\rm out,vir}$ is much lower than the peak of $\dot{M}_{\rm out,2kpc}$ at $t \approx$ 80 Myr.
While this is partly due to the gravitational pull that slows down the outflowing gas,
the more important reason is that not all of the gas which passed through the $z=\pm$ 2kpc planes eventually made it to $r_{\rm vir}$.
Some would later fall back onto the disc (galactic fountain) and some would fill the space in the halo region.
No inflows at $r_{\rm vir}$ are detected ($\dot{M}_{\rm in,vir}$ = 0 at all times) so $\dot{M}_{\rm out,vir}$ can be regarded as a net outflow.
The mass loading $\eta_{\rm out,vir}$ first reaches about the order of unity and then decreases as the gas accumulates in the halo and hinders the subsequent outflows.

In Fig. \ref{fig:oifr} panel (d) we show the averaged velocity $v_{\rm out, vir}$ at $r_{\rm vir}$ and $v_{\rm out/in, 2kpc}$ at $z=\pm$ 2kpc, respectively.
The averaged velocity is mass-weighted over the particles within the measuring shells.
It may seem inconsistent that $v_{\rm out, vir}$ is much larger than $v_{\rm out, 2kpc}$.
This is not because the outflowing particles are accelerated,
but is again due to the fact that not all particles crossing the $z\pm$ 2kpc planes have enough kinetic energy to eventually escape the halo, and the averaged velocities are thus much smaller.
The initial high $v_{\rm out, vir}$ is due to the vacuum initial conditions in the halo region,
and over time $v_{\rm out, vir}$ decreases as the halo gas hinders the subsequent outflows.

In Fig. \ref{fig:oifr} panel (e) we show the total gas mass in different spatial regions as a function of time.
The disc region is defined as $R <$ 6 kpc and $|z| <$ 2 kpc,
and the halo is defined as $r < r_{\rm vir}$ excluding the disc region.
We denote $M_{\rm disc}$, $M_{\rm halo}$, and $M_{\rm escape}$ as the total gas mass in the disc, in the halo, and outside of the halo, respectively.
The total mass of stars that have formed in the simulations is denoted as $M_{\rm star}$ and is well confined within the disc region.
Except for the initial phase of strong outflow, 
there is roughly the same amount of gas ejected into the halo as the amount of stars formed in the simulations ($M_{\rm halo}\approx M_{\rm star}$), both of which are more than one order of magnitude lower than $M_{\rm disc}$.
Therefore, $M_{\rm disc}$ hardly changes throughout the simulations,
and as a consequence the system is able to settle into a quasi-steady
state.
The escaped mass $M_{\rm escape}$ is about 10-30\% of $M_{\rm halo}$ (and $M_{\rm star}$), which is consistent with $\eta_{\rm vir} <$ 1 shown in panel (c).

In Fig. \ref{fig:oifr} panel (f) we show the mass weighted mean metallicity of the gas in the disc, Z$_{\rm disc}$, in the halo, Z$_{\rm halo}$, and for gas that has escaped the halo, Z$_{\rm escape}$.
Starting from 0.1 Z$_{\odot}$, the disc metallicity slowly increases due to metal enrichment from the stellar feedback.
At the end of simulations ($t$ = 1 Gyr),
Z$_{\rm disc}$ only increased to about 0.15 Z$_{\odot}$ in \textit{G1D001} and \textit{G01D01} and about 0.12 Z$_{\odot}$ in \textit{G1D01}.
The slightly slower enrichment in \textit{G1D01} is due to its lower ${\rm SFR}$,
a consequence of more efficient photo-electric heating that suppresses gas cooling and star formation. 
The metallicity in the halo Z$_{\rm halo}$ is only slightly higher (about 20\%) than Z$_{\rm disc}$,
as not only the highly enriched SN-ejecta but also the low-metallicity gas in the ISM is pushed out of the disc, and Z$_{\rm halo}$ is thus diluted by the low-metallicity gas.
The metallicity of the escaped gas Z$_{\rm escape}$ shows a sharp peak as high as 0.3 Z$_{\odot}$ initially, but also drops to about the same level as Z$_{\rm halo}$ very quickly (less than 50 Myr).
In general, the outflows in our simulations are enriched winds, moderately.

\begin{figure*}   
	\begin{minipage}[b]{1.\linewidth}
		\begin{center}
			\includegraphics[trim = 20mm 10mm 0mm 0mm, clip, width=180mm]{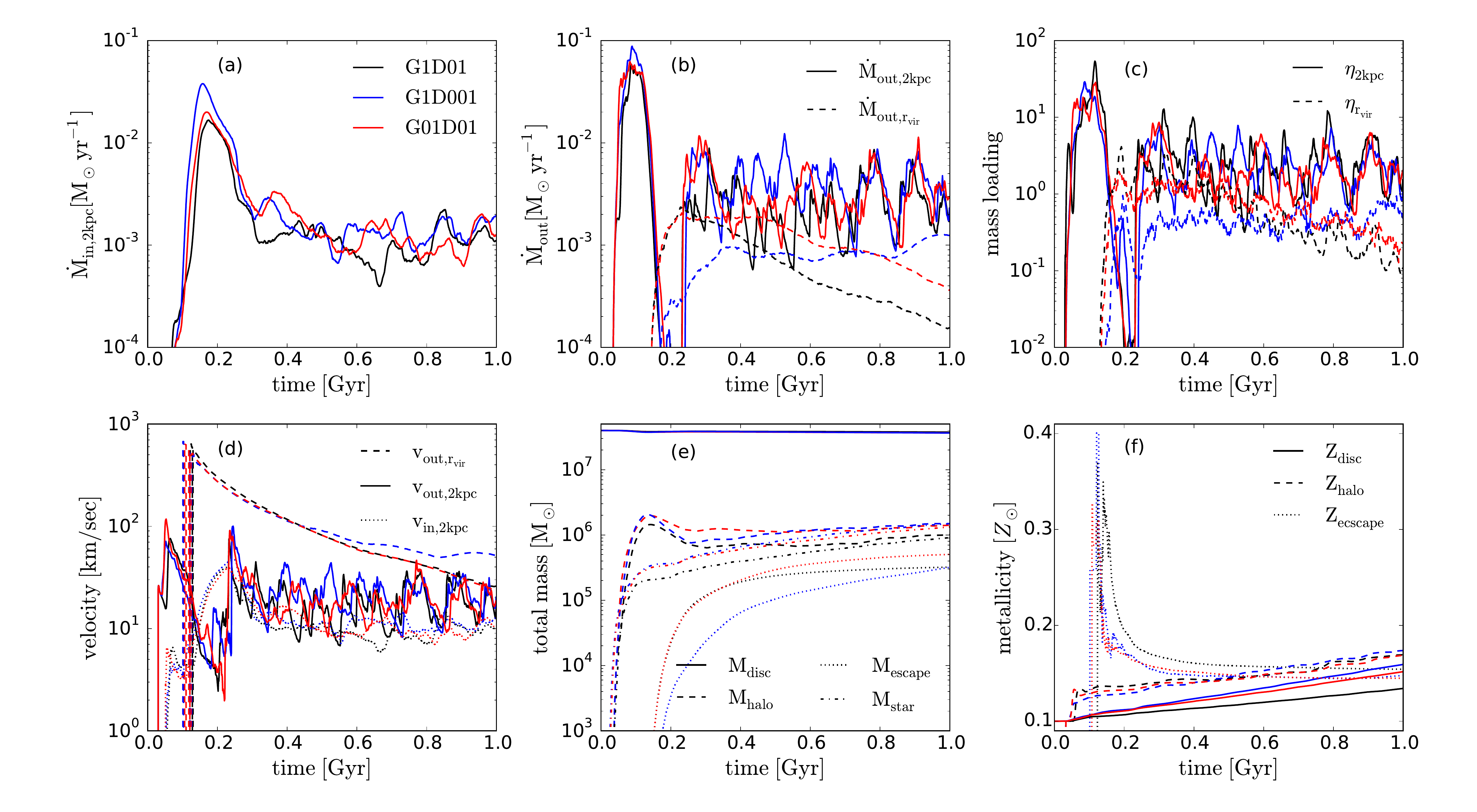} 
		\end{center}
		\caption{
			Time evolution of outflow-related physical quantities in \textit{G1D01} (black), \textit{G1D001} (blue) and \textit{G01D01} (red). 
			\textit{Panel (a)}: the inflow rate at $|z|$ = 2 kpc ($\dot{M}_{\rm in,2kpc}$). \textit{Panel (b)}: the outflow rate at $|z|$ = 2 kpc ($\dot{M}_{\rm out,2kpc}$) and at $r$ = $r_{\rm vir}$ ($\dot{M}_{\rm out,vir}$). 
			\textit{Panel (c)}: the mass loading factor at $|z|$ = 2 kpc ($\eta_{\rm out,2kpc}$) and at $r$ = $r_{\rm vir}$ ($\eta_{\rm out,vir}$). 
			\textit{Panel (d)}: the mass-weighted outflow velocity at $|z|$ = 2 kpc ($v_{\rm out, 2kpc}$) and at $r$ = $r_{\rm vir}$ ($v_{\rm out, vir}$), and inflow velocity at $|z|$ = 2 kpc ($v_{\rm in, 2kpc}$). 
			\textit{Panel (e)}: the total gas mass in the disc ($M_{\rm disc}$), in the halo ($M_{\rm halo}$), outside of the halo ($M_{\rm escape}$), and the total mass of stars formed in the simulations ($M_{\rm star}$). 
			\textit{Panel (f)}: the mass-weighted metallicity in the disc (Z$_{\rm disc}$), in the halo (Z$_{\rm halo}$) and outside of the halo (Z$_{\rm escape}$).
		} \label{fig:oifr}
	\end{minipage}
\end{figure*}

\subsection{ISM Properties}
Having described the global properties of the simulated galaxies,
we now turn to the local properties of the ISM and its multi-phase structure.

\subsubsection{Thermal properties of the ISM}\label{sec:thermal}

In Fig. \ref{fig:PD_plus_Coolrates} we show the phase diagrams (temperature vs. density, left panels) and the corresponding cooling and heating rates of different processes (median value, right panels) in the ISM region ($R <$ 2 kpc and $|z| < $ 1 kpc) at $t$ = 300 Myr in run \textit{G1D01}, \textit{G1D001}, \textit{G01D01} and \textit{G1D01\_noFB} (from top to bottom).
The black curves in the left panels show the (thermal-)equilibrium temperature-density relation as shown in Fig. \ref{fig:PD_thermal_eq}.
The dashed lines show the contours where $M_{\rm J} = M_{\rm ker}$.
Particles lying below the dashed lines are unresolved in terms of self-gravity.
Except for \textit{G1D1\_noFB} where the gas has undergone runaway collapse,
the majority of gas lies in the Jeans-resolved region up to $n_{\rm H} \approx $ 200 cm$^{-3}$.
In the right panels,
the thick dashed black curves represent the shock-heating rate, i.e., the viscous heating.
\textbf{
Note that this shock heating rate does not include the thermal injection of the SN feedback, which instantaneously heats up the nearest 100 gas particles as described in Section \ref{sec:feedback}.
}
The thick solid blue and dashed brown lines show respectively the cooling and heating rate caused by adiabatic expansion and compression $\dot{u}_{\rm PdV} = - (2/3) u \nabla\cdot \vec{v}$ where $u$ is the specific internal energy and $\nabla\cdot \vec{v}$ is the velocity divergence.

For $n_{\rm H} \geq $ 100 cm$^{-3}$,
the gas is approximately in thermal equilibrium at around few tens of Kelvin as the cooling time is much shorter than its local dynamical time (can be estimated by $\rho / \dot{\rho} = (3/2) u / \dot{u}_{\rm PdV} = - (\nabla\cdot \vec{v})^{-1}$).
A small bump at the highest densities is due to H$_2$ formation heating.
Below $n_{\rm H}$ = 100 cm$^{-3}$, the scatter starts to increase due to feedback-driven turbulent stirring.
For $n_{\rm H} <$ 1 cm$^{-3}$, the adiabatic heating/cooling dominates over other radiative processes,
and the gas spends several dynamical times before it can cool. Once the gas cools below $10^4$ K, it is likely to be shock heated to $10^4$ K in a dynamical time (gas being shock heated to $T > 10^4$ K is possible but at such temperatures the gas will quickly cool down to $10^4$ K).
Therefore,
turbulence drives the gas temperatures away from the equilibrium curve and keeps them at $10^4$ K at $n_{\rm H} \lesssim $ 1 cm$^{-3}$ (see also \citealp{2011ApJ...733...47W} and \citealp{2012MNRAS.426..377G}).

The distribution of gas in the phase diagram is not very sensitive to the choice of $G_0$ and $D$.
This is because SN feedback, which 
enhances turbulence and drives the gas out of thermal equilibrium in the range of $0.1 < n_{\rm H} < 10$ cm$^{-3}$
where the equilibrium curve is most sensitive to $G_0$ and $D$ (Fig. \ref{fig:PD_thermal_eq}),
does not depend explicitly on $G_0$ and $D$.
In \textit{G1D01} the gas distribution follows the equilibrium curve better as the photo-electric heating is more efficient and dominates over shock heating.
In \textit{G1D1\_noFB} the gas also follows the equilibrium curve and shows the least scatter, as the turbulent stirring is much weaker without SN feedback.
There is no hot gas in \textit{G1D1\_noFB} because the hot gas is only generated by the SN feedback.


\begin{figure*}   
	\begin{minipage}[b]{1.\linewidth}
		\begin{center}
			\includegraphics[trim = 0mm 0mm 0mm 0mm, clip, width=180mm]{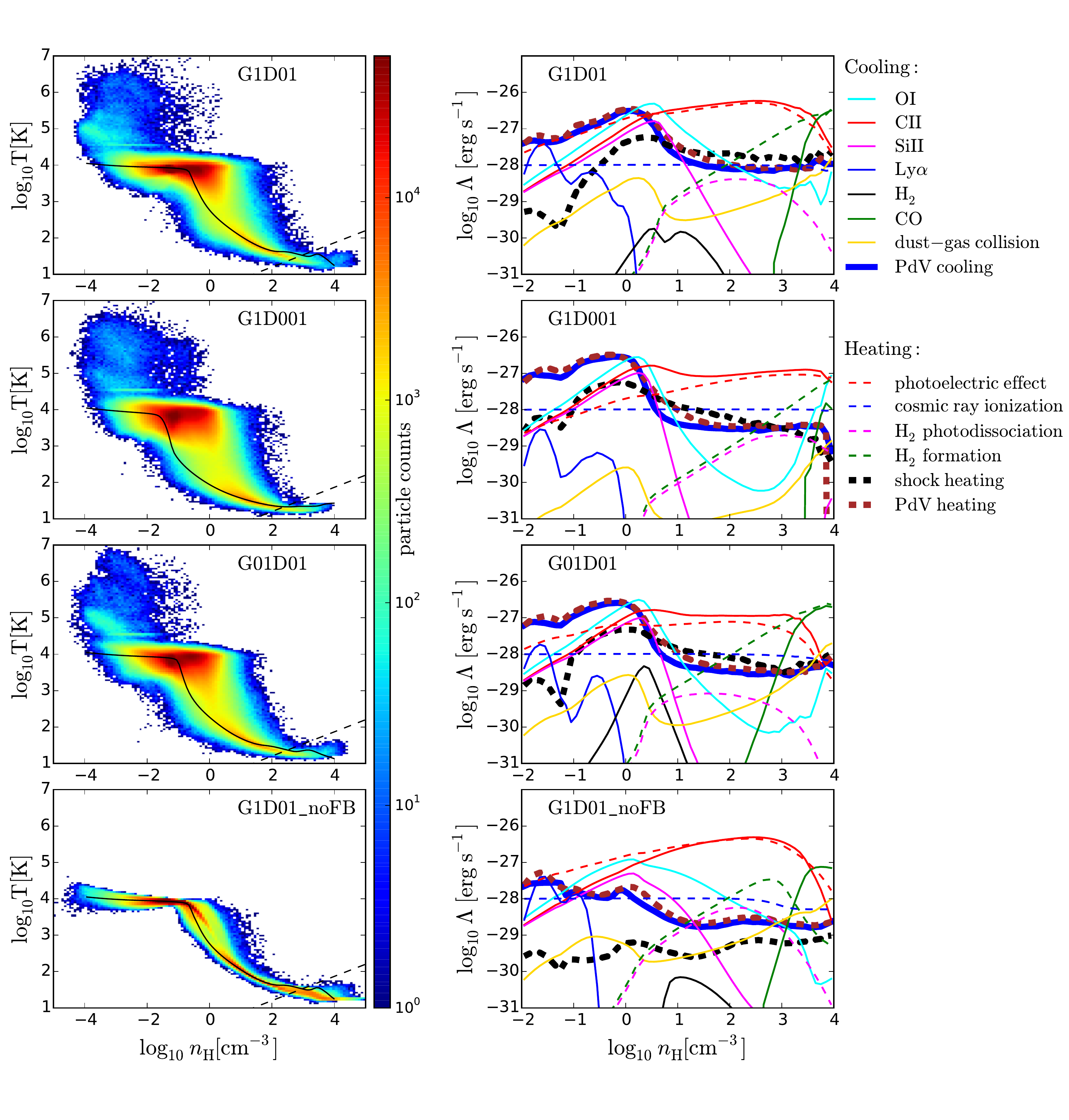} 
		\end{center}
		\caption{
			Phase diagrams (temperature vs. density, left panels) and the cooling and heating rates of different processes (median value, right panels) in the ISM region ($R <$ 2 kpc and $|z| < $ 1 kpc) at $t$ = 300 Myr in run \textit{G1D01}, \textit{G1D001}, \textit{G01D01} and \textit{G1D01\_noFB}.
			The black solid lines shows the (thermal-)equilibrium temperature-density relation as shown in Fig. \ref{fig:PD_thermal_eq}.
			The dashed line shows the contour where $M_{\rm J} = M_{\rm ker}$.
			SN feedback keeps the gas temperatures at $10^4$ K in the range of $0.1 < n_{\rm H} < 10$ cm$^{-3}$ even if the equilibrium temperatures are much lower.
			As a result, the distribution is not sensitive to $G_0$ and $D$, both of which affect photo-electric heating but not SN feedback.
		} \label{fig:PD_plus_Coolrates}
	\end{minipage}
\end{figure*}

\subsubsection{Molecular hydrogen}\label{sec:h2_noneq}

In the top row of Fig. \ref{fig:h2_eq_vs_noneq_alpha_nH}, we show the H$_2$ mass fraction $f_{\rm H_2}$ vs. the number density $n_{\rm H}$.
The dashed lines represent the star formation threshold $n_{\rm H,th}$ = 100 cm$^{-3}$.
Since most gas particles with $n_{\rm H} > n_{\rm th}$ are also cooler than 100 K (cf. Fig. \ref{fig:PD_plus_Coolrates}),
we can regard those located to the right of dashed lines as star-forming gas while those to the left are not star forming.
In general, $f_{\rm H_2}$ increases with $n_{\rm H}$ as higher density leads to a higher H$_2$ formation rate and usually implies more shielding.
Except for \textit{G1D01\_noFB},
a large fraction of star-forming gas has low $f_{\rm H_2}$.
The H$_2$ mass fraction of gas with $n_{\rm H} >$ 100 cm$^{-3}$ is $F_{\rm H_2,sfg}$ = 10\%, 0.7\%, 18\% and 70\% in \textit{G1D01}, \textit{G1D001}, \textit{G01D01} and \textit{G1D01\_noFB}, respectively,
though the precise values depend on our definition of star-forming gas ($n_{\rm H,th}$).
This implies that most star formation actually occurs in regions dominated
by atomic hydrogen (except for \textit{G1D01\_noFB}), in agreement with previous theoretical predictions \citep{2012MNRAS.426..377G}.

On the other hand,
there are also particles whose densities are too low to be star forming but that have $f_{\rm H_2} \approx$ 1.
The fraction of the total H$_2$ mass that resides in $n_{\rm H} <$ 100 cm$^{-3}$ is 42\%, 25\%, 70\% and 4\% in \textit{G1D01}, \textit{G1D001}, \textit{G01D01} and \textit{G1D01\_noFB}, respectively.
This demonstrates that, just as in our own Galaxy, H$_2$-dominated regions are not necessarily star-forming\footnote{
\textbf{We note that if we were able to follow the collapse to much smaller scales and much higher densities (e.g. prestellar cores), eventually all star formation will occur in H$_2$.
Our definition of star-forming gas corresponds to the reservoir gas that correlates (both spatially and temporary) with star formation.}}.
\textbf{
This is in line with \citet{2015arXiv150304370E} who also concluded that there should be a significant amount of diffuse H$_2$ in dwarf galaxies.
}

As discussed in Section \ref{sec:timescale},
the formation and destruction timescales of H$_2$ are long compared to the free-fall time.
In the presence of feedback,
turbulence further reduces the local dynamical time.
Therefore,
the gas can easily be out of chemical equilibrium.
In the bottom row of Fig. \ref{fig:h2_eq_vs_noneq_alpha_nH},
we show $f_{\rm H_2}$ vs. the dimensionless quantity $\alpha \equiv n_1 D /f_{\rm sh} G_0$ (see Section \ref{sec:chem_eq} for the definition).
The black curves indicate the $f_{\rm H_2}$-$\alpha$ relation in chemical equilibrium (Eq. \ref{eq:xeq}).
Gas particles with under/over-abundant $f_{\rm H_2}$ lie below/above the black curves.
The gas would follow the equilibrium curves if the chemical timescales were much shorter than the local dynamical timescales.
The large scatter of the distribution indicates the opposite:
the gas is far from chemical equilibrium locally.
However,
since there are both significant amount of particles under-abundant and over-abundant in H$_2$,
the total H$_2$ mass of the galaxy is not very different from the equilibrium predictions (cf. Fig. \ref{fig:newIC_sfr_sc_h2_time}).

As shown in Fig. \ref{fig:newIC_sfr_sc_h2_time}, the global value of $F_{\rm H_2}$ is sensitive to the assumed $G_0$ and $D$.
This is also true locally:
the particle distribution in Fig. \ref{fig:h2_eq_vs_noneq_alpha_nH} differs dramatically when adopting different $G_0$ and $D$.
This is in contrast to the particle distribution in the phase diagram (Fig. \ref{fig:PD_plus_Coolrates}) which is rather insensitive to variations of $G_0$ and $D$.
The largest scatter is in \textit{G01D01}, 
as the weak ISRF leads to a lower H$_2$ destruction rate.
A significant fraction of over-abundant H$_2$ can therefore survive for a longer time.
In \textit{G1D001},
the low DGR makes it difficult to have high $f_{\rm H_2}$ even for the densest gas.
Low $f_{\rm H_2}$ means little self-shielding (dust-shielding is almost absent given its DGR) and thus shorter H$_2$ destruction time.
In addition, although its H$_2$ formation time is the longest among all runs,
it takes much less time to reach the low equilibrium mass fraction $f_{\rm H_{2,eq}}$.
The scatter in \textit{G1D001} is therefore the smallest.
Despite the absence of feedback,
the \textit{G1D1\_noFB} run also shows some scatter as a result of its turbulent motions
due to thermal-gravitational instabilities.
Nevertheless,
there is a large fraction of H$_2$-dominated ($f_{\rm H_2} \approx$ 1) gas that follows the equilibrium curve quite well.
This explains the excellent agreement between the global $f_{\rm H_2}$ and the equilibrium prediction in Fig. \ref{fig:newIC_sfr_sc_h2_time}.

\begin{figure*}   
	\begin{minipage}[b]{1.\linewidth}
		\begin{center}
			\includegraphics[trim = 35mm 0mm 0mm 0mm, clip, width=190mm]{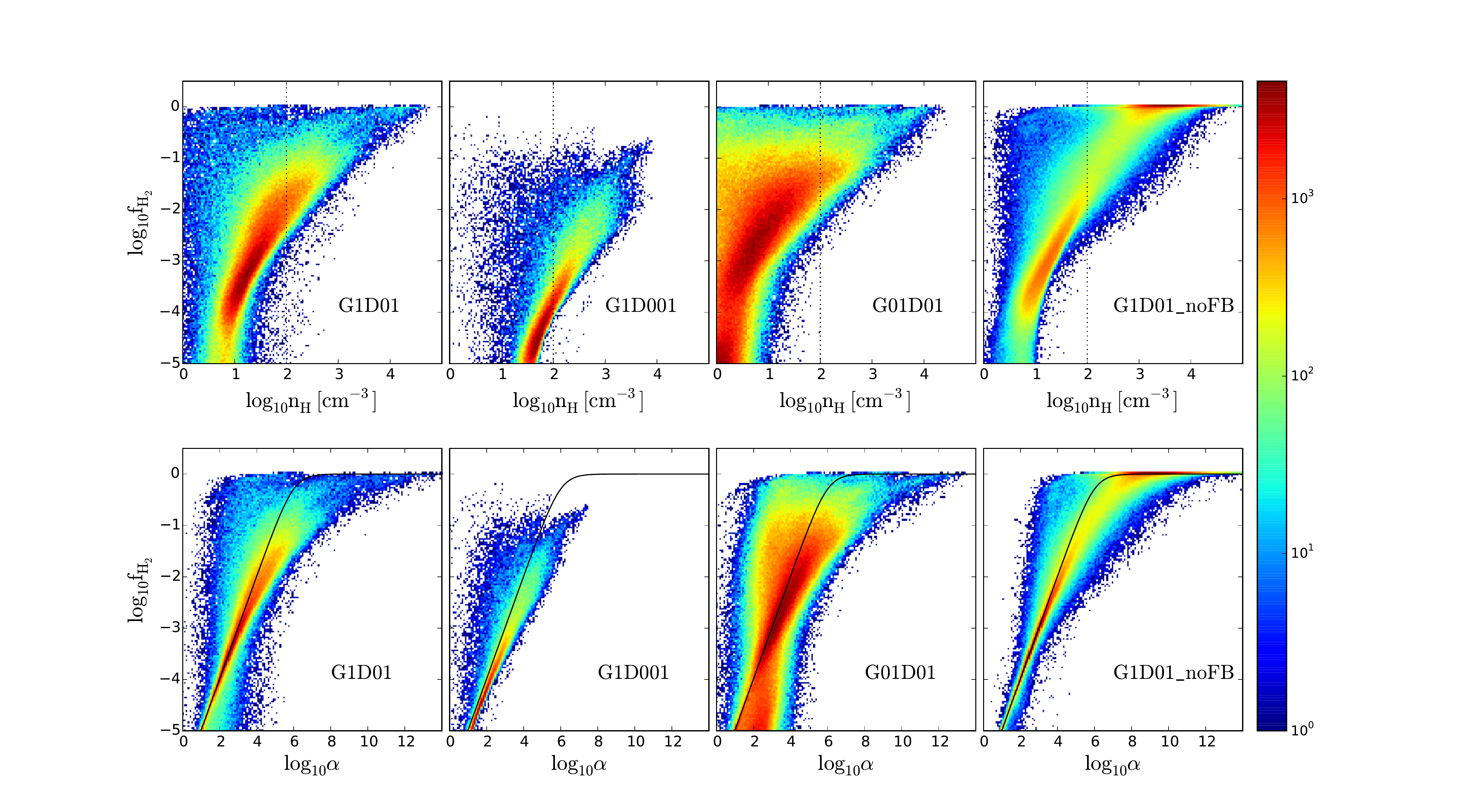} 
		\end{center}
		\caption{
			\textit{Top panels}: the H$_2$ mass fraction $f_{\rm H_2}$ vs. the number density $n_{\rm H}$ at $t$ = 300 Myr. The dotted line indicates the star formation threshold density $n_{\rm H,th}$ = 100 cm$^{-3}$.
			Most star formation actually occurs in regions dominated by atomic hydrogen, while H$_2$-dominated regions are not necessarily star-forming.
			\textit{Bottom panels}: $f_{\rm H_2}$ vs. the dimensionless quantity $\alpha \equiv n_1 D /f_{\rm sh} G_0$ at $t$ = 300 Myr.
			The solid line shows the $f_{\rm H_2}$-$\alpha$ relation in chemical equilibrium (cf. Eq. \ref{eq:xeq}).
			Particles with under/over-abundant $f_{\rm H_2}$ lie below/above the black curves.
			The gas is far from chemical equilibrium locally.
		} \label{fig:h2_eq_vs_noneq_alpha_nH}
	\end{minipage}
\end{figure*}

To see in which phase most of the H$_2$ can be found, we show in Fig. \ref{fig:whereH2_newIC} the phase diagram weighted by the H$_2$ mass.
\textbf{
The projected histograms of density and temperature (also weighted by $f_{\rm H_2} m_{\rm gas}$) are shown on the upper and right-hand sides of each panel, respectively.
}
Most of H$_2$ can be found in the cold and dense phase of the ISM.
In \textit{G1D01} and \textit{G01D01} there is some H$_2$ in the warmer (T $\lesssim$ 1000 K) and less dense phase.
This is the H$_2$-over-abundant, high $f_{\rm H_2}$ gas shown in Fig. \ref{fig:h2_eq_vs_noneq_alpha_nH},
which would not be predicted by the equilibrium approach.
Note that the H$_2$ distribution does not necessarily reside in the star forming regimes (T $<$ 100 K and $n_{\rm H}$ $>$ 100 cm$^{-3}$).
Again, this implies that non-star-forming regions can also be H$_2$-dominated.

\begin{figure*}
	\centering
	\includegraphics[trim = 0mm 0mm 0mm 0mm, clip, width=165mm]{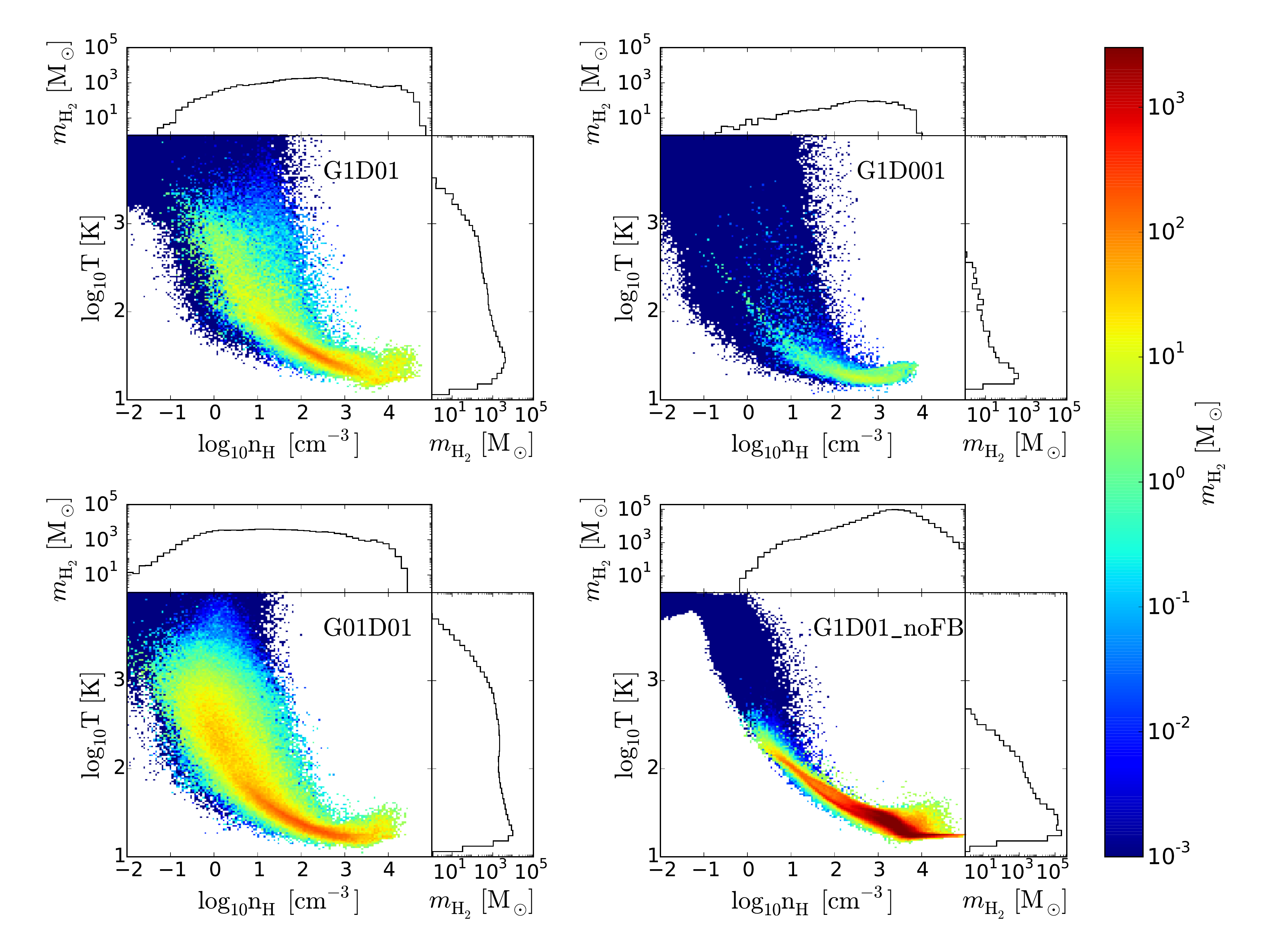}
	\caption{
		Phase diagrams weighted by the H$_2$ mass ($f_{\rm H_2} m_{\rm gas}$) at $t$ = 300 Myr.
		The projected histograms of density and temperature (also weighted by $f_{\rm H_2} m_{\rm gas}$) are shown on the upper and right-hand sides of each panel, respectively.
		Most H$_2$ is located in the cold (T $<$ 100 K) gas. 
		A significant fraction of H$_2$ is located in the warm and diffuse phase in \textit{G1D01} and \textit{G01D01} due to non-equilibrium effects.}
	\label{fig:whereH2_newIC} 
\end{figure*}

\subsection{Radial Variations}

\subsubsection{Scale height of the gas disc}
We define the scale height $h_{\rm gas, 75\%}$ as the altitude where 75\% of the gas mass between $z=\pm h_{\rm max}$ is enclosed between $z=\pm h_{\rm gas, 75\%}$ in each radial bin.
The choice of $h_{\rm max}$ = 2 kpc is to discard the particles that clearly do not belong to the disc in any reasonable definition.
Setting $h_{\rm max}$ = 1 kpc gives essentially identical results.	
The definition of 75\% enclosed mass is motivated by the ${\rm sech}^2 (z)$ distribution for a self-gravitating disc ($\tanh(-1) \approx 0.75$).
Similarly, we define $h_{\rm gas, 95\%}$ in the same way as $h_{\rm gas, 75\%}$ but for 95\% of enclosed mass (as $\tanh(-2) \approx 0.95$).
We also define the scale height of the cold ($T <$ 100 K) gas and the H$_2$ in the same fashion, denoted as $h_{\rm cold, 75(95)\%}$ and $h_{\rm H_2, 75(95)\%}$, respectively.

In Fig. \ref{fig:SH_3by1} we show $h_{\rm gas, 75\%/95\%}$ (green), $h_{\rm cold, 75\%/95\%}$ (blue) and $h_{\rm H_2, 75\%/95\%}$ (red) in the solid/dashed curves as a function of $R$,
(arithmetic-)averaged over time from $t$ = 250 Myr to $t$ = 1 Gyr with a time interval of 1 Myr.
The filled areas show the $\pm 1 \sigma$ bands where $\sigma$ is the standard deviation of $h_{\rm gas/cold/H_2, 75\%}$ in each radial bin.
The size of each radial bin is 0.1 kpc.
The scale height of the total gas $h_{\rm gas, 75\%}$ and $h_{\rm gas, 95\%}$ both increase with $R$,
especially at larger radii.
On the other hand,
75\%(95\%) of the cold gas and H$_2$ is found within a thin layer $|z| <$ 0.1(0.2) kpc without flaring,
\textbf{in line with what \citet{2015arXiv150304370E} inferred.}
There is some apparent flaring of $h_{\rm cold}$ and $h_{\rm H_2}$ for the $1 \sigma$ band (but not the mean value) at $R \approx$ 2 kpc in both \textit{G1D001} and \textit{G01D01}.
However,
very little cold gas and H$_2$ can be found at such radii (cf. Fig. \ref{fig:profile_SFRsc_3by1}).
Star formation mainly takes place in the mid-plane $|z| <$ 0.1 kpc,
as this is where the gas density is high enough to cool efficiently and trigger gravitational collapse.
The cold and dense environment in the mid-plane also favors H$_2$ formation.
Once the H$_2$ is pushed above the disc by SN feedback,
it would be destroyed by the unshielded ISRF very quickly.
Therefore, the H$_2$ is also confined to the mid-plane.

\begin{figure*}   
	\begin{minipage}[b]{1.\linewidth}
		\begin{center}
			\includegraphics[trim = 10mm 0mm 0mm 0mm, clip, width=170mm]{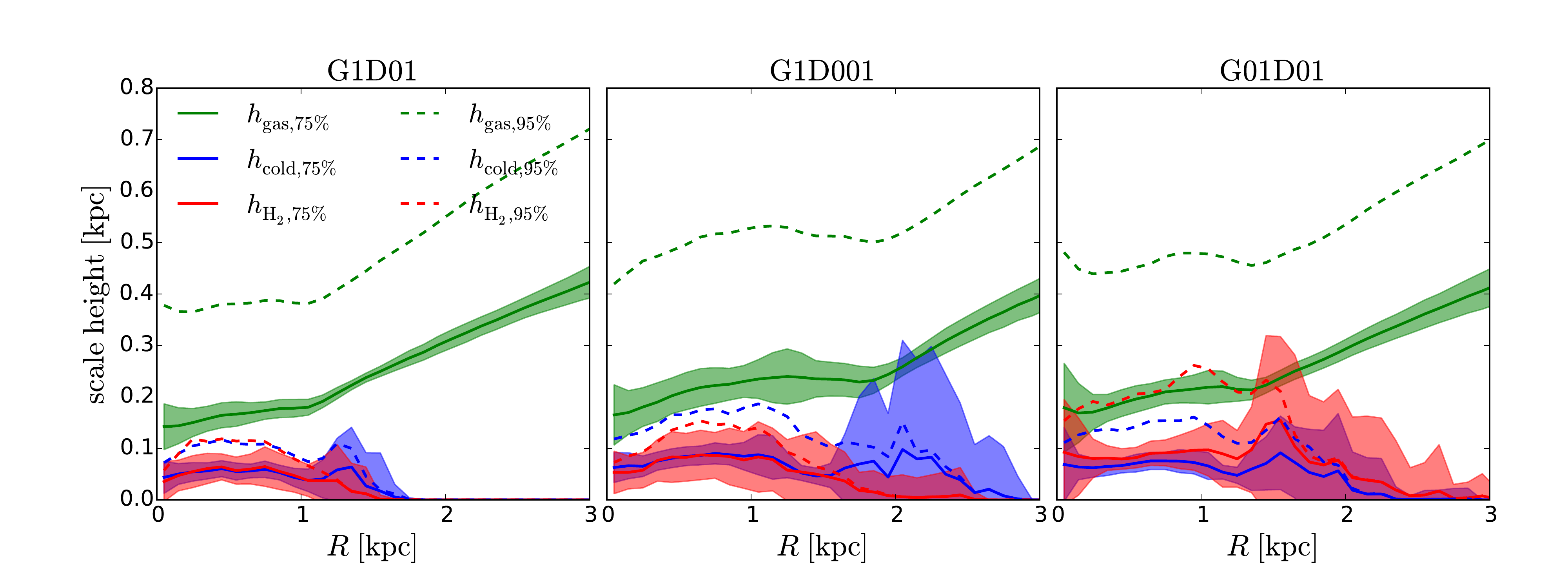} 
		\end{center}
		\caption{
			Scale height of the disc as a function of $R$.
			$h_{\rm gas, 75\%(95\%)}$ is defined as the altitude where 75\% (95\%) of the gas mass between $z=\pm h_{\rm max}$ is enclosed between $z=\pm h_{\rm gas, 75\%(95\%)}$.
			The scale height of the cold ($T <$ 100 K) gas $h_{\rm cold, 75\%(95\%)}$ and the H$_2$ $h_{\rm H_2, 75\%(95\%)}$ are defined in a similar way. 
			Results are averaged over time from $t$ = 250 Myr to $t$ = 1 Gyr with a time interval of 1 Myr.
			The solid lines represent the average values in each radial bin and the filled areas show the $\pm 1 \sigma$ bands.
			The disc thickens at large radii.
			On the contrary, the cold atomic and molecular gas are confined within a thin layer in the mid-plane without flaring.
		} \label{fig:SH_3by1}
	\end{minipage}
\end{figure*}

\subsubsection{Radial profile}\label{sec:profile}
In Fig. \ref{fig:profile_SFRsc_3by1} we show the radial profile of the surface density of the total gas ($\Sigma_{\rm gas}$), star-forming gas ($\Sigma_{\rm sfg}$), H$_2$ ($\Sigma_{\rm H_2}$), star formation rate ($\Sigma_{\rm SFR}$), 
as well as the mid-plane number density ($n_{\rm 0}$) and mid-plane pressure ($P_{\rm 0}$), averaged over time from $t$ = 250 Myr to $t$ = 1 Gyr with a time interval of 1 Myr. 
The solid lines are the arithmetic average and the filled areas are $\pm0.5\sigma$ bands in the given radial bins ($\pm 1 \sigma$ bands become negative in some bins and thus can not be shown in log-scale).
The mid-plane is defined as a thin layer within $|z| <$ 0.1 kpc.

An interesting feature in Fig. \ref{fig:profile_SFRsc_3by1} is that $\Sigma_{\rm SFR}$ drops much faster than $\Sigma_{\rm gas}$ as $R$ increases,
i.e., 
the gas depletion time $t_{\rm dep} \equiv \Sigma_{\rm gas}/\Sigma_{\rm SFR}$ increases rapidly as $R$ increases.
\citet{2015arXiv150304370E} explored a possible explanation that the flaring of the disc (cf. Fig. \ref{fig:SH_3by1}) makes the density of the mid-plane, where star formation takes place, drop faster than the projected surface density.
Our results do not favor this explanation as the radial gradient of $n_{\rm 0}$ and $\Sigma_{\rm gas}$ do not differ much.
Therefore, disc flaring is not responsible for the gradient difference between $\Sigma_{\rm SFR}$ and $\Sigma_{\rm gas}$.
Since the majority of the gas is in the warm phase of $T \approx 10^4$ K,
the mid-plane pressure $P_{\rm 0}$ also shows a similar radial gradient as that of $n_{\rm 0}$.
On the other hand,
$\Sigma_{\rm SFR}$, $\Sigma_{\rm sfg}$ and $\Sigma_{\rm H_2}$ all show much steeper radial gradients.
Roughly speaking,
these three quantities all trace the cold gas, while $\Sigma_{\rm gas}$, $n_{\rm 0}$ and $P_{\rm 0}$ trace the total gas.
Therefore, we can infer that the gradient difference between the former and the latter stems from the fact that the cold gas fraction declines as $R$ increases.

The radius where the cold gas fraction drops to zero (as do $\Sigma_{\rm SFR}$, $\Sigma_{\rm sfg}$ and $\Sigma_{\rm H_2}$), denoted as $R_{\rm cool}$, is about 1.4 kpc, 2.1 kpc and 1.7 kpc in \textit{G1D01}, \textit{G1D001} and \textit{G01D01}, respectively. 
The mid-plane density $n_{\rm 0}$ at $R_{\rm cool}$ roughly corresponds to $n_{\rm cool}$, the maximum density that keeps the equilibrium temperature at about $10^4$ K (cf. Fig. \ref{fig:PD_thermal_eq}).
The gas is unable to cool below $10^4$ K at $R > R_{\rm cool}$
because the mid-plane density is too low to trigger thermal-gravitational instability, which leads to an abrupt cutoff of $\Sigma_{\rm SFR}$, $\Sigma_{\rm sfg}$ and $\Sigma_{\rm H_2}$.
SN feedback is expected to scale with $\Sigma_{\rm SFR}$ and is therefore insignificant around $R_{\rm cool}$ compared to photo-electric heating,
which does not vary with $\Sigma_{\rm SFR}$ in our model.
As a result, 
although star formation in the inner region is regulated by the SN feedback,
$R_{\rm cool}$ is controlled by the photo-electric heating and thus varies with $G_0$ and $D$.
This is a consequence of our over-simplified assumption of spatially constant $G_0$.

In the inner part of the disc,
the values of $\Sigma_{\rm SFR}$ and $\Sigma_{\rm sfg}$ are similar for the three runs,
while $\Sigma_{\rm H_2}$ differs dramatically,
demonstrating again that H$_2$ formation is much more sensitive to variations of $G_0$ and $D$ than star formation is.
The excellent match between $\Sigma_{\rm sfg}$ and $\Sigma_{\rm H_2}$ in \textit{G01D01} is coincidental.
As shown in both Fig. \ref{fig:h2_eq_vs_noneq_alpha_nH} and Fig. \ref{fig:whereH2_newIC},
the star-forming gas is not necessarily H$_2$-dominated and vice versa.

\begin{figure*}   
	\begin{minipage}[b]{1.\linewidth}
		\begin{center}
			\includegraphics[trim = 0mm 0mm 0mm 0mm, clip, width=180mm]{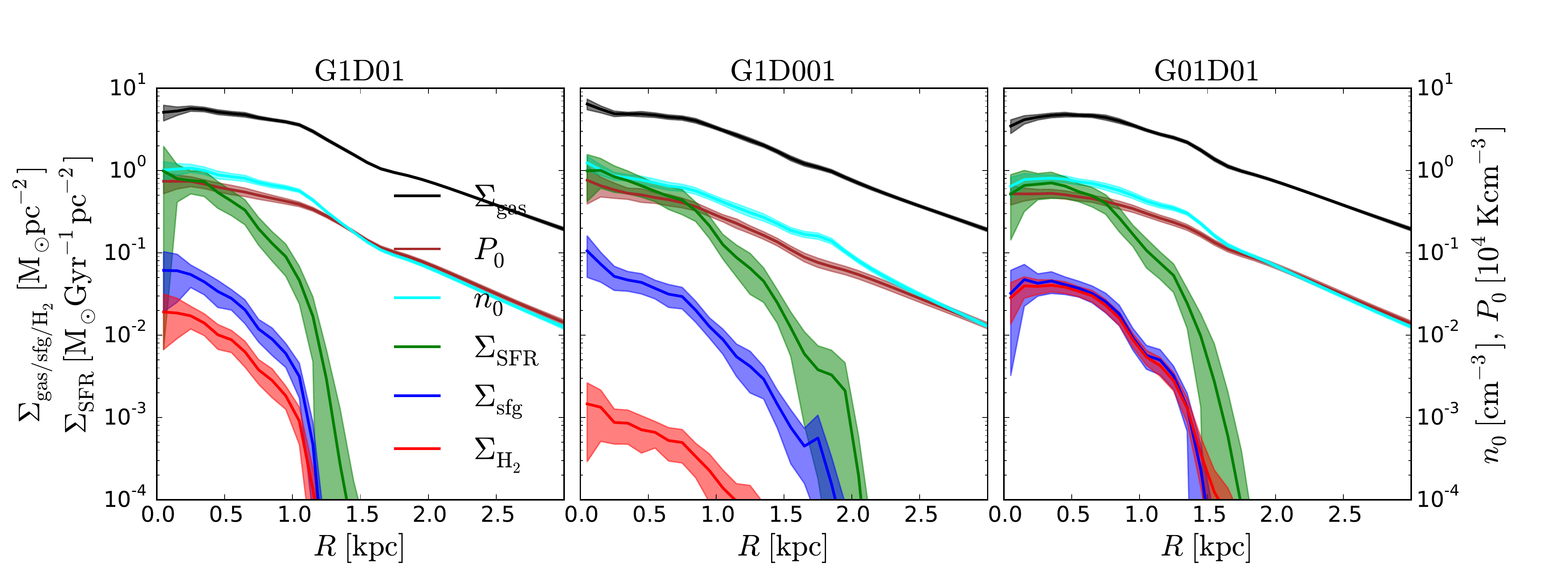} 
		\end{center}
		\caption{
			The radial profile of the surface density of the total gas ($\Sigma_{\rm gas}$, black), star-forming gas ($\Sigma_{\rm sfg}$, blue), H$_2$ ($\Sigma_{\rm H_2}$, red), star formation rate ($\Sigma_{\rm SFR}$, green), mid-plane number density ($n_{\rm 0}$, cyan) and mid-plane pressure ($P_{\rm 0}$, brown). 
			Results are averaged over time from $t$ = 250 Myr to $t$ = 1 Gyr with a time interval of 1 Myr.
			The solid lines represent the average values in each radial bin and the filled areas show the $\pm 0.5\sigma$ bands.
			The radial gradients of $\Sigma_{\rm sfg}$, $\Sigma_{\rm H_2}$ and $\Sigma_{\rm SFR}$ are much steeper than that of $\Sigma_{\rm gas}$, $n_{\rm 0}$ and $P_{\rm 0}$, as a result of the decreasing cold gas fraction as $R$ increases.
		} \label{fig:profile_SFRsc_3by1}
	\end{minipage}
\end{figure*}

\subsection{Density Distribution Function}\label{sec:rhoPDF}
In Fig. \ref{fig:rhoPDF_4by1_g1d01} we show the density distribution function $N(n_{\rm H})$ of the hot ($T >3 \times 10^4$ K, in red), warm ($100 < T < 3 \times 10^4$ K, in green) and cold ($T < 100$ K, in blue) gas within the mid-plane $|z| <$ 0.2 kpc in four different radial regions in \textit{G1D01}, 
averaged over time from $t$ = 250 Myr to $t$ = 1 Gyr with a time interval of 1 Myr.
It is binned in the range of $n_{\rm H} = 10^{-5} - 10^6$ cm$^{-3}$ with 110 density bins in log-space.
The solid curves are the average values and the filled areas are the $\pm 1\sigma$ bands
which indicate the fluctuations of the density distributions.

The cold gas fraction clearly decreases as $R$ increases,
in agreement with what can be inferred from Fig. \ref{fig:profile_SFRsc_3by1}.
The hot gas fraction also decreases as $R$ increases since the SN feedback occurs less frequently at large radii.
At $R >$ 1.5 kpc,
both the cold and hot gas are absent and the warm gas has a rather uniform (narrow) density distribution.
The mean density of the warm gas decreases as $R$ increases, which reflects the radial profile of the mid-plane density.
On the other hand,
the mean and the width of the cold gas density distribution do not change much with $R$ (except in the last panel where there is no cold gas).
This means that gas starts to cool efficiently above a certain density which is not a sensitive function of $R$.
Since the mean density is higher in the inner part of the disc,
a larger fraction of gas would have high enough density to cool and form cold gas,
which explains the decreasing cold gas fraction as $R$ increases.

Another feature is that the cold gas distribution above the star formation threshold density $n_{\rm H,th}$ (dotted lines) drops faster than $n_{\rm H}^{-0.5}$ (shown in dashed line segments).
Since the star formation rate of a gas particle is $\epsilon_{\rm sf} m_{\rm gas} / t_{\rm ff} \propto n_{\rm H}^{0.5}$,
gas particles with densities just above $n_{\rm H,th}$ would have a dominant contribution to the total star formation rate (SFR = $\int_{n_{\rm H,th}}^{\infty} \epsilon_{\rm sf} m_{\rm gas} / t_{\rm ff}(n_{\rm H}) N(n_{\rm H}) dn_{\rm H}$).
This leads to a linear relation between the surface density of gas and star formation rate, with a depletion time roughly equals $t_{\rm ff}(n_{\rm H,th})/\epsilon_{\rm sf}$ (see Section \ref{sec:sfr}),
even though the assumed relation is super-linear locally ($\propto n_{\rm H}^{1.5}$).

\begin{figure}
	\centering
	\includegraphics[trim = 0mm 10mm 10mm 10mm, clip, width=3.2in]{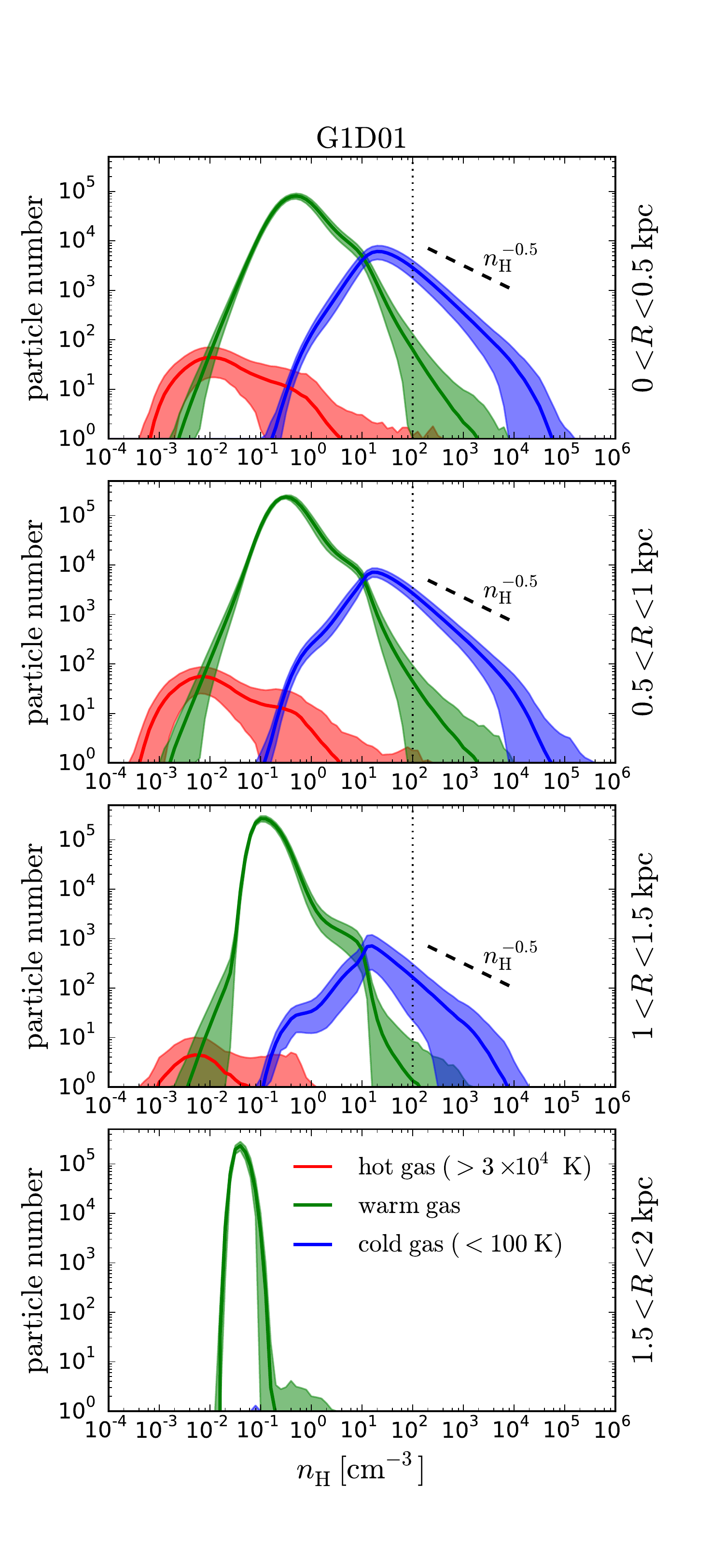}
	\caption{
		The density distribution of the hot ($T >3 \times 10^4$ K, in red), warm ($100 < T < 3 \times 10^4$ K, in green) and cold ($T < 100$ K, in blue) gas in \textit{G1D01} in the regions of $|z| <$ 0.2 kpc and, from top to bottom, 0 $< R <$ 0.5 kpc, 0.5 $< R <$ 1 kpc, 1 $< R <$ 1.5 kpc and 1.5 $< R <$ 2 kpc, respectively.		
		The results are averaged over time from $t$ = 250 Myr to $t$ = 1 Gyr with a time interval of 1 Myr.
		The solid lines represent the average values in each (log-)density bin and the filled areas show the $\pm 1 \sigma$ bands.
		The vertical dotted lines indicate the star formation threshold density and the dashed line segments show the slope of $N \propto n_{\rm H}^{-0.5}$ for reference.
		}
	\label{fig:rhoPDF_4by1_g1d01} 
\end{figure}

\subsection{Star Formation}\label{sec:sfr}

In Fig. \ref{fig:KS_pix05_compare} we show the star formation rate surface density $\Sigma_{\rm SFR}$ vs. the surface density of the total gas ($\Sigma_{\rm gas}$, in black), the star-forming gas ($\Sigma_{\rm sfg}$, in blue) and the H$_2$ ($\Sigma_{\rm H_2}$, in red), respectively.
All quantities are measured with aperture size $l_{\rm ap}$ = 500 pc,
as opposed to the radially averaged ones in Section \ref{sec:profile}.
Each dot represents one aperture and 
the dots are the results from $t$ = 250 Myr to $t$ = 1 Gyr with a time interval of 1 Myr (only a randomly chosen 4\% are shown for clarity).
The solid line represents the conventional Kennicutt-Schmidt (KS) relation \citep{1998ApJ...498..541K}.
The five dashed lines mark five different gas depletion times $t_{\rm dep}$ = 0.01, 0.1, 1, 10 and 100 Gyr.
The horizontal dashed lines show the lower limit for $\Sigma_{\rm SFR}$ for the given $l_{\rm ap}$ where one aperture only contains one star particle with age smaller than 5 Myr, i.e., $\Sigma_{\rm SFR,min} = m_{\rm star} / (t_{\rm SF}~ l_{\rm ap}^2 )$.

For comparison with observations,
the green dots are from \citet{2010AJ....140.1194B} with $l_{\rm ap} \approx$ 750 pc.
The purple line is the average value taken from the FIGGS survey from \citet{2015MNRAS.449.3700R} with $l_{\rm ap} \approx$ 400 pc, while the filled area encloses the fifth and ninety-fifth percentile of the data.
The cyan dots are the ring-averaged quantities and the cyan line is a fit to all the cyan points from \citet{2015arXiv150304370E}.

The KS relation for the total gas surface density (black dots) lies below the black lines with a very steep slope, 
which is consistent with observations in this regime.
On the other hand,
the relation between $\Sigma_{\rm sfg}$ and $\Sigma_{\rm SFR}$ is almost linear,
which implies a constant $t_{\rm dep}$ around 0.1 Gyr. 
As discussed in Section \ref{sec:rhoPDF},
gas with $n_{\rm H}$ just above the star formation threshold density $n_{\rm H, th}$ = 100 cm$^{-3}$ contributes the most to the total SFR.
The depletion time of star-forming gas $t_{\rm dep, sf} \equiv \Sigma_{\rm sfg} / \Sigma_{\rm SFR}$ is therefore the free-fall time over the star-formation efficiency $t_{\rm ff}(n_{\rm H,th}) /\epsilon_{\rm sf} \approx$ 0.1 Gyr. 
Note that the value of $t_{\rm dep, sf}$ merely reflects our choice of $n_{\rm H,th}$ and therefore is not a prediction of our model.
The fact that the $\Sigma_{\rm sfg}$-$\Sigma_{\rm SFR}$ relation is approximately linear while the $\Sigma_{\rm gas}$-$\Sigma_{\rm SFR}$ relation shows a much steeper slope is in fact already foreseeable in Fig. \ref{fig:profile_SFRsc_3by1},
as $\Sigma_{\rm SFR}$ and $\Sigma_{\rm sfg}$ show similar radial gradients much steeper than $\Sigma_{\rm gas}$.
The steep slope of the $\Sigma_{\rm gas}$-$\Sigma_{\rm SFR}$ relation can be viewed as a dilution effect, as only a fraction of $\Sigma_{\rm gas}$ is actually participating in star formation.

The relation between $\Sigma_{\rm H_2}$ and $\Sigma_{\rm SFR}$ is also approximately linear. 
However, defining the H$_2$ depletion time, $t_{\rm dep, H_2} \equiv \Sigma_{\rm H_2} / \Sigma_{\rm SFR}$, is not particularly useful and somewhat misleading,
because the H$_2$ is not necessary for star formation,
at least in our adopted star formation model.
Moreover,
the H$_2$ is not a good tracer for the star-forming gas as shown in Fig. \ref{fig:h2_eq_vs_noneq_alpha_nH} and Fig. \ref{fig:whereH2_newIC}.
A small $t_{\rm dep, H_2}$ gives the wrong impression that the star formation in H$_2$ is very efficient,
though in reality it is simply due to the low H$_2$ fraction 
\textbf{
(under the assumption that H$_2$ is not a necessary condition for star formation).
}

Comparing \textit{G1D01}, \textit{G1D001} and \textit{G01D01} in Fig. \ref{fig:KS_pix05_compare},
$\Sigma_{\rm H_2}$ varies a lot horizontally while $\Sigma_{\rm sfg}$ is almost unchanged.
Again, this indicates that the H$_2$ formation is much more sensitive to the variation of $G_0$ and $D$ than the thermal properties of the gas (and hence star formation).
In the no-feedback run \textit{G1D01\_noFB}, 
the total gas KS-relation is orders of magnitude above the solid line,
which is obviously inconsistent with observations in this regime.

\begin{figure*}
	\centering
	\includegraphics[trim = 10mm 10mm 0mm 0mm, clip, width=7in]{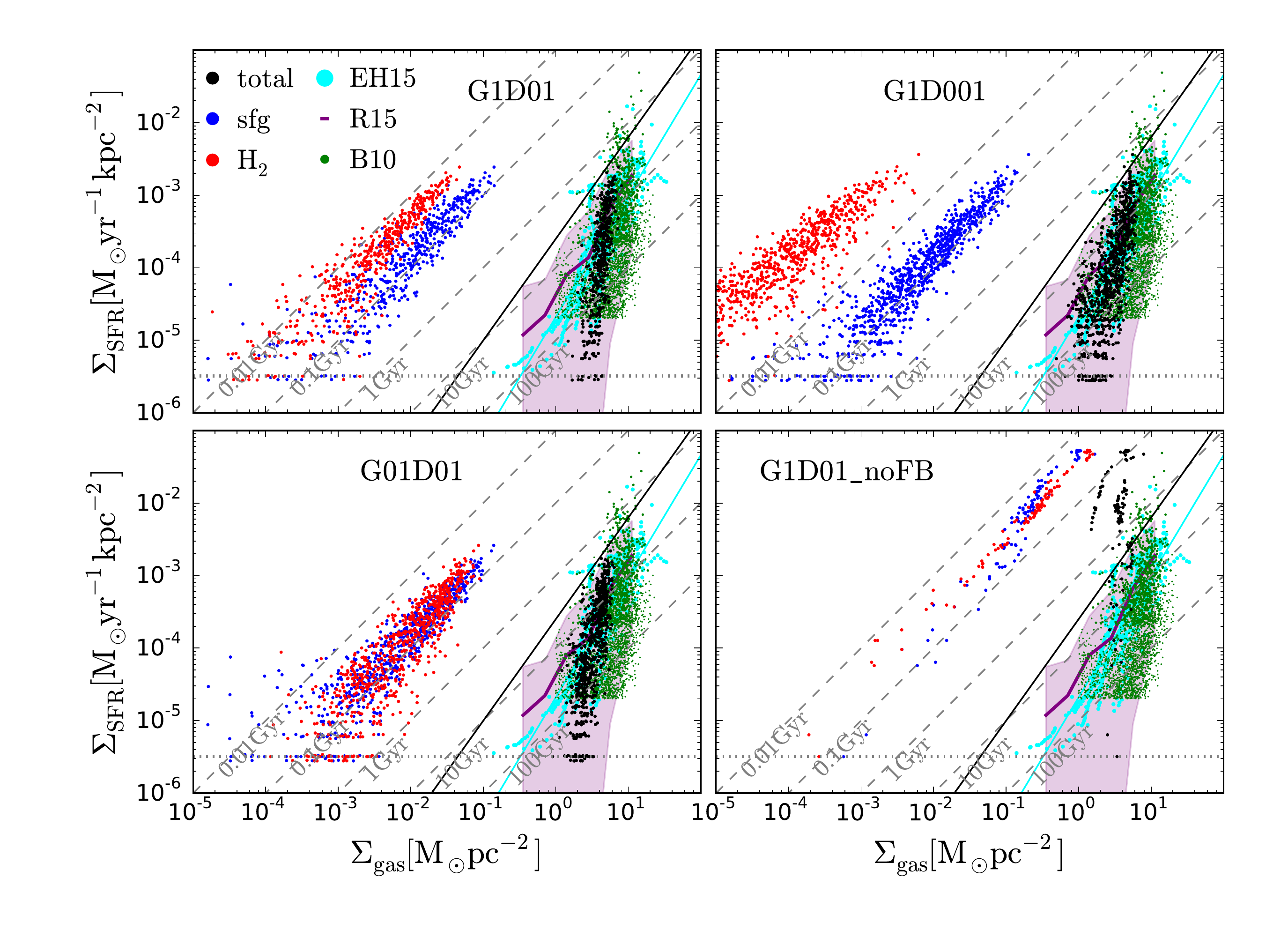}
	\caption{
		Star formation rate surface density vs. surface density of the total (black), cold (blue) and H$_2$ (red) gas, measured with aperture size $l_{\rm ap}$ = 500 pc.
		The results from $t$ = 250 Myr to $t$ = 1 Gyr with a time interval of 1 Myr are shown
		(only a randomly chosen 4\% are shown for clarity).
		The solid line represents the conventional KS-relation and the dashed lines represent five different gas depletion times $t_{\rm dep}$ = 0.01, 0.1, 1, 10 and 100 Gyr.
		The horizontal dotted lines are the minimum surface density allowed by our particle mass.
		The green and cyan dots are observational results from \citet{2010AJ....140.1194B} and \citet{2015arXiv150304370E}, respectively.
		The cyan line is a fit to all the cyan points.
		The purple line is the average value from \citet{2015MNRAS.449.3700R} and the filled area encloses the fifth and ninety-fifth percentile.
		The steep slope in the total gas KS-relation is due to the decreasing cold gas fraction at low gas surface density.}
	\label{fig:KS_pix05_compare} 
\end{figure*}

\subsubsection{The effect of aperture size}\label{sec:KS}
In Fig. \ref{fig:KS_vary_pix_g1d01} we show the relation between $\Sigma_{\rm sfg}$ and $\Sigma_{\rm SFR}$ in \textit{G1D01} with four different aperture sizes ($l_{\rm ap}$ = 1000 pc, 500 pc, 200 pc and 100 pc from left to right).
Results from $t$ = 250 Myr to $t$ = 1 Gyr with a time interval of 1 Myr are over-plotted in each panel (only a randomly chosen 4\% are shown for clarity).
The y-axis in the top row $\Sigma_{\rm SFR,gas}$ is the instantaneous star formation rate of the gas particles, 
while in the bottom row $\Sigma_{\rm SFR}$ is the star formation rate calculated by counting newly formed (age $<$ 5 Myr) stars (see Section \ref{sec:sfrmodel}).
The horizontal and vertical dashed lines show the lower limit for $\Sigma_{\rm SFR}$ and $\Sigma_{\rm sfg}$, respectively, for the given $l_{\rm ap}$.

The top row of Fig. \ref{fig:KS_vary_pix_g1d01} shows a very tight $\Sigma_{\rm sfg}$-$\Sigma_{\rm SFR,gas}$ relation that is very close to linear.
The relation remains tight when $l_{\rm ap}$ shrinks.
This suggests that the underlying cold gas density distribution does not change much from pixel to pixel even at the scale of 100 pc.

As discussed in Section \ref{sec:sfrmodel},
the instantaneous star formation rate of gas is not an observable and is only a conveniently defined input quantity for our model to estimate the future star formation activities.
If we look at how many young stars really formed in the simulations (the bottom row of Fig. \ref{fig:KS_vary_pix_g1d01}),
which is observable,
the relation becomes less tight, with a trend of increased scatter at lower $\Sigma_{\rm sfg}$.
The overall scatter increases significantly as $l_{\rm ap}$ shrinks.
This is due to the time evolution effect of the star formation process that is discussed in detail in \citet{2010ApJ...722.1699S} and \citet{2014MNRAS.439.3239K}.
This phenomenon would not be visible if we were to look at the instantaneous star formation rate.

\begin{figure*}   
	\begin{minipage}[b]{1.\linewidth}
		\begin{center}
			\includegraphics[trim = 10mm 0mm 0mm 0mm, clip, width=190mm]{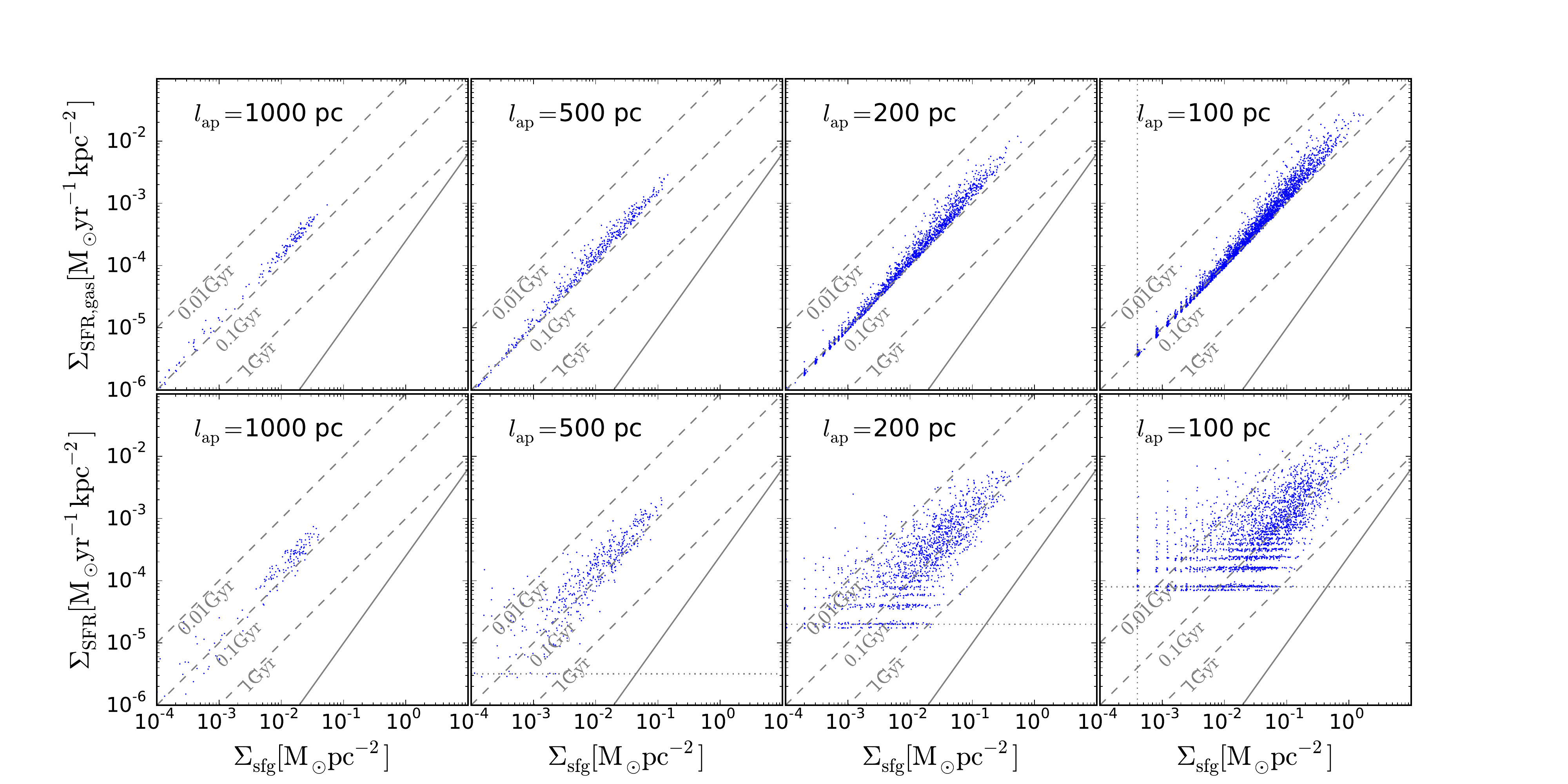} 
		\end{center}
		\caption{
			The relation between $\Sigma_{\rm sfg}$ and $\Sigma_{\rm SFR}$ in \textit{G1D01} with four different aperture sizes ($l_{\rm ap}$ = 1000 pc, 500 pc, 200 pc and 100 pc from left to right). Only a randomly chosen 4\% are shown for clarity. In the top panels $\Sigma_{\rm SFR,gas}$ are the instantaneous star formation rate from the gas particles while in the bottom panels $\Sigma_{\rm SFR}$ are obtained by counting the young stars formed in the last 5 Myr. 
			The vertical and horizontal dotted lines are the minimum surface density allowed by particle mass.
		} \label{fig:KS_vary_pix_g1d01}
	\end{minipage}
\end{figure*}


\section{Discussion}\label{sec:discussion}

\subsection{Non-equilibrium H$_2$ Formation}
In dwarf galaxies,
the timescale for H$_2$ formation is longer than that in normal spiral galaxies because both 
the densities and the DGR are lower (Eq. \ref{eq:timeH2form}).
Due to the long chemical timescales compared to the dynamical times,
the gas is far out of chemical equilibrium locally (cf. Fig. \ref{fig:h2_eq_vs_noneq_alpha_nH}).
It is therefore essential to incorporate a time-dependent chemistry network in simulations to correctly predict the spatial distribution of H$_2$.
On galactic scales,
as there are both significant amounts of H$_2$ over- and under-abundant gas, the total H$_2$ mass is only slightly over-predicted by the equilibrium prediction (cf. Fig. \ref{fig:newIC_sfr_sc_h2_time}).
However,
$F_{\rm H_2,eq}$ shown in Fig. \ref{fig:newIC_sfr_sc_h2_time} is calculated using the self-shielding factor that is obtained from the non-equilibrium $f_{\rm H_2}$.
If one uses an equilibrium recipe in simulations,
the H$_2$ fraction is likely to be even more severely over-estimated than in Fig. \ref{fig:newIC_sfr_sc_h2_time} because of the nonlinearity that enters Eq. \ref{eq:xeq} via $f_{\rm sh}$.
Therefore,
a time-dependent chemistry network is still necessary even if one is merely interested in the total H$_2$ mass in a galaxy.

\subsection{Star Formation in HI-dominated Gas}
Fig. \ref{fig:h2_eq_vs_noneq_alpha_nH} and \ref{fig:whereH2_newIC} demonstrate that the reservoir for star formation is not necessarily H$_2$-dominated (and vice versa),
which means that the correlation between star formation and H$_2$ breaks down in our simulated dwarf galaxy.
This is in line with \citet{2012MNRAS.426..377G} who simulated an isolated $10^4$ M$_\odot$ cloud with much higher resolution ($m_{\rm gas} = 5\times 10^{-3}$ M$_\odot$, $N_{\rm ngb}$ = 100).
On the other hand,
\citet{2012ApJ...759....9K} compared the timescales of H$_2$ formation, radiative cooling, and gravitational free-fall,
and predicted the correlation would break down only when the metallicity is below a few percent of Z$_{\odot}$, which is much lower than Z = 0.1 Z$_{\odot}$ adopted in this work.
One obvious reason for the discrepancy is the assumption of a linear Z-DGR relation in \citet{2012ApJ...759....9K},
as $R_{\rm form}$ mainly depends on $D$ and not on Z per se.
However, the star-forming gas is HI-dominated even in \textit{G1D01} and \textit{G01D01} where we do assume a linear Z-DGR relation.
One possible explanation is that feedback-driven turbulence constantly disrupts high density clouds where most H$_2$ formation takes place.
Indeed, the star-forming gas becomes H$_2$-dominated if we switch off stellar feedback (\textit{G1D01\_noFB}).
In addition,
\citet{2012ApJ...759....9K} boosted the H$_2$ formation rate coefficient, $R_{\rm form}$, by a clumping factor $C$ = 10 to account for the density distribution below the scale the density is defined.
The value of $C$ is however very uncertain, and the prescription that simply takes it as a multiplication factor for $R_{\rm form}$ is likely to over-predict the H$_2$ formation rate \citep{2012MNRAS.421.2531M}.
On the other hand, 
our simulations do not adopt the clumping factor (or $C$ = 1, equivalently) and thus may under-estimate the H$_2$ fraction if there is significant clumping below our resolution scales.
In the turbulent ISM picture, the gas is thought to be clumpier at large scales and more uniform at small scales.
\citet{2012ApJ...746..135M} conducted high resolution ($\approx$ 0.1 pc) turbulence simulations and reported an effective clumping factor $C \approx$ 2 for turbulent clouds at the length scale of 10 pc.
Since our effective spatial resolution is about 2 pc,
we do not expect that the H$_2$ fraction in our simulations would be severely under-estimated.

In massive spiral galaxies,
the use of H$_2$-dependent star formation recipes in hydrodynamical simulations is justified (though probably unnecessary) since most star-forming gas is also expected to be H$_2$-dominated.
In dwarf galaxies, however,
the star-forming gas is not necessarily H$_2$-dominated and vice versa.
The H$_2$-dependent star formation recipes would therefore be inappropriate.
This is especially true when the DGR falls below the linear Z-DGR relation.
The star formation rate in our simulated galaxy would be much lower if we restricted star formation to occur only in H$_2$
\footnote{
\textbf{
More precisely,
if we suddenly switch to a H$_2$-dependent star formation model during our simulations,
the star formation rate would be much lower for a short period of time, which could lead to further collapse of the gas (but see \citealp{2009ApJ...707..954P}).
The system may still eventually reach a new quasi-equilibrium state with a much clumpier density structure that cannot be resolved with our resolution.
}
}.

\subsection{The KS-relation for Dwarf Galaxies}

Observations of nearby star-forming galaxies have shown a roughly constant gas depletion time $t_{\rm dep} \approx$ 1 Gyr at $\Sigma_{\rm gas} \gtrsim$ 10 M$_\odot$ pc$^{-2}$,
and a clear transition to much longer $t_{\rm dep}$ at $\Sigma_{\rm gas} \lesssim$ 10 M$_\odot$ pc$^{-2}$, the regimes where most dwarf galaxies reside (see e.g., \citealp{2008AJ....136.2846B, 2010AJ....140.1194B, 2011AJ....142...37S}).
The transition leads to a steepening of the slope in the KS-relation, often referred to as the star formation threshold (not to be confused with $n_{\rm H,th}$).
Our simulations suggest that 
such a transition does not correspond to the H$_2$-to-HI transition, as the H$_2$ fraction differs a lot while the KS relation for the total gas remains almost unchanged in different runs with feedback.
Instead, this is a consequence of the dilution effect,
as only a fraction of gas is cold and dense enough to form stars.
Even though the mid-plane density decreases at almost the same rate as $\Sigma_{\rm gas}$,
the cold gas fraction also decreases with $\Sigma_{\rm gas}$ because a smaller fraction of gas would be dense enough to cool.
If the density distribution of the cold gas does not change its shape but only its normalization factor from pixel to pixel,
the $\Sigma_{\rm sfg}$-$\Sigma_{\rm SFR}$ relation should be linear,
and so the $\Sigma_{\rm gas}$-$\Sigma_{\rm SFR}$ would be super-linear.
This dilution effect of the warm gas would not be visible at higher $\Sigma_{\rm gas}$ where the system is already cold-gas dominated (by mass), as is the case in most spiral galaxies.
Our simulations do not cover high enough $\Sigma_{\rm gas}$ to see the transition,
but we can speculate that the transition from a linear KS-relation to a much steeper slope corresponds to a transition from a cold-gas dominated system to a warm-gas dominated one (by mass).


On the other hand,
the slopes of the $\Sigma_{\rm gas}$-$\Sigma_{\rm SFR}$ relation shown in Fig. \ref{fig:KS_pix05_compare} are probably steeper than observed due to our simplistic assumption of constant $G_0$.
Physically,
the strength of the diffuse ISRF is expected to scale with the average star formation rate,
and therefore should also decrease as $R$ increases until it drops to the level of the extragalactic UV background.
However, it is unclear what spatial scales are appropriate to define the averaged star formation rate that contributes to the local $G_0$.
Our assumption of constant $G_0$ is likely to be a severe over-estimation at large radii.
As discussed in Section \ref{sec:profile},
this leads to an abrupt truncation at $R_{\rm cool}$,
which in turn leads to very steep slopes in Fig. \ref{fig:KS_pix05_compare}.
In reality, 
$G_0$ should decreases as $R$ increases,
and the cold gas fraction would not drop as abruptly as $R$ increases.
This is because while the mean density is lower at large $R$,
$n_{\rm cool}$ is also lower
as photo-electric heating is weaker.
The gas would therefore still be able to cool below $10^4$ K at much larger radii up to the point where the gas is in thermal equilibrium at $10^4$ K due to the extragalactic UV background \citep{2004ApJ...609..667S} or it has a cooling time longer than the Hubble time.
As such, the slope in the KS-relation should be flatter than in Fig. \ref{fig:KS_pix05_compare}.
Note that the slope in Fig. \ref{fig:KS_pix05_compare} is slightly flatter for the runs with less efficient photo-electric heating,
as the SN feedback which controls the cold gas fraction in these runs naturally scales with $R$ in the inner region.
Our result is consistent with \citet{2015arXiv150608829R} who simulated slightly more massive isolated galaxies with a more sophisticated chemistry network and found that a weaker ISRF leads to a flatter slope and a lower threshold surface density.
It is also in line with \citet{2010AJ....140.1194B} where they suggested an "S-shape" KS-relation where the slope flattens at disc outskirts.
Interestingly,
\citet{2015MNRAS.449.3700R} reported a much flatter slope of 1.5 for the average SFR in a given gas surface density bin,
arguing that the individual measurement can be severely biased at low surface densities due to the stochasticity of star formation and thus only the average SFR can be faithfully obtained.

\subsection{Spatial Variations of ISRF and DGR}
Our assumption of constant a $G_0$ and $D$ is obviously an over-simplification.
As already discussed,
$G_0$ is expected to scale with the averaged star formation rate. 
Moreover,
at smaller scales
in the vicinity of the young stars,
the radiation field would easily be orders of magnitude stronger than the diffuse background,
forming the photon-dominated regions (PDRs) around young stars which is not included in our model.
Although the radiation field would drop rapidly away from the sources (geometrical effect),
it can still potentially destroy a substantial amount of H$_2$,
which makes the H$_2$ fraction even lower.
A more sophisticated radiative transfer is required to properly model the small-scale spatial variations of $G_0$.

On the other hand,
the assumption of a constant DGR might actually be reasonable as the observed dust in our Galaxy seems to be well mixed with gas \citep{1996A&A...312..256B}.
Though locally dust is continuously created and destroyed via various physical processes \citep{2014A&A...562A..76Z,2015MNRAS.449.3274F},
this inhomogeneity can be effectively smoothed out through turbulent mixing.
A dust evolution model is required to investigate such effects.

\section{Summary and Conclusions}\label{sec:summary}

We have conducted high-resolution ($m_{\rm gas}$ = 4 M$_\odot$, $N_{\rm ngb}$ = 100) hydrodynamical simulations for an isolated star-forming dwarf galaxy run for 1 Gyr.
Our model includes self-gravity, non-equilibrium cooling and chemistry, shielding from \textbf{a uniform and constant} ISRF, star formation, stellar feedback and metal enrichment self-consistently.
We have investigated the physical properties of the ISM in low metallicity environments on galactic scales. 
Our main results can be summarized as follows:

\begin{itemize}
\item
The ISM in dwarf galaxies is dominated by the warm gas (100 K $< T \leqslant 3\times 10^4$ K) both in mass and in volume (Fig. \ref{fig:VMfrac_time}).
The hot gas ($T > 3\times 10^4$ K) occupies about 10\% of volume but contributes little of the mass ($\approx$ 0.1\%).
The cold gas ($T <$ 100 K) contributes up to 1\%-10\% of the mass but occupies little volume ($<$ 0.1\%)

\item
We found SN-driven galactic outflows in our simulated galaxy.
The mass-loading measured at $|z| = \pm$ 2 kpc is slightly larger than unity.
A small fraction of outflowing gas at $|z| = \pm$ 2 kpc eventually escapes the halo.
The outflowing gas is slightly enriched:
the metallicity in the halo (Z$_{\rm halo}$) is about 20\% higher than that in the disc (Z$_{\rm disc}$).

\item
The ISM in dwarf galaxies is a hostile environment for H$_2$ formation
because of its low DGR, strong ISRF, and low densities in general.
The H$_2$ mass fraction of the galaxy is very low (less than 0.1\%) in all runs except for \textit{G1D01\_noFB} (Fig. \ref{fig:newIC_sfr_sc_h2_time}).
The H$_2$ mass fraction is very sensitive to the variations of $G_0$ and $D$
while the gas thermal properties and hence the total star formation rate are insensitive to both $G_0$ and $D$.

\item
Feedback-driven turbulence keeps the gas from cooling below $10^4$ K at $n_{\rm H} \lesssim$ 1 cm$^{-3}$ by reducing the local dynamical time to be smaller than the cooling time.
Once the gas cools below $10^4$ K it would be shocked heated to $10^4$ K in a dynamical time,
and so the gas is out of thermal equilibrium in the presence of SN feedback (Fig. \ref{fig:PD_plus_Coolrates}).
SN feedback (which does not depend explicitly on both $G_0$ and $D$) therefore determines the gas distribution in the phase diagram and regulates star formation.
However, the size of the star-forming region $R_{\rm cool}$ is controlled by the photo-electric heating (Fig. \ref{fig:profile_SFRsc_3by1}), as it controls the ability to cool and form stars in the first place.

\item
Since the chemical timescales are much longer than the local dynamical times,
the gas is far out of chemical equilibrium locally (Fig. \ref{fig:h2_eq_vs_noneq_alpha_nH}).
The star-forming gas is HI-dominated in all runs except for \textit{G1D01\_noFB}.
There is also a significant amount of H$_2$-dominated gas that resides in the diffuse and warm phase and that is not star forming (Fig. \ref{fig:whereH2_newIC}).
This suggests that the correlation between H$_2$ and star formation breaks down at low metallicity.


\item
The scale height of the total gas increases with $R$.
On the other hand,
most of the cold gas (T $<$ 100 K) and H$_2$ are confined within a thin layer ($|z| <$ 0.1 kpc) in the mid-plane with no sign of flaring (Fig. \ref{fig:SH_3by1}).
However,
the radial gradient of the mid-plane density $n_0$ is only slightly steeper than $\Sigma_{\rm gas}$,
both of which are much shallower than the radial gradient of $\Sigma_{\rm SFR}$ (Fig. \ref{fig:profile_SFRsc_3by1}).
Therefore, disc flaring does not account for the steep slopes in the $\Sigma_{\rm gas}$-$\Sigma_{\rm SFR}$ relation (Fig. \ref{fig:KS_pix05_compare}).

\item
The cold gas fraction decreases with the mid-plane density because a lesser fraction of gas is dense enough to cool and form cold gas if the mid-plane (mean) density is lower.
Meanwhile,
the cold gas density distribution does not vary much from region to region, and so a constant fraction of the cold gas will be star-forming.
Therefore, the steep slopes in the KS-relation are due to the dilution effect of warm gas: only a small fraction of gas is forming stars and this fraction decreases as $\Sigma_{\rm gas}$ decreases.


\end{itemize}

\section*{Acknowledgements}
We thank the referee for the useful and constructive comments on the paper.
We thank Bruce Elmegreen and Sambit Roychowdhury for providing the observational data from \citet{2015arXiv150304370E} and \citet{2015MNRAS.449.3700R}.
TN acknowledges support from DFG priority program SPP 1573 ``Physics of the interstaller medium''.
SW acknowledges support by the DFG via SFB 956 ``Conditions \& impact of star formation'' and SPP 1573, as well as by the Bonn-Cologne Graduate School.
SCOG acknowledges financial support from the Deutsche Forschungsgemeinschaft  via SFB 881, 
``The Milky Way System'' (sub-projects B1, B2 and B8) and SPP 1573, ``Physics of the Interstellar Medium''
(grant number GL 668/2-1). SCOG further acknowledges support from the European Research Council under 
the European Community’s Seventh Framework Programme (FP7/2007-2013) via the ERC Advanced Grant 
STARLIGHT (project number 339177).

\bibliographystyle{mn2e}
\bibliography{dwf}   

\appendix

\section{Energy conservation in pressure-based SPH}\label{app:pesph}

We investigate the energy conservation property of two different SPH formulations: pressure-entropy SPH and pressure-energy SPH \citep{2013MNRAS.428.2840H}.
We model a non-radiative blastwave in a uniform medium with the number density $n$ = 100 cm$^{-3}$ and the initial temperature $T$ = 1000 K.
A total energy of $10^{51}$ erg is injected into the neighboring 100 particles in terms of thermal energy and is distributed by the smoothing kernel.
The radiative cooling is turned off.
A global timestep is used to ensure that the violation of energy conservation is not due to the adaptive timesteps.
The particle mass is 1 M$_{\odot}$.

In Fig. \ref{fig:Sedov} we show the time evolution of the total energy (black), thermal energy (red) and kinetic energy (blue).
The solid and dotted lines represent the results of using pressure-energy SPH and pressure-entropy SPH, respectively.
The horizontal dashed lines are the exact solution for a non-radiative blastwave, the so-called Sedov solution \citep{1959sdmm.book.....S}: the total energy is conserved and the fraction of thermal and kinetic energy are about 73 \% and 27 \%, respectively.

When using pressure-energy SPH, the total energy is conserved and the thermal and kinetic energy converge to the exact solution.
On the other hand,
when using pressure-entropy SPH, the total energy increases over time by up to about 10\% and then gradually decreases even if a global timestep is adopted.
This is because when converting entropy into energy in pressure-entropy SPH,
an estimate of density is required.
There are two different ways to estimate the density: the entropy-weighted density $\rho^{e}$
and the traditional mass-weighted density $\rho^{m}$ (see e.g. \citealp{2014MNRAS.443.1173H}).
In the Lagrangian formulation \citep{2013MNRAS.428.2840H}, $\rho^{e}$ is a natural choice for the density estimate.
Choosing $\rho^{m}$ leads to inconsistencies and so compromises the conservation property in dissipationless systems.
However,
when dissipation (artificial viscosity) is included,
$\rho^{e}$ causes a large error at entropy discontinuities due to its entropy-weighting and thus also violates energy conservation \citep{2014MNRAS.443.1173H}.
This leads to a dilemma when converting entropy into energy in pressure-entropy SPH.
Therefore, in this work we adopt pressure-energy SPH which shows much better energy conservation property.

\begin{figure}
	\centering
	\includegraphics[trim = 0mm 0mm 0mm 0mm, clip, width=3.2in]{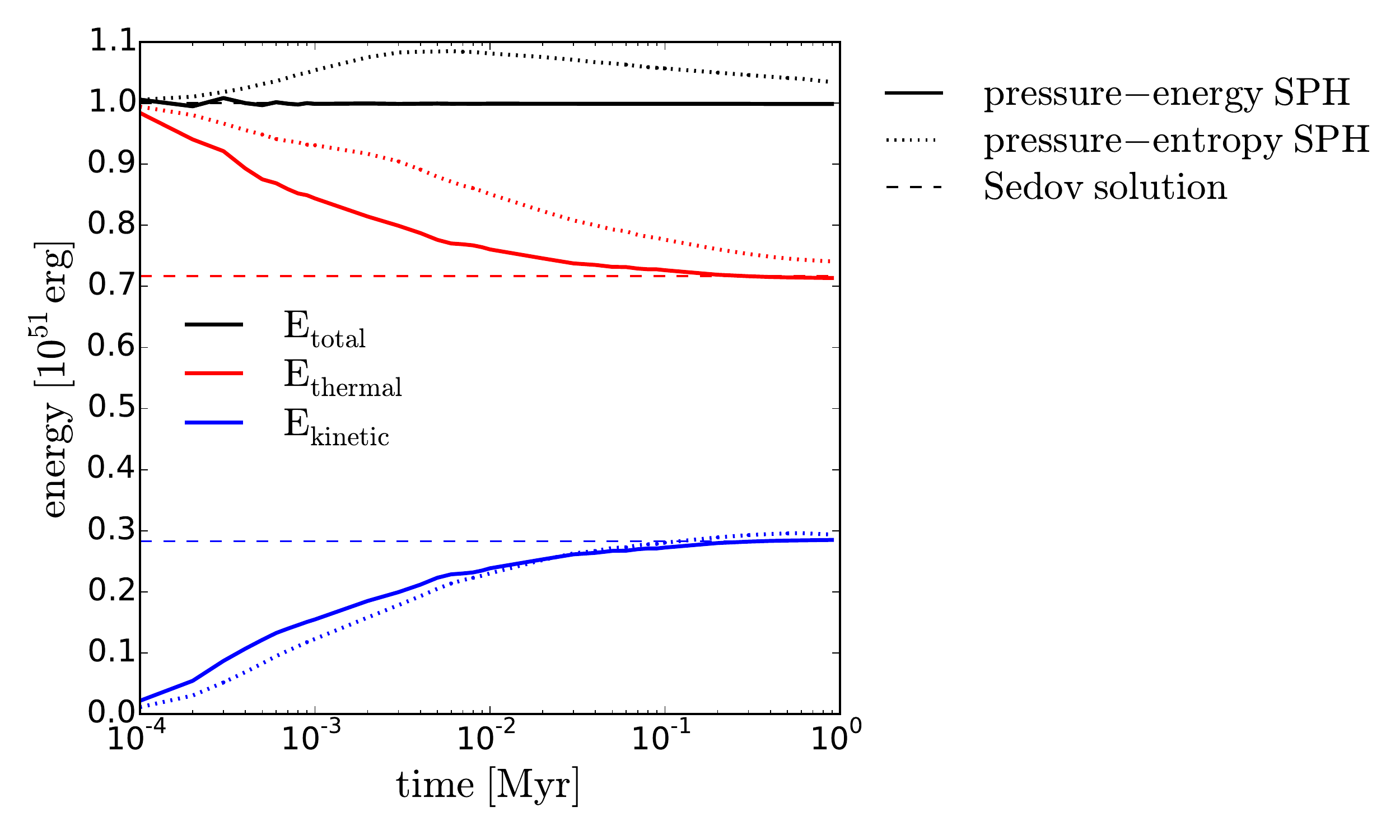}
	\caption{Time evolution of the total energy (black), thermal energy (red) and kinetic energy (blue) for a non-radiative blastwave in a uniform medium.
	The solid and dotted lines represent the results of using pressure-energy SPH and pressure-entropy SPH, respectively.
	The horizontal dashed lines are the exact solution.
	Pressure-energy SPH conserves total energy and the thermal and kinetic energy converge to the exact solutions, while pressure-entropy SPH suffers from cumulative error of energy conservation.}
	\label{fig:Sedov} 
\end{figure}

\section{Resolution Study of the Supernova Feedback} \label{app:fbtest}

In order to investigate what resolution is required to fully resolve the supernova feedback without suffering from the numerical over-cooling problem,
we perform a resolution study of a single supernova explosion in a uniform medium.
A total energy of $10^{51}$ erg is injected into the neighboring 100 particles in terms of thermal energy and is distributed by the smoothing kernel.
The medium has similar ISM properties as our fiducial run in Section \ref{sec:sim} with initial temperature $T$ = 1000 K.
In Fig. \ref{fig:SNtest_1pccm} we show the time evolution of the supernova feedback with five runs with different particle mass (0.01, 0.1, 1, 10 and 100 M$_\odot$).
Panel (a) shows the total linear momentum and panel (b) shows the total energy (in black), thermal energy (in red) and kinetic energy (in blue).
Panels (c) and (d) show the shell velocity and shell mass, respectively.
The shell mass is defined as the total mass of particles whose velocity $v >$ 0.1 km/s, i.e., the shock-accelerated particles.
The shell velocity is defined as the total momentum divided by the shell mass.
The medium number density $n$ = 1 cm$^{-3}$.

In the initial non-radiative phase (the so-called Sedov phase) the total energy is conserved, 
and the fraction of thermal and kinetic energy will be about 73 \% and 27 \%, respectively (see e.g. \citealp{2015MNRAS.451.2757W}), 
as we show by the two horizontal dashed lines.
If the kinetic energy remains constant, then as more material is swept up,
the total mass increases and the system gains momentum while decreasing its velocity.

The total energy is no longer conserved once the shell material starts to cool ($t \approx$ 0.07 Myr) and loses thermal energy.
In the limit of the complete removal of thermal energy, the shock will enter a momentum-conserving phase where the advance of the shock relies entirely on the inertia of the shell.
However, most of the shell material would first cool rapidly down to $T \lesssim 10^4$ K where the cooling rate drops significantly below the Lyman alpha peak, and then cools much more gradually afterwards.
Therefore,
there is first a rapid drop in the total energy ($t \approx$ 0.07 - 0.5 Myr) followed by a much more gradual one ($t \approx$ 0.5 - 5 Myr).
There is also a slight increase in thermal energy between $t \approx$ 0.5 - 5 Myr as the conversion from kinetic to thermal energy counteracts the cooling.
This residual thermal energy provides extra fuel for further momentum gain,
though not as efficient as in the Sedov phase.

The numerical over-cooling problem can be clearly seen in the energy evolution.
At worse resolution,
the conversion from thermal to kinetic energy becomes slower.
If the system starts to cool before the blastwave is fully developed,
the kinetic energy will be underestimated.
It is interesting to note that the total momentum is actually not as sensitive to resolution.
Even at our worst resolution the momentum evolution seems to agree with high resolution runs well.
However,
the swept-up mass differs quite dramatically.
The low resolution runs seem to sweep up too much mass due to their inability of resolving the thin shell.
The over-estimated mass gives rise to an under-estimated velocity and kinetic energy,
even though the total momentum generation is approximately correct.

Fig. \ref{fig:SNtest_100pccm} shows another test similar to Fig. \ref{fig:SNtest_1pccm} except in a denser medium of $n$ = 100 cm$^{-3}$.
The cooling time $t_{\rm cool}$ is about ten times shorter than in the $n$ = 1 cm$^{-3}$ medium, 
which is consistent with the scaling $t_{\rm cool} \propto n^{-9/17}$ reported in \citet{1998ApJ...500..342B}.
The general features are quite similar to the low-density case.
The momentum again shows only weak dependence on resolution,
except for the lowest resolution run, where the momentum is under-estimated by about a factor of two.
The energy (both kinetic and thermal) evolution shows a slightly more under-estimation for given resolution,
suggesting that better resolution is required to avoid numerical over-cooling in denser medium.

In summary,
the results suggest that for a $10^{51}$ erg SN explosion, it requires a resolution of about 1 M$_\odot$ to recover the reasonably converged velocities.
In the unresolved case where too much mass is swept up,
the ISM would only be accelerated to a much under-estimated velocity,
even if the right amount of momentum were generated.
This has a direct impact on the capability of driving galactic outflows as it is the outflowing velocity that determines whether a cloud is able to escape the potential well.

\begin{figure}
	\centering
	\includegraphics[trim = 0mm 0mm 0mm 0mm, clip, width=3.6in]{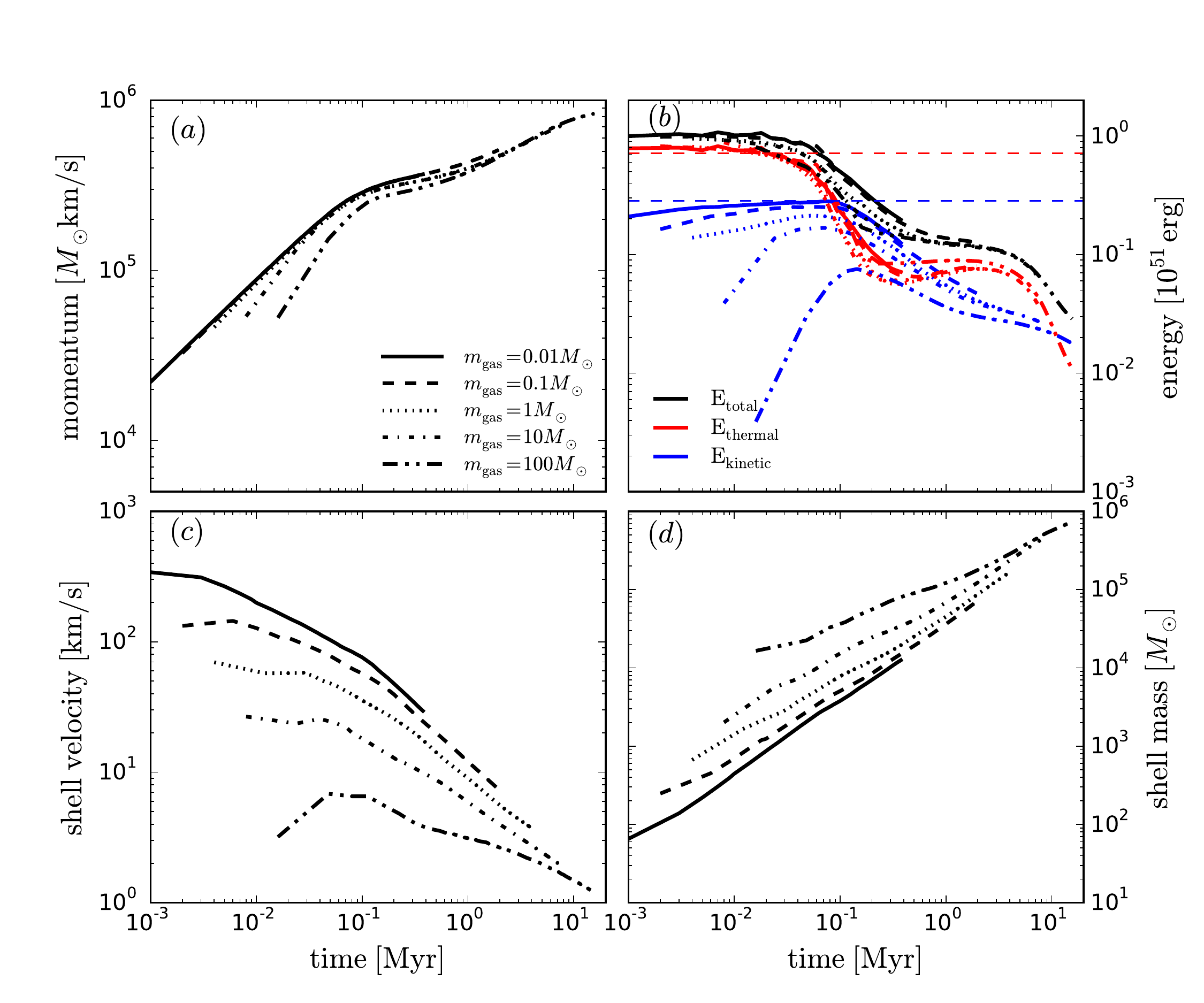}
	\caption{Evolution of a supernova remnant (SNR) of energy $E$ = $10^{51}$ erg with five different mass resolutions $m_{\rm gas}$ = 0.01 (solid), 0.1 (dashed), 1 (dotted), 10 (dash-dotted), and 100 (dash-double-dotted) M$_\odot$ in a medium of number density $n$ = 1 cm$^{-3}$.
	Panel (a): the total radial momentum; panel (b): the total energy (black), thermal energy (red) and kinetic energy (blue), with the dashed lines showing the energy partition (73 \% thermal and 27 \% kinetic) in the Sedov phase; panel (c): the shell velocity, defined as the total momentum divided by the shell mass; panel (d): the shell mass. The SNR is defined by all particles whose velocity $>$ 0.1 km/s.}
	\label{fig:SNtest_1pccm} 
\end{figure}
\begin{figure}
	\centering
	\includegraphics[trim = 0mm 0mm 0mm 0mm, clip, width=3.6in]{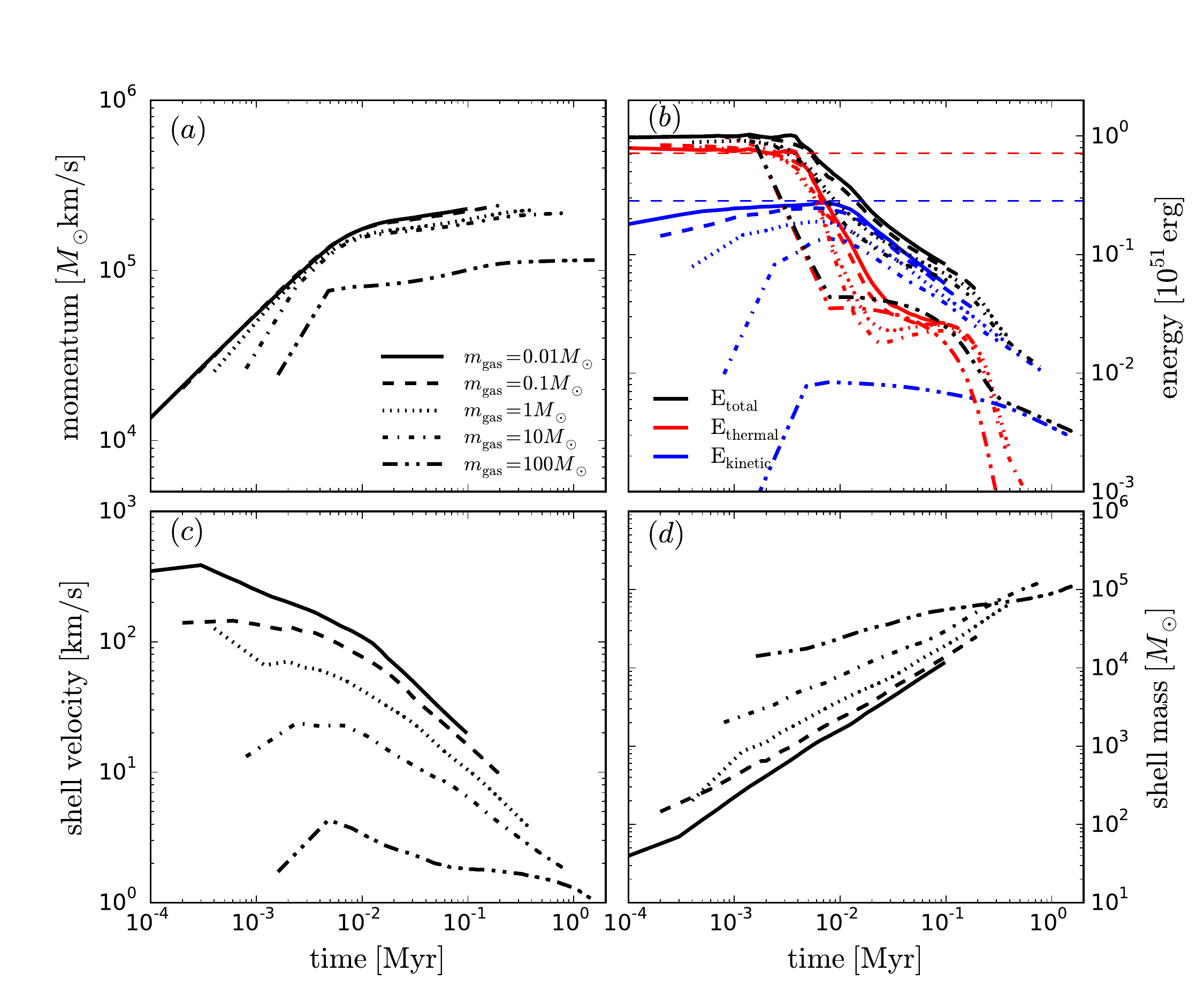}
	\caption{Same as Fig. \ref{fig:SNtest_1pccm} but in a medium of number density $n$ = 100 cm$^{-3}$.}
	\label{fig:SNtest_100pccm} 
\end{figure}

\section{Parameter dependence}

We explore the effects of varying the values of the physical parameters in our models.
Instead of starting from the initial conditions described in Sec. \ref{sec:IC}, all runs in this section are started from the snapshot at $t$ = 300 Myr of the \textit{G1D01} run.

\subsection{Star formation threshold density}\label{app:nth}

Fig. \ref{fig:nth} shows the time evolution of the total star formation rate (\textit{upper panel}) and the H$_2$ mass fraction (\textit{lower panel}) in the ISM ($R <$ 2 kpc and $|z| <$ 1 kpc),
with the star formation threshold density $n_{\rm th}$ = 100 (black), 10 (blue) and 1000 cm$^{-3}$ (red) respectively.
While the SFR is insensitive to the choice of $n_{\rm th}$,
the H$_2$ mass fraction does show significant differences with different $n_{\rm th}$,
as a higher $n_{\rm th}$ leads to a more clumpy ISM structure (broader density distribution) and more dense gas which increases the H$_2$ mass fraction.
However, even in the $n_{\rm th} = 1000 \: {\rm cm^{-3}}$ run,
the H$_{2}$ mass fraction remains very small, $F_{\rm H_{2}}
\sim 10^{-3}$.
We note that with $n_{\rm th}$ = 1000 cm$^{-3}$ the star-forming gas would be strongly unresolved and therefore is not an appropriate choice for our resolution.
Setting $n_{\rm th}$ = 10 cm$^{-3}$ makes the system even more H$_2$-poor,
though the star formation efficiency $\epsilon$ = 2\% may be too high for such a choice.

\begin{figure}
	\centering
	\includegraphics[trim = 0mm 0mm 0mm 0mm, clip, width=3.in]{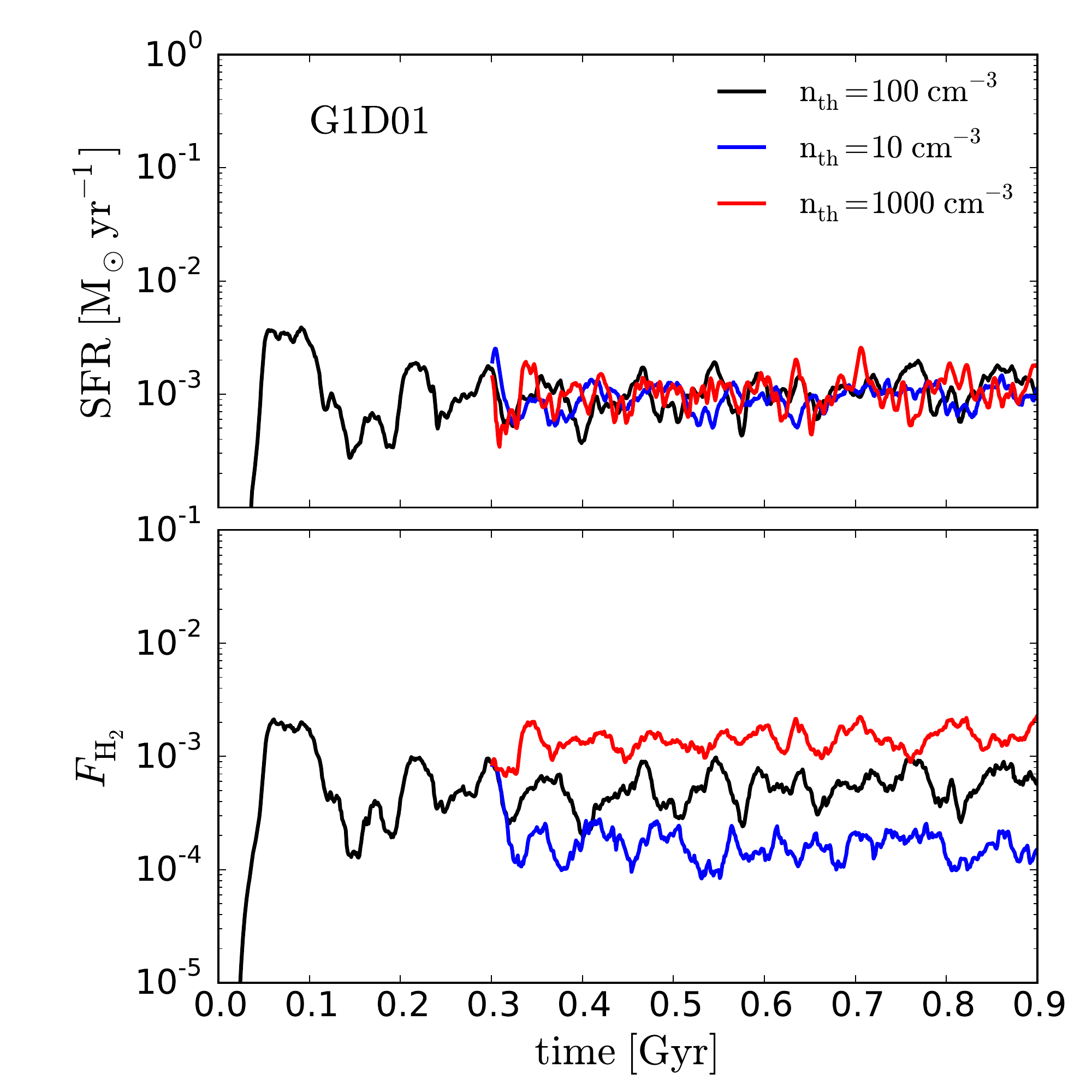}
	\caption{Time evolution of the total star formation rate (\textit{upper panel}) and the H$_2$ mass fraction (\textit{lower panel}) in the ISM ($R <$ 2 kpc and $|z| <$ 1 kpc) with the \textit{G1D01} model, with the star formation threshold density $n_{\rm th}$ = 100 (black), 10 (blue) and 1000 cm$^{-3}$ (red).
	}
	\label{fig:nth} 
\end{figure}

\subsection{Shielding length}\label{app:Lsh}

Fig. \ref{fig:Lsh} shows the time evolution of the total star formation rate (\textit{1st panel}), 
the H$_2$ mass fraction (\textit{2nd panel}),
the H$_2$ mass fraction for the dense ($n > n_{\rm th}$) gas (\textit{3nd panel}),
and the fraction of the H$_2$ mass that resides in the diffuse ($n < n_{\rm th}$) gas (\textit{4th panel})
in the ISM ($R <$ 2 kpc and $|z| <$ 1 kpc).
Different lines represent different choices of the shielding length $L_{\rm sh}$ = 20 pc (green), 50 pc (black), 100 pc (red), and 200 pc (blue), respectively.
None of these quantities are sensitive to the choice of $L_{\rm sh}$ (see the discussion in Section \ref{sec:shielding}).

\begin{figure}
	\centering
	\includegraphics[trim = 0mm 0mm 0mm 0mm, clip, width=3.in]{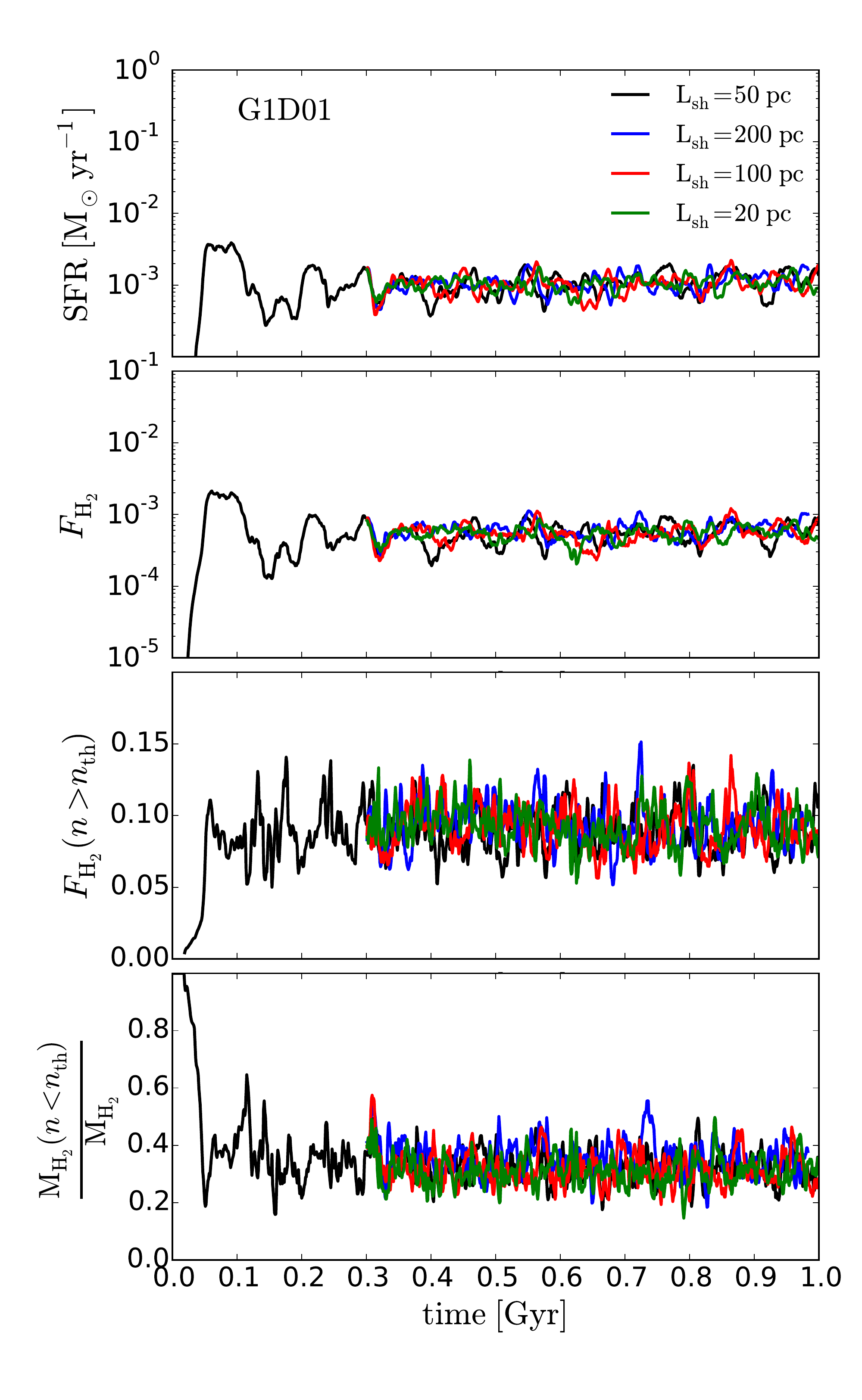}
	\caption{Time evolution of the total star formation rate (\textit{1st panel}), 
		the H$_2$ mass fraction (\textit{2nd panel}),
		the H$_2$ mass fraction for the dense ($n > n_{\rm th}$) gas (\textit{3nd panel}),
		and the fraction of the H$_2$ mass that resides in the diffuse ($n < n_{\rm th}$) gas (\textit{4th panel})
		in the ISM ($R <$ 2 kpc and $|z| <$ 1 kpc) with the \textit{G1D01} model.
		None of these quantities are sensitive to the choice of $L_{\rm sh}$.}
	\label{fig:Lsh} 
\end{figure}

\end{document}